\documentclass[preprint,12pt]{elsarticle}

\usepackage{hyperref}
\usepackage{graphicx}
\graphicspath{{pics/case/}{pics/method/}}
\usepackage{amsmath}
\usepackage{amssymb}
\usepackage[version=4]{mhchem}
\usepackage{siunitx}
\usepackage{longtable,tabularx}
\usepackage[caption=false]{subfig}
\usepackage{multirow}
\usepackage{threeparttable}
\usepackage{float}
\usepackage[top=2.5cm, bottom=2.5cm, left=2cm, right=2cm]{geometry}
\renewcommand{\vec}[1]{\boldsymbol{#1}}
\journal{}

\begin{document}

\begin{frontmatter}

\title{On the Applicability of the Gas-Kinetic Scheme with Kinetic Boundary Conditions for Near-Continuum Hypersonic Flows}

\author[a]{Wenpei Long}
\ead{wlongab@connect.ust.hk}
\author[a]{Junzhe Cao}
\ead{jcaobb@connect.ust.hk}
\author[a]{Yue Zhang}
\ead{yzhangnl@connect.ust.hk}
\author[a,b,c]{Kun Xu\corref{cor1}}
\ead{makxu@ust.hk}

\cortext[cor1]{Corresponding author.}

\address[a]{Department of Mathematics, Hong Kong University of Science and Technology, Hong Kong, China}
\address[b]{Department of Mechanical and Aerospace Engineering, Hong Kong University of Science and Technology, Hong Kong, China}
\address[c]{HKUST Shenzhen Research Institute, Shenzhen, 518057, China}

\begin{abstract}

         Rarefied gas effects are of critical importance for the aerodynamic performance of hypersonic vehicles operating at high altitudes. In these scenarios, conventional computational fluid dynamics (CFD) solvers break down as the linear constitutive relations underlying the Navier–Stokes equations cease to be valid. Based on direct modeling, the unified gas-kinetic scheme (UGKS) and the unified gas-kinetic wave-particle (UGKWP) method successfully capture non-equilibrium physics across all Knudsen numbers, yet they incur substantially higher computational costs than continuum solvers. Within the same kinetic framework, the gas-kinetic scheme (GKS) employs the Chapman–Enskog expansion for near-equilibrium flow physics and adopts the same kinetic boundary conditions as UGKS and UGKWP. This formulation naturally permits velocity slip and temperature jump, thereby extending the applicability of GKS into the slip and transitional regimes. By utilizing this natural kinetic slip boundary condition, the GKS provides a more physically faithful representation of non-equilibrium wall interactions than conventional CFD solvers equipped with Maxwell-type slip conditions, ultimately yielding more accurate aerodynamic predictions. To determine the applicability of the GKS in near-continuum flow regimes, we first examine a simple circular cylinder geometry, comparing surface quantities and distribution functions in detail. Furthermore, we investigate a
$9^\circ$ blunted cone, a $70^\circ$  blunted cone with a cylindrical sting, and the Apollo 6 command module. This analysis focuses on integrated aerodynamic predictions, which are validated against experimental data, Direct Simulation Monte Carlo (DSMC) simulations, and other kinetic methods.

\end{abstract}


\end{frontmatter}

\section{Introduction}\label{sec:intro}

Multiscale flows are frequently encountered in aerospace engineering, particularly during the design and optimization of re-entry spacecraft~\cite{ivanov1998hypersonic,boyd2017nonequilibrium}. Because the experimental replication of hypersonic, high-altitude flight conditions is both challenging and prohibitively expensive, numerical simulation has emerged as an efficient and cost-effective alternative. However, conventional computational fluid dynamics (CFD) methods based on the Navier–Stokes (NS) equations break down in rarefied flow regimes characterized by strong non-equilibrium phenomena~\cite{boyd2007hypersonic_continuum,bird1994molecular}. Conversely, while kinetic methods can accurately capture non-equilibrium physics, they must resolve high-dimensional gas distribution functions, making them computationally demanding. Consequently, systematically investigating the operational limits of these methods regarding both accuracy and efficiency is essential to guide the selection of appropriate numerical schemes for practical engineering applications.

Rooted in the direct modeling methodology~\cite{xu2015}, the unified gas-kinetic scheme (UGKS) was proposed as an effective multiscale framework capable of describing all flow regimes~\cite{xu2010unified}. Traditional kinetic methods, such as the Direct Simulation Monte Carlo (DSMC) method and the discrete velocity method (DVM)\cite{bird1970direct,broadwell1964study}, are often constrained by their reliance on operator splitting. In contrast, the UGKS overcomes these limitations by constructing the cell-interface flux from the integral solution of the kinetic equation. This formulation couples free transport and molecular collision within a single time step at the observation scale, naturally recovering the original gas-kinetic scheme (GKS)\cite{xu2001gas} in the near-continuum regime. To broaden its applicability, the UGKS has been extended to incorporate the thermal non-equilibrium of rotational and vibrational degrees of freedom~\cite{liu2014diatomic,wei2024adaptiveugks}, as well as chemically reacting flows~\cite{wei2026reactive}. 
However, deterministic velocity-space discretization still imposes prohibitive memory and computational overheads, particularly for high-speed, non-equilibrium flows. Consequently, significant research has focused on acceleration and memory-reduction techniques, including implicit iteration and multigrid convergence acceleration~\cite{zhu_implicit_2016,zhu2017multigrid,zhu_implicit_2019}, adaptive velocity-space meshes~\cite{chen2012moving,xiao2020velocity,wei2024adaptiveugks,long2024implicit}, and highly efficient, memory-saving programming paradigms~\cite{zhang2025memory}.

To alleviate the computational overhead and memory footprint of the UGKS, the unified gas-kinetic wave-particle (UGKWP) method was developed~\cite{zhu2019unified,liu2020unified,chen2020three,long2024nonequilibrium}. Within the same framework, the UGKWP method employs stochastic particles to efficiently represent the non-equilibrium portion of the distribution function. Following a similar developmental trajectory as the UGKS, the UGKWP method has been successfully extended to photon transport~\cite{li2020unified}, plasmas~\cite{liu2021unified,pu2025plasma,pu2026electromagnetic}, real gases~\cite{xu2021ugkwp_diatomic,wei2024diatomic}, gas-particle multiphase flows~\cite{yang2022ugkwp_disperse,yang2022ugkwp_twophase,yang2023ugkwp_fluidized,yang2024polydisperse}, radiative transport~\cite{liu2023ugkwp_radiative,yang2025ugkwp_radiation}, and phonon transport~\cite{liu2025ugkwp_phonon}. Recent work has also extended the UGKWP framework to turbulence modeling~\cite{yang2025waveparticle}. To further enhance computational efficiency, several studies have proposed adaptive wave-particle decomposition criteria~\cite{wei2023adaptive,cao2026adaptive}.

Despite their broad validity, kinetic multiscale methods incur a substantially higher computational cost than macroscopic solvers, such as the GKS~\cite{xu2001gas}, due to their explicit resolution of discrete, non-equilibrium distribution functions. Consequently, a rigorous comparison of these diverse computational approaches is essential to guide method selection in engineering practice. For flows in the slip regime, a common compromise is to employ a Navier–Stokes (NS) solver equipped with Maxwell boundary conditions, which successfully recover velocity slip and temperature jump at the wall. In the classical Maxwell boundary model~\cite{maxwell_1879}, the gas adjacent to the solid boundary is not assumed to be in thermal or mechanical equilibrium with the wall. Instead, a fraction of the incoming molecules undergo full momentum and energy exchange with the surface and are re-emitted in equilibrium with the wall, while the remaining fraction undergo specular reflection. This mechanism yields a non-zero slip velocity and a temperature jump at the surface, thereby extending the validity of continuum solvers. This kinetic-based formulation is highly compatible with gas-kinetic methods, which naturally describe gas behavior through statistical distribution functions. In the GKS, the kinetic boundary condition is constructed such that the incident gas distribution function is derived from the Chapman–Enskog (CE) expansion~\cite{li2005kinetic}, while the reflected distribution function is determined by the wall Maxwellian. Similarly, the UGKS and UGKWP methods apply the same kinetic boundary condition, though their incident distribution functions are represented in a discrete velocity space and by stochastic particles, respectively.

In this paper, we systematically investigate the applicability limits of the GKS equipped with kinetic boundary conditions for external flows. All numerical results are compared against those obtained from the UGKS or UGKWP under identical kinetic boundary conditions, as well as other kinetic methods and experimental data.  Notably, our findings reveal that even at an altitude of 85 km, where the GKS predicts a significantly sharper shock wave than the physical flow, it still yields highly reliable wall aerodynamic forces. This capability represents a remarkable advantage over classical NS solvers. 

The remainder of this paper is organized as follows. Section~\ref{sec:method} introduces the governing gas-kinetic theory, followed by the formulations of the GKS, UGKS, and UGKWP methods, as well as the implementation of the kinetic boundary condition. Section~\ref{sec:result} presents various numerical test cases and comparative analyses against the UGKS and UGKWP, supplemented by experimental data and DSMC simulations where appropriate. Finally, Section~\ref{sec:conclusion} concludes the paper by summarizing the applicability limits of the GKS equipped with the kinetic boundary condition.

\section{Numerical method}\label{sec:method}

In this study, the GKS serves as the near-continuum flow solver, while the UGKS and the UGKWP
method are briefly reviewed for comparison.
All three schemes share the same finite volume framework and construct the numerical flux from the
integral solution of the Bhatnagar--Gross--Krook (BGK) model at the cell interface.
They differ in the representation of the distribution function, where the GKS uses a continuous distribution function
based on the CE expansion for near-equilibrium flows, while the UGKS and the UGKWP use either a
discrete velocity space or a wave-particle decomposition to fully recover nonequilibrium flow physics.

\subsection{Gas kinetic theory}
In the gas-kinetic theory, the physics of the flow is described by the microscopic gas distribution
function $f(\vec{r}, \vec{u}, \boldsymbol{\xi}, t)$, where $\vec{r}=(x, y, z)$ is the location, $\vec{u}=(u, v, w)$ is
the particle velocity, and $\boldsymbol{\xi}$ represents the internal degrees of freedom.
The evolution of $f$ is governed by the Boltzmann equation with the BGK model
\begin{equation}\label{eq:BGK}
\frac{\partial f}{\partial t} + \vec{u} \cdot \nabla f = \frac{g - f}{\tau},
\end{equation}
where $\tau = \mu/p$ is the collision time and $g$ is the equilibrium state. The Maxwellian distribution is
\begin{equation}\label{eq:g}
	g = \rho \left(\frac{\lambda}{\pi}\right)^{\frac{K+3}{2}}
	e^{-\lambda\left({\vec{c}}^{2}+\vec{\xi}^{2}\right)},
\end{equation}
where $\rho$ is the density, $\lambda = m_0/(2k_B T)$ with $m_0$ and $k_B$ the molecular
mass and Boltzmann constant, $\vec{c} = \vec{u} - \vec{U}$ is the peculiar velocity,
$\vec{U}=(U, V, W)$ is the macroscopic fluid velocity, and
$\vec{\xi}=(\xi_1,\xi_2,\dots,\xi_K)$ accounts for the internal degrees of freedom with
$K = (5-3\gamma)/(\gamma-1)$ and $\gamma$ the specific heat ratio.

The collision term satisfies the compatibility condition
\begin{equation*}
	\int \frac{g - f}{\tau}\,\boldsymbol{\Psi}\,\mathrm{d}\Xi = 0,
\end{equation*}
where $\boldsymbol{\Psi} = \left(1, \vec{u}, \frac{1}{2}\left(\vec{u}^2+\vec{\xi}^2\right)\right)^T$ and $\mathrm{d}\Xi = \mathrm{d}\vec{u}\mathrm{d}\vec{\xi}$.
The macroscopic conservative variables $\vec{W}=(\rho, \rho U, \rho V, \rho W, \rho E)^T$
and the flux are
\begin{equation*}
	\vec{W} = \int f\,\boldsymbol{\Psi}\,\mathrm{d}\Xi, \qquad
	\vec{F} = \int (\vec{u} \cdot \vec{n})\, f\,\boldsymbol{\Psi}\,\mathrm{d}\Xi,
\end{equation*}
where $\vec{n}$ is the normal direction.

To recover the correct Prandtl number, the Shakhov model~\cite{shakhov1968} is used
in the collision term $(g^S - f)/\tau$ for all schemes below with the Shakhov equilibrium distribution
\begin{equation*}
	g^S = g\left[1 + (1-\mathrm{Pr})\, \frac{\vec{c} \cdot \vec{Q}}{5pRT}\,
	\left(\frac{|\vec{c}|^2}{RT} - 5\right)\right],
\end{equation*}
where $p = \rho RT$ and $\vec{Q}$ is the heat flux.

\subsection{Gas-kinetic scheme}

The presented GKS is constructed in the finite volume framework. The computational
domain is divided into cells $\Omega_i$, and the cell-averaged conservative
variables $\vec{W}_i$ are updated by
\begin{equation}\label{eq:FVM}
	\vec{W}_i^{n+1} = \vec{W}_i^n
	- \frac{\Delta t}{\Omega_i} \sum_{j \in N(i)} \vec{F}_{ij}\, \mathcal{A}_{ij},
\end{equation}
where $\vec{F}_{ij}$ is the numerical flux across the interface between cell $i$
and neighbor $j$, $\mathcal{A}_{ij}$ is the interface area, and $N(i)$ is the
set of neighbors.

Along the characteristic line, the integral solution of Eq.~\eqref{eq:BGK} at the
cell interface $\vec{r}_{ij}$ is
\begin{equation}\label{eq:integral}
	f(\vec{r}_{ij},t) = \frac{1}{\tau}\int_0^t e^{-(t-t')/\tau} g(\vec{r}',t')\,\mathrm{d}t'
	+ e^{-t/\tau} f_0(\vec{r}_{ij}-\vec{u}t), \quad \vec{r}' = \vec{r}_{ij} - \vec{u}(t-t'),
\end{equation}
where $\vec{r}'$ is the particle trajectory and $f_0$ is the initial distribution.

To fully recover the NS solution and ensure spatial and temporal
second-order accuracy, the initial distribution function $f_0$ is constructed
by combining the CE expansion with a Taylor expansion of the
Maxwellian distributions around the cell interface. In a local frame at the
cell interface, let $u_n = \vec{u}\cdot\vec{n}_{ij}$ be the normal particle
velocity, and $v_t$, $w_t$ be the two orthogonal tangential velocities.
Evaluated along the particle trajectory $\vec{r}_{ij}-\vec{u}t$, $f_0$ is
given by:
\begin{equation}\label{eq:GKS-f0}
	f_0(\vec{r}_{ij}-\vec{u}t) = \begin{cases}
		g^l\left[1 - t(a^l u_n + b^l v_t + c^l w_t) - \tau\bigl(a^l u_n + b^l v_t + c^l w_t + A^l\bigr)\right], & u_n > 0, \\[2pt]
		g^r\left[1 - t(a^r u_n + b^r v_t + c^r w_t) - \tau\bigl(a^r u_n + b^r v_t + c^r w_t + A^r\bigr)\right], & u_n \leq 0,
	\end{cases}
\end{equation}
where $g^l$ and $g^r$ are the Maxwellian equilibrium distributions at the left and right sides.
The microscopic derivatives $a^{l,r},b^{l,r},c^{l,r},A^{l,r}$ are obtained from the
reconstructed macroscopic variables at the respective sides~\cite{xu2001gas}.
The interface equilibrium distribution $g_0$ is determined by the compatibility
condition at the cell interface at $t=0$:
\begin{equation*}
	\int g_0 \boldsymbol{\Psi} \mathrm{d}\Xi = \int_{u_n>0} g^l \boldsymbol{\Psi} \mathrm{d}\Xi + \int_{u_n<0} g^r \boldsymbol{\Psi} \mathrm{d}\Xi,
\end{equation*}
which provides the macroscopic variables at the interface to construct $g_0$.
Using the Taylor expansion in space and time, the equilibrium state
$g$ evaluated at $\vec{r}' = \vec{r}_{ij}-\vec{u}(t-t')$ is expanded to first
order around the interface as
\begin{equation*}
	g(\vec{r}', t') = g_0\left[1 - (t-t')(\bar{a}\, u_n + \bar{b}\, v_t + \bar{c}\, w_t) + \bar{A}\, t'\right].
\end{equation*}
The tangential equilibrium derivatives $\bar{b},\bar{c}$
follow the kinetic weighted average, while the normal derivative
$\bar{a}$ is supplemented with a penalty term~\cite{zhang2023cgkssliding}
to suppress odd--even decoupling:
\begin{equation}\label{eq:GKS-abar}
	\int \bar{a}\, g_0\,\boldsymbol{\Psi}\,\mathrm{d}\Xi
	= \int_{u_n>0} a^l g^l \boldsymbol{\Psi}\,\mathrm{d}\Xi
	+ \int_{u_n<0} a^r g^r \boldsymbol{\Psi}\,\mathrm{d}\Xi
	+ \frac{\vec{W}^{\,r} - \vec{W}^{\,l}}{(\vec{x}_r - \vec{x}_l)\cdot\vec{n}_{ij}},
\end{equation}
where $\vec{x}_l,\vec{x}_r$ are the centroids of the left and right cells.
The temporal derivative $\bar{A}$ follows from the compatibility
condition~\cite{xu2001gas}.
To suppress spurious gradients in shock regions, a discontinuity feedback
factor is introduced~\cite{ji2021cgks3d,zhang2023cgkssliding}.
First, at each cell interface $ij$, a face-based discontinuity feedback factor
$\alpha_{ij}$ is computed from the pressure jump and Mach number differences:
\begin{equation*}
	\alpha_{ij} = \frac{1}{1 + A_{ij}^2}, \qquad
	A_{ij} = \frac{|p^l-p^r|}{p^l} + \frac{|p^l-p^r|}{p^r}
	       + \bigl(\mathrm{Ma}_n^l-\mathrm{Ma}_n^r\bigr)^2
	       + \bigl(\mathrm{Ma}_t^l-\mathrm{Ma}_t^r\bigr)^2,
\end{equation*}
where the pressure-difference terms capture discontinuities and the
squared Mach number differences improve robustness~\cite{zhang2023cgkssliding}.
Then, a cell-based factor $\alpha_i$ is synthesized as the product of the
face factors over all interfaces of cell $i$~\cite{ji2021cgks3d}:
\begin{equation*}
	\alpha_i = \prod_{j\in\mathcal{N}_f(i)} \alpha_{ij}, \qquad \alpha_i\in(0,1],
\end{equation*}
where $\mathcal{N}_f(i)$ denotes the set of faces of cell $i$.
The cell-averaged gradient is then compressed as
\begin{equation*}
	\widetilde{\nabla\vec{W}}_i = \alpha_i \, \nabla\vec{W}_i,
\end{equation*}
and the reconstruction at cell interfaces $ij$ subsequently yields the
compressed left and right gradients $\widetilde{\nabla\vec{W}}^{l,r}$,
from which the microscopic derivatives $a^{l,r},b^{l,r},c^{l,r}$ are evaluated.
Substituting these expansions (with $\bar{a}$ from
Eq.~\eqref{eq:GKS-abar}) into the integral solution, the second-order gas
distribution function at the cell interface becomes
\begin{equation}\label{eq:GKS-f-interface}
	\begin{aligned}
	f_{ij}(t) &= \left(1 - e^{-t/\tau}\right) g_0 + \left(t + \tau(e^{-t/\tau}-1)\right) \bar{A}\, g_0 \\
	&\quad + \left(t\, e^{-t/\tau} + \tau(e^{-t/\tau}-1)\right)\left(\bar{a}\, u_n + \bar{b}\, v_t + \bar{c}\, w_t\right) g_0 \\
	&\quad + e^{-t/\tau}\Bigl\{\left[1 - (t+\tau)(a^l u_n + b^l v_t + c^l w_t) - \tau A^l\right] H[u_n]\, g^l \\
	&\quad + \left[1 - (t+\tau)(a^r u_n + b^r v_t + c^r w_t) - \tau A^r\right]\bigl(1-H[u_n]\bigr)\, g^r\Bigr\},
	\end{aligned}
\end{equation}
where $H[u_n]$ is the Heaviside function defined by $H[u_n]=1$ when $u_n>0$
(particles from the left) and $H[u_n]=0$ when $u_n \leq 0$ (particles from the right).
The detailed formulations and determination of the microscopic derivatives can be
found in Refs.~\cite{xu2001gas,xu2015}.

The macroscopic flux $\vec{F}_{ij}$ is obtained by taking the moment of the
distribution function $f_{ij}(t)$ at the interface and averaging in time:
\begin{equation*}
	\vec{F}_{ij} = \frac{1}{\Delta t}\int_0^{\Delta t}
	\int (\vec{u}\cdot\vec{n}_{ij})\, f_{ij}(t)\, \boldsymbol{\Psi}\,\mathrm{d}\Xi\,
	\mathrm{d}t,
\end{equation*}
where $f_{ij}(t)$ is given by Eq.~\eqref{eq:GKS-f-interface}. The conservative
variables are updated by Eq.~\eqref{eq:FVM}.

\subsection{Unified gas-kinetic scheme}

Based on the direct modeling of the cell size and time step scale,
the unified gas-kinetic scheme (UGKS)~\cite{xu2010unified,xu2015} extends the
GKS from the continuum to all flow regimes by adopting a discrete velocity space.
As in the GKS, the interface flux is constructed from the same integral solution
Eq.~\eqref{eq:integral}; the essential difference is that the initial distribution
function $f_0$ is discretized in a velocity space for a more accurate representation
of the nonequilibrium distribution function.

At the cell interface, the time-dependent distribution function at discrete
velocity $\vec{u}_k$ is
\begin{equation}\label{eq:UGKS-integral}
	\hat{f}(\vec{r}_{ij}, t, \vec{u}_k) = \frac{1}{\tau}\int_{0}^t g(\vec{r}', t', \vec{u}_k)\,
	e^{-(t-t')/\tau}\,\mathrm{d}t'
	+ e^{-t/\tau}\, f_0\bigl(\vec{r}_{ij} - \vec{u}_k t, \vec{u}_k\bigr),
\end{equation}
where $\vec{r}' = \vec{r}_{ij} - \vec{u}_k(t-t')$ is the particle trajectory.
The initial distribution $f_0$ at the interface is reconstructed from the left
and right cells with second-order accuracy. With $u_{k,n} = \vec{u}_k \cdot
\vec{n}_{ij}$ and $v_{k,t}$, $w_{k,t}$ the two tangential components of
$\vec{u}_k$, $f_0$ along the trajectory $\vec{r}_{ij} - \vec{u}_k t$ is
\begin{equation*}
	f_{0,k}(\vec{r}_{ij} - \vec{u}_k t)
	= \bigl(f_k^l - t\,\vec{u}_k \cdot \nabla f_k^l\bigr)\, H[u_{k,n}]
	+ \bigl(f_k^r - t\,\vec{u}_k \cdot \nabla f_k^r\bigr)\, \bigl(1 - H[u_{k,n}]\bigr),
\end{equation*}
where $H[u_{k,n}]$ is the Heaviside function ($H=1$ for $u_{k,n}>0$, $H=0$ for
$u_{k,n} \leq 0$). Here, $f_k^l$ and $f_k^r$ are the distribution functions
interpolated to the interface from the left and right cell centers, and
$\nabla f_k^l$, $\nabla f_k^r$ are obtained from reconstruction.

The equilibrium $g$ in Eq.~\eqref{eq:UGKS-integral} is expanded in space and time as in the GKS; the interface distribution reads
\begin{equation*}
	\begin{aligned}
	\hat{f}_{ij,k}(t) &= \left(1 - e^{-t/\tau}\right) g_{0,k}
	+ \left(t + \tau(e^{-t/\tau}-1)\right) \bar{A}\, g_{0,k} \\
	&\quad + \left(t\, e^{-t/\tau} + \tau(e^{-t/\tau}-1)\right)
	\Bigl[\bigl(\bar{a}^l u_{k,n} + \bar{b}^l v_{k,t} + \bar{c}^l w_{k,t}\bigr)\, H[u_{k,n}] \\
	&\quad + \bigl(\bar{a}^r u_{k,n} + \bar{b}^r v_{k,t} + \bar{c}^r w_{k,t}\bigr)\,
	\bigl(1 - H[u_{k,n}]\bigr)\Bigr] g_{0,k} \\
	&\quad + e^{-t/\tau}\,\Bigl[\bigl(f_k^l - t\,\vec{u}_k \cdot \nabla f_k^l\bigr)\, H[u_{k,n}]
	+ \bigl(f_k^r - t\,\vec{u}_k \cdot \nabla f_k^r\bigr)\, \bigl(1 - H[u_{k,n}]\bigr)\Bigr].
	\end{aligned}
\end{equation*}
The time-averaged microscopic flux at the interface for velocity $\vec{u}_k$ is
\begin{equation*}
	\mathcal{F}_{ij,k} = \frac{1}{\Delta t}\int_0^{\Delta t}
	\vec{u}_k \cdot \vec{n}_{ij}\, \hat{f}_{ij,k}(t)\,\mathrm{d}t.
\end{equation*}
The discrete velocity distribution function is updated in the finite volume framework by the time-averaged microscopic flux and the collision term:
\begin{equation*}
	f_{i,k}^{n+1} = f_{i,k}^n
	- \frac{\Delta t}{\Omega_i} \sum_{j \in N(i)} \mathcal{F}_{ij,k}\,
	\mathcal{A}_{ij}
	+ \frac{1}{\Omega_i} \int_{\Omega_i} \int_0^{\Delta t} \frac{g_{i,k} - f_{i,k}}{\tau_i}\,\mathrm{d}t\,\mathrm{d}\vec{x},
\end{equation*}
where the collision term is usually solved semi-implicitly.
The macroscopic flux $\vec{F}_{ij}$ is the moment of the time-averaged
microscopic flux over the discrete velocity space:
\begin{equation*}
	\vec{F}_{ij} = \sum_k \omega_k\,\boldsymbol{\Psi}(\vec{u}_k)\, \mathcal{F}_{ij,k},
\end{equation*}
where $\omega_k$ are the velocity space weights. The macroscopic update for
the conservative variables $\vec{W}_i$ is given by the same finite volume
formula Eq.~\eqref{eq:FVM}.
Detailed numerical treatment of the UGKS can be found in~\cite{xu2010unified,long2024implicit}.

\subsection{Unified gas-kinetic wave-particle method}

Following the UGKS, the unified gas-kinetic wave-particle (UGKWP)
method~\cite{liu2020unified,zhu2019unified} employs stochastic
particles together with deterministic hydrodynamic waves to represent the
microscopic gas distribution, thus avoiding a deterministic velocity space.
This particle-wave decomposition
can fully recover the gas-kinetic scheme without particles in the continuum limit,
where the particle efficiently describes the nonequilibrium distribution function
without dimensionality problem. The deterministic hydrodynamic waves
account for the equilibrium evolution, while the nonequilibrium
free-transport part is realized by stochastic particles.

According to the integral solution
Eq.~\eqref{eq:UGKS-integral}, the distribution function after $\Delta t$ is a
combination of the initial state and the equilibrium state. For a particle from
the initial state, the probability of remaining collisionless over $\Delta t$
is $e^{-\Delta t/\tau}$; otherwise, it experiences collision and contributes to
the local equilibrium part. By taking a random number $\epsilon$ uniformly distributed
in $(0,1)$, the free-transport time $t_{f,k}$ of a particle labeled by $k$ can
be calculated by
\begin{equation*}
	t_{f,k}=\min\left(-\tau_k\ln\epsilon,\Delta t\right).
\end{equation*}
Then the particle position is renewed as
\begin{equation*}
	\vec{x}_{p,k}^{*}=\vec{x}_{p,k}^{n}+\vec{u}_k\, t_{f,k}.
\end{equation*}
The contribution of all particles to the macroscopic conserved variables is
assembled by their trajectory crossings of cell interfaces. After the free
transport, if $t_{f,k}<\Delta t$, the particle is called a collisional
particle: it is deleted and its mass, momentum, and energy are transferred to
the hydrodynamic wave in the corresponding cell. Otherwise, if
$t_{f,k}=\Delta t$, the particle is called a collisionless particle and is
kept to the next time step.

Besides the particle transport flux,
the interface flux is decomposed into
equilibrium-wave and analytical free-streaming-wave contributions:
\begin{equation*}
	\vec{F}_{ij}=\vec{F}_{ij}^{eq}+\vec{F}_{ij}^{fr,h}.
\end{equation*}
The equilibrium-wave flux $\vec{F}_{ij}^{eq}$ is computed in the same way as in
the GKS subsection. For the free-streaming wave part,
the free-streaming distribution function at the interface is first written as
\begin{equation}\label{eq:UGKWP-ffr}
	\begin{aligned}
	f_{ij}^{fr}(t)
	=\Bigl[&
	\bigl(1-t(a^l u_n+b^l v_t+c^l w_t)\bigr)g_l^h\,H[u_n] \\
	&+\bigl(1-t(a^r u_n+b^r v_t+c^r w_t)\bigr)g_r^h\,(1-H[u_n])\Bigr].
	\end{aligned}
\end{equation}
Here $g_l^h$ and $g_r^h$ are the left and right equilibrium states at the
interface, reconstructed from $\vec{W}_l^h$ and $\vec{W}_r^h$. The coefficients
$a^{l,r}$, $b^{l,r}$, and $c^{l,r}$ are the corresponding microscopic spatial
derivative coefficients in the local coordinates.
Using Eq.~\eqref{eq:UGKWP-ffr}, the UGKS free-streaming flux is obtained by
time-averaging the corresponding interface moment:
\begin{equation*}
	\vec{F}_{UGKS}^{fr}\!\left(\vec{W}^{h}\right)
	=\frac{1}{\Delta t}\int_0^{\Delta t}\int
	(\vec{u}\cdot\vec{n}_{ij})\,e^{-t/\tau}f_{ij}^{fr}(t)\,\boldsymbol{\Psi}\,\mathrm{d}\Xi\,\mathrm{d}t,
\end{equation*}
For the sampled collisionless part
$\vec{W}^{hp}=e^{-\Delta t/\tau_i}\vec{W}^{h}$, the DVM free-streaming flux is
\begin{equation*}
	\vec{F}_{DVM}^{fr}\!\left(\vec{W}^{hp}\right)
	=\frac{e^{-\Delta t/\tau}}{\Delta t}\int_0^{\Delta t}\int
	(\vec{u}\cdot\vec{n}_{ij})\,f_{ij}^{fr}(t)\,\boldsymbol{\Psi}\,\mathrm{d}\Xi\,\mathrm{d}t.
\end{equation*}
Then the analytical wave free-streaming flux is given by
\begin{equation*}
	\vec{F}_{ij}^{fr,h}
	=\vec{F}_{UGKS}^{fr}\!\left(\vec{W}^{h}\right)
	-\vec{F}_{DVM}^{fr}\!\left(\vec{W}^{hp}\right).
\end{equation*}
This decomposition is necessary because
$\vec{F}_{DVM}^{fr}(\vec{W}^{hp})$ is already carried by tracked particles;
subtracting it avoids double counting and keeps consistency between the
analytical wave evolution and particle transport.

For the hydrodynamic wave itself, only the fraction that would stay collisionless
over $\Delta t$ must be realized as new simulation particles
\begin{equation*}
	\vec{W}^{hp}=e^{-\Delta t/\tau_i}\vec{W}^{h}.
\end{equation*}
The net free-transport
contribution assembled in cell $i$ from particle crossings is denoted by
$\vec{W}_i^{fr,p}$.

The macroscopic update is
\begin{equation*}
	\vec{W}_i^{n+1}
	=\vec{W}_i^n
	-\frac{\Delta t}{\Omega_i}\sum_{j\in N(i)}\left(\vec{F}_{ij}^{eq}+\vec{F}_{ij}^{fr,h}\right)\mathcal{A}_{ij}
	+\vec{W}_i^{fr,p}.
\end{equation*}
After the update, $\vec{W}_i^h=\vec{W}_i^{n+1}-\vec{W}_i^p$ is reconstructed and
the wave-particle decomposition is repeated at the next step. Detailed
algorithmic procedures are given in
Refs.~\cite{liu2020unified,zhu2019unified,chen2020three,wei2023adaptive,cao2026adaptive}.

\subsection{Kinetic boundary condition}

Since the GKS, UGKS, and UGKWP are all constructed on the particle
distribution function, the wall boundary condition follows naturally from
Maxwell's diffuse scattering model~\cite{maxwell_1879,li2005kinetic}.
In the UGKS, the incident distribution is obtained by one-sided
extrapolation of the multiscale flux from the interior to the wall.
In the UGKWP, the nonequilibrium part is carried by stochastic particles
that hit the wall directly and bounce back with random velocities sampled from
the wall Maxwellian.
In the GKS, the CE expansion of the distribution function is
extrapolated to the wall from one side. Taking $\vec{n}_w$ as the unit
normal directed into the flow, the time-dependent incident distribution
$f^i$ (for $\vec{u} \cdot \vec{n}_w < 0$) takes the form
\begin{equation}\label{eq:wall-incident}
	f^i = g^i\left[1 - \tau(\bar{a}\, u_n + \bar{A}) + t\, \bar{A}\right], \quad u_n = \vec{u} \cdot \vec{n}_w,
\end{equation}
where $g^i$ is the Maxwellian at the wall reconstructed from the interior, and
$\bar{a}$, $\bar{A}$ are the corresponding microscopic spatial and temporal
derivatives.

The reflected particles are assumed to follow the wall Maxwellian at the
prescribed wall temperature $T_w$:
\begin{equation*}
	g^w = \rho_w \left(\frac{\lambda_w}{\pi}\right)^{\frac{K+3}{2}}
	\exp\left(-\lambda_w \left(\vec{u}^2 + \vec{\xi}^2\right)\right), \quad \lambda_w = \frac{m_0}{2k_B T_w},
\end{equation*}
with $\boldsymbol{\xi}$ the internal degrees of freedom as in
Eq.~\eqref{eq:g}.

Imposing zero net mass flux across the wall over $\Delta t$ determines
$\rho_w$:
\begin{equation*}
	\int_0^{\Delta t} \int_{\vec{u}\cdot\vec{n}_w<0} (\vec{u}\cdot\vec{n}_w)\, f^i\,\mathrm{d}\Xi\,\mathrm{d}t
	= -\int_0^{\Delta t} \int_{\vec{u}\cdot\vec{n}_w>0} (\vec{u}\cdot\vec{n}_w)\, g^w\,\mathrm{d}\Xi\,\mathrm{d}t.
\end{equation*}
Solving gives
\begin{equation}\label{eq:wall-density}
	\rho_w = -\frac{2\sqrt{\pi\lambda_w}}{\Delta t}
	\int_0^{\Delta t} \int_{\vec{u}\cdot\vec{n}_w<0} (\vec{u}\cdot\vec{n}_w)\, f^i\,\mathrm{d}\Xi\,\mathrm{d}t,
\end{equation}
which uniquely specifies the wall Maxwellian. Under full diffuse reflection the
wall distribution is $f = f^i$ for $\vec{u}\cdot\vec{n}_w < 0$ and
$f = g^w$ for $\vec{u}\cdot\vec{n}_w > 0$.

With the complete wall distribution, the macroscopic flux of the solid wall
boundary, averaged over the time step $\Delta t$, is obtained by taking moments:
\begin{equation}\label{eq:wall-flux}
	\vec{F}_w = \frac{1}{\Delta t}\int_0^{\Delta t}
	\biggl(\,
	\int_{\vec{u}\cdot\vec{n}_w<0} \psi\, (\vec{u}\cdot\vec{n}_w)\, f^i\,\mathrm{d}\Xi
	+ \int_{\vec{u}\cdot\vec{n}_w>0} \psi\, (\vec{u}\cdot\vec{n}_w)\, g^w\,\mathrm{d}\Xi
	\biggr)\,\mathrm{d}t,
\end{equation}
where $\psi = (\,1,\;\vec{u},\;\frac{1}{2}(\vec{u}^2+\vec{\xi}^2)\,)^\mathrm{T.pdf}$
is the vector of collision invariants.
Since $f^i$ is known from the interior reconstruction with CE expansion and $g^w$ is
fully determined by $T_w$ and the $\rho_w$ obtained from
Eq.~\eqref{eq:wall-density}, no additional modeling is required.
This kinetic formulation reduces to
the no-slip isothermal wall in the continuum flow regime.

Comparing with the kinetic boundary condition,
the Maxwell slip boundary condition used in the conventional CFD solver also originates from
Maxwell's molecular description of gas--surface interaction~\cite{greenshields2012rarefied}.
Because they share the same conceptual basis, the conventional NS approach with Maxwell slip boundary
can accurately model the wall velocity and temperature jump in the
slip regime.
However, the two approaches have notable differences when producing the flux induced by the wall.
As depicted by the Eq.~\eqref{eq:wall-flux}, the kinetic boundary condition of the GKS is built on an
explicit incident-reflected picture, in which the incident CE expansion together with
the reflected wall Maxwellian forms the actual nonequilibrium distribution at the wall.
The NS slip approach, by contrast, simply adopts the linear consititution relation to construct flux at the wall.
Actually, it is a full-moment of CE expansion, thus smear the nonequilibrium information caused by the wall reflection.

\subsection{Relaxation-time limit}

\begin{figure}[htbp]
	\centering
	\includegraphics[width=0.85\textwidth]{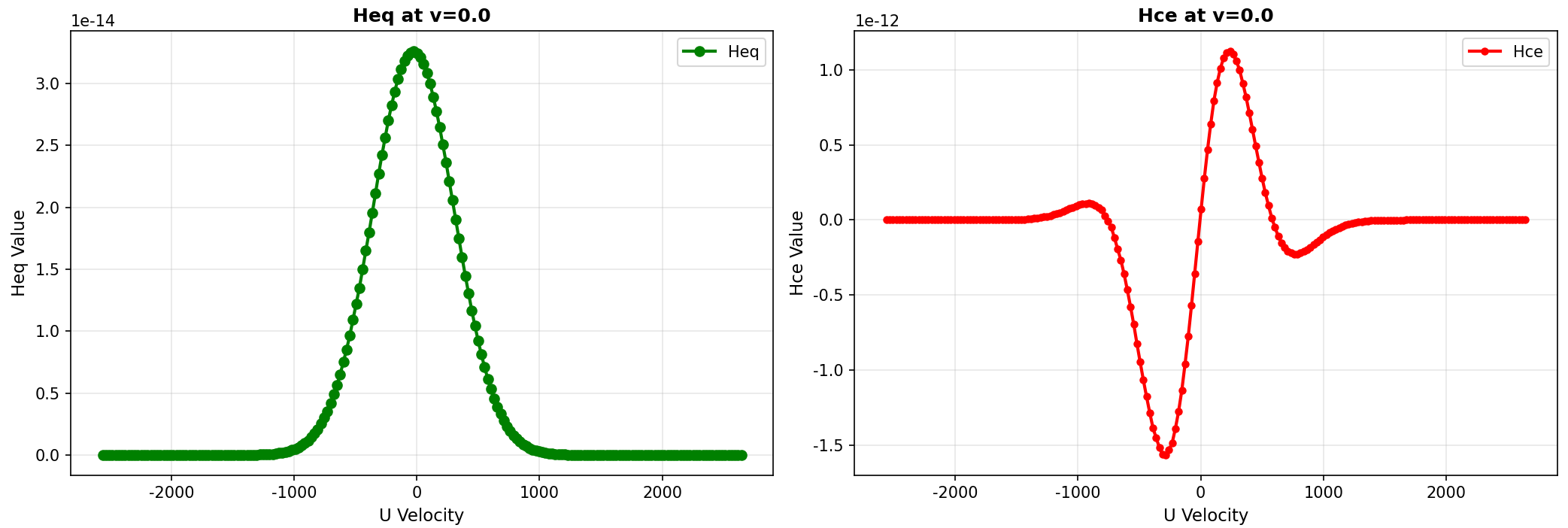}
	\caption{Comparison of equilibrium and CE expanded reduced distribution functions in the strongly rarefied cylinder wake.}
	\label{fig:CE-error}
\end{figure}

The CE expansion used in the construction of $f_0$ in
Eq.~\eqref{eq:GKS-f0} is valid only for small perturbations. When the
relaxation time $\tau$ is large (e.g.\ in the slip regime or more rarefied
flow), the resulting nonequilibrium distribution can become unphysical. For
instance, the distribution function may take negative values.
Figure~\ref{fig:CE-error} compares the equilibrium and CE
expanded distributions in the cylinder wake, where the flow is strongly
rarefied; the CE-expanded distribution
clearly exhibits negative values, illustrating the need to limit
$\tau$ in such regions.
To improve the robustness of the GKS in these regimes, $\tau$ in the
CE expansion is limited. An effective relaxation time
$\tau_{\mathrm{eff}}$ is defined by
\begin{equation*}
	\tau_{\mathrm{eff}} = \begin{cases}
		\tau, & \tau \le C_{\tau,\mathrm{limit}}\, \Delta t, \\
		C_{\tau,\mathrm{limit}}\, \Delta t, & \tau > C_{\tau,\mathrm{limit}}\, \Delta t,
	\end{cases}
\end{equation*}
where $C_{\tau,\mathrm{limit}}$ is a constant (a typical value in this work is
$C_{\tau,\mathrm{limit}} = 2000$) and $\Delta t$ is the local time step of
the local cell (e.g.\ $\Delta t_i$ for cell $i$ when evaluating the interface
between cells $i$ and $j$, or the time step of the cell adjacent to the wall for
the wall boundary). The effective relaxation time $\tau_{\mathrm{eff}}$ replaces
$\tau$ only in the terms that come from the CE expansion; $\tau$ is
still used in the integral solution (e.g.\ $e^{-t/\tau}$) and in the collision
term.

At the cell interface, the CE-limited initial distribution is
Eq.~\eqref{eq:GKS-f0} with $\tau$ replaced by $\tau_{\mathrm{eff}}$ in the
coefficients of the expansion:
\begin{equation*}
	f_0(\vec{r}_{ij}-\vec{u}t) = \begin{cases}
		g^l\left[1 - t(a^l u_n + b^l v_t + c^l w_t) - \tau_{\mathrm{eff}}\bigl(a^l u_n + b^l v_t + c^l w_t + A^l\bigr)\right], & u_n > 0, \\[2pt]
		g^r\left[1 - t(a^r u_n + b^r v_t + c^r w_t) - \tau_{\mathrm{eff}}\bigl(a^r u_n + b^r v_t + c^r w_t + A^r\bigr)\right], & u_n \leq 0.
	\end{cases}
\end{equation*}
For the kinetic wall boundary, the limited incident distribution is
Eq.~\eqref{eq:wall-incident} with $\tau$ replaced by $\tau_{\mathrm{eff}}$:
\begin{equation*}
	f^i = g^i\left[1 - \tau_{\mathrm{eff}}(\bar{a}\, u_n + \bar{A}) + t\, \bar{A}\right], \quad u_n = \vec{u} \cdot \vec{n}_w.
\end{equation*}
This keeps the scheme robust in the slip and more rarefied regimes while minimizing the effect on
the near-continuum regime.
\section{Flow studies and discussion}\label{sec:result}

This section presents numerical simulations to investigate the applicability limit of the GKS with the kinetic boundary condition
in typical hypersonic flow scenarios.
Leveraging the discrete velocity space description of the UGKS, the nonequilibrium distribution function comparison
is performed for the two-dimensional circular cylinder case to demonstrate the
cause of the discrepancy between the GKS and multiscale methods.
For the remaining cases ($9^\circ$ blunted cone, $70^\circ$ blunted cone with sting, and
Apollo~6 command module), the GKS results are compared with available simulation by multiscale methods and DSMC,
as well as the experimental data to quantify the applicability limit of the
current GKS implementation.

In the following subsections, the degree of rarefaction is characterized by the Knudsen number.
For the initial condition, the freestream Knudsen number ${\rm Kn}_\infty$ is used to define each case:
\begin{equation*}
	{\rm Kn}_\infty = \frac{l_\infty}{L_{ref}},
\end{equation*}
where $l_\infty$ is the mean free path of the freestream flow and $L_{ref}$ is the reference length.
The freestream density is obtained from the freestream Knudsen number as
\begin{equation*}
	\rho_\infty=\frac{4 \alpha(5-2 \omega)(7-2 \omega)}{5(\alpha+1)(\alpha+2)} \sqrt{\frac{m}{2 \pi k_B T_\infty}} \frac{\mu_\infty}{L_{ref} \mathrm{Kn}_\infty},
\end{equation*}
where $m$ is the molecular mass.
The dynamic viscosity is given by a power law
\begin{equation*}
	\mu = \mu_{ref}\left(\frac{T.pdf}{T_{ref}}\right)^\omega.
\end{equation*}
The internal degrees of freedom in the Maxwellian Eq.~\eqref{eq:g} are given by $K = 2/(\gamma-1)$.
Table~\ref{tab:gas} lists the gas parameters used in this study for the cases included in that table; the reference dynamic viscosity is
defined at $T_{ref} = 273$~K.
For the Apollo~6 command module, the working gas is air; the altitude-dependent gas constant $R$, viscosity exponent $\omega$, and other
freestream and wall quantities are given in Table~\ref{tab:apollo-altitude}.

To compare surface quantities with the reference data, the non-dimensionalized pressure,
shear stress, and heat flux coefficients are used:
\begin{equation*}
	C_p = \frac{p_s}{\frac{1}{2}\rho_\infty \|\vec{U}_\infty\|^2}, \quad
	C_\tau = \frac{f_s}{\frac{1}{2}\rho_\infty \|\vec{U}_\infty\|^2}, \quad
	C_h = \frac{h_s}{\frac{1}{2}\rho_\infty \|\vec{U}_\infty\|^3},
\end{equation*}
where $p_s$, $f_s$, and $h_s$ are the surface pressure, friction, and heat flux.

The aerodynamic force and moment coefficients are normalized by the dynamic pressure,
the reference area $A_{ref}$, and reference length $L_{ref}$:
\begin{equation*}
	C_d = \frac{F_d}{\frac{1}{2}\rho_\infty U_\infty^2 A_{ref}}, \quad
	C_l = \frac{F_l}{\frac{1}{2}\rho_\infty U_\infty^2 A_{ref}}, \quad
	C_m = \frac{M}{\frac{1}{2}\rho_\infty U_\infty^2 A_{ref}\, L_{ref}},
\end{equation*}
where $F_d$ and $F_l$ are the drag and lift forces in the wind-axis coordinate system,
and $M$ is the pitching moment.
In addition, the axial and normal force coefficients in the body-axis system are defined as
\begin{equation*}
	C_A = \frac{F_A}{\frac{1}{2}\rho_\infty U_\infty^2 A_{ref}}, \quad
	C_N = \frac{F_N}{\frac{1}{2}\rho_\infty U_\infty^2 A_{ref}},
\end{equation*}
where $F_A$ and $F_N$ are the force components aligned with and normal to the vehicle axis.

\begin{table}[htbp]
	\centering
	\caption{Gas and physical properties for each case; $\mu_{ref}$ is defined at $T_{ref} = 273$~K.}
	\begin{tabular}{lcccccccc}
		\hline
		Case & Gas & $m$, kg & $\mu_{ref}$, $\rm{N}\rm{s}\rm{m}^{-2}$ & $\omega$ & $\alpha$ & $\gamma$ & $\mathrm{Pr}$
		\\ \hline
		Circular cylinder & Ar & $6.63\times 10^{-26}$ & $2.12\times 10^{-5}$ & 0.81 & 1.0 & 1.67 & 1.0\\
		$9^\circ$ blunted cone & ${\rm N}_2$ & $4.65\times 10^{-26}$ & $1.656\times 10^{-5}$ & 0.74 & 1.0  & 1.4 & 0.72\\
		$70^\circ$ blunted cone & ${\rm N}_2$ & $4.65\times 10^{-26}$ & $1.656\times 10^{-5}$ & 0.74 & 1.0  & 1.4 & 0.72\\ \hline
	\end{tabular}
	\label{tab:gas}
\end{table}

\subsection{Hypersonic flow around a circular cylinder}\label{sec:case-cylinder}

The circular cylinder is recognized as a well-established two-dimensional benchmark with sufficient numerical reference data.
In this study, hypersonic flow of argon at ${\rm Ma}_\infty = 5$ is simulated at two freestream Knudsen
numbers, first ${\rm Kn}_\infty = 0.02$ and then the more rarefied ${\rm Kn}_\infty = 0.4$.
The reference length $L_{ref}$ is taken as the cylinder radius $R_c = 1$ m.
The same kinetic boundary condition (diffuse reflection with full accommodation) and isothermal wall
temperature are applied in the GKS and UGKS simulations.
The first cell height $\Delta h_{wall}$ at the wall is set to $0.5\times 10^{-3}\,L_{ref}$ for ${\rm Kn}_\infty = 0.02$
and $10^{-2}\,L_{ref}$ for ${\rm Kn}_\infty = 0.4$ to resolve the boundary layer; the relaxation-time limit is applied in the wake region.

\begin{figure}[H]
	\centering
	\subfloat[]{\includegraphics[width=0.33\textwidth]{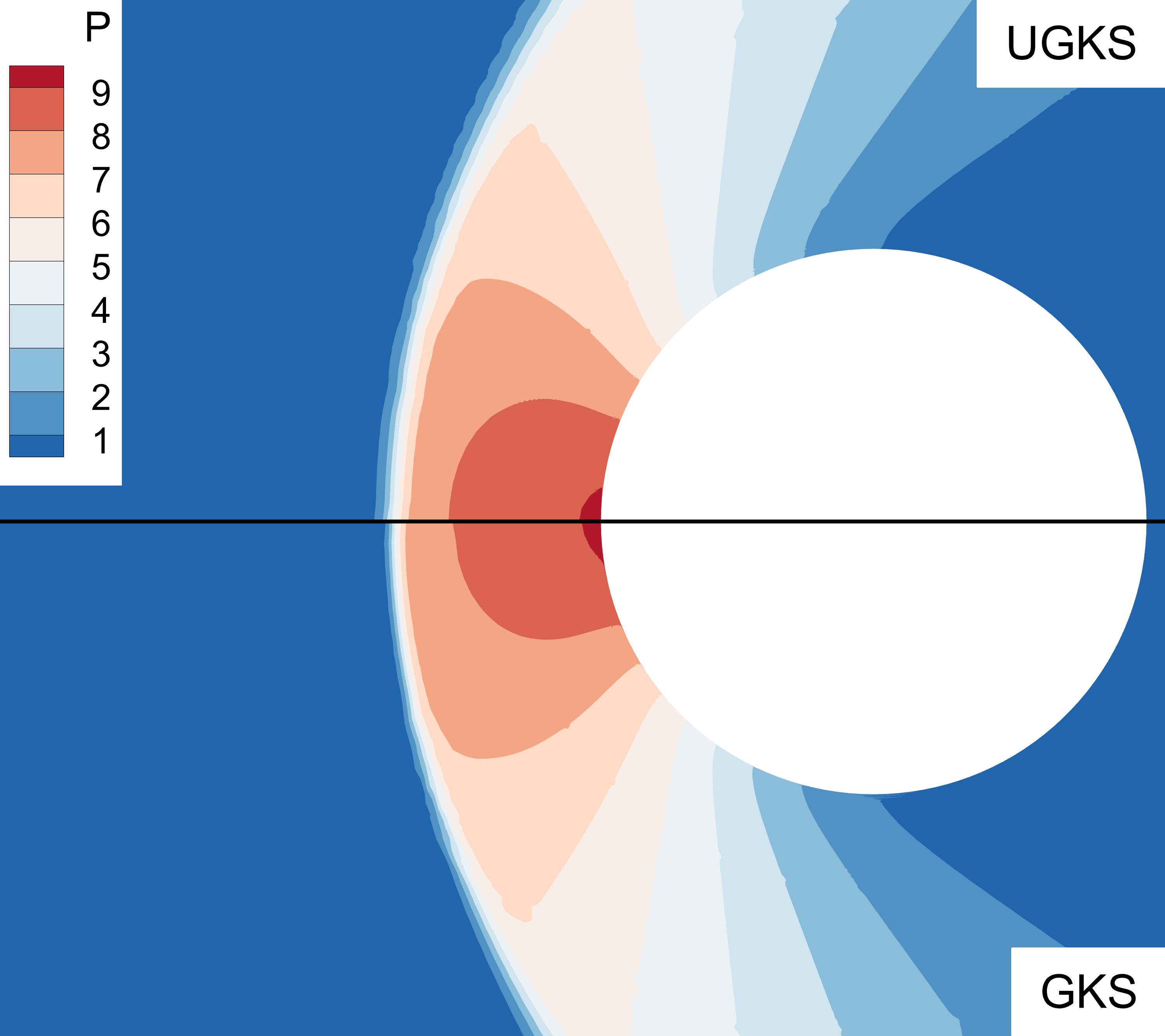}}~
	\subfloat[]{\includegraphics[width=0.33\textwidth]{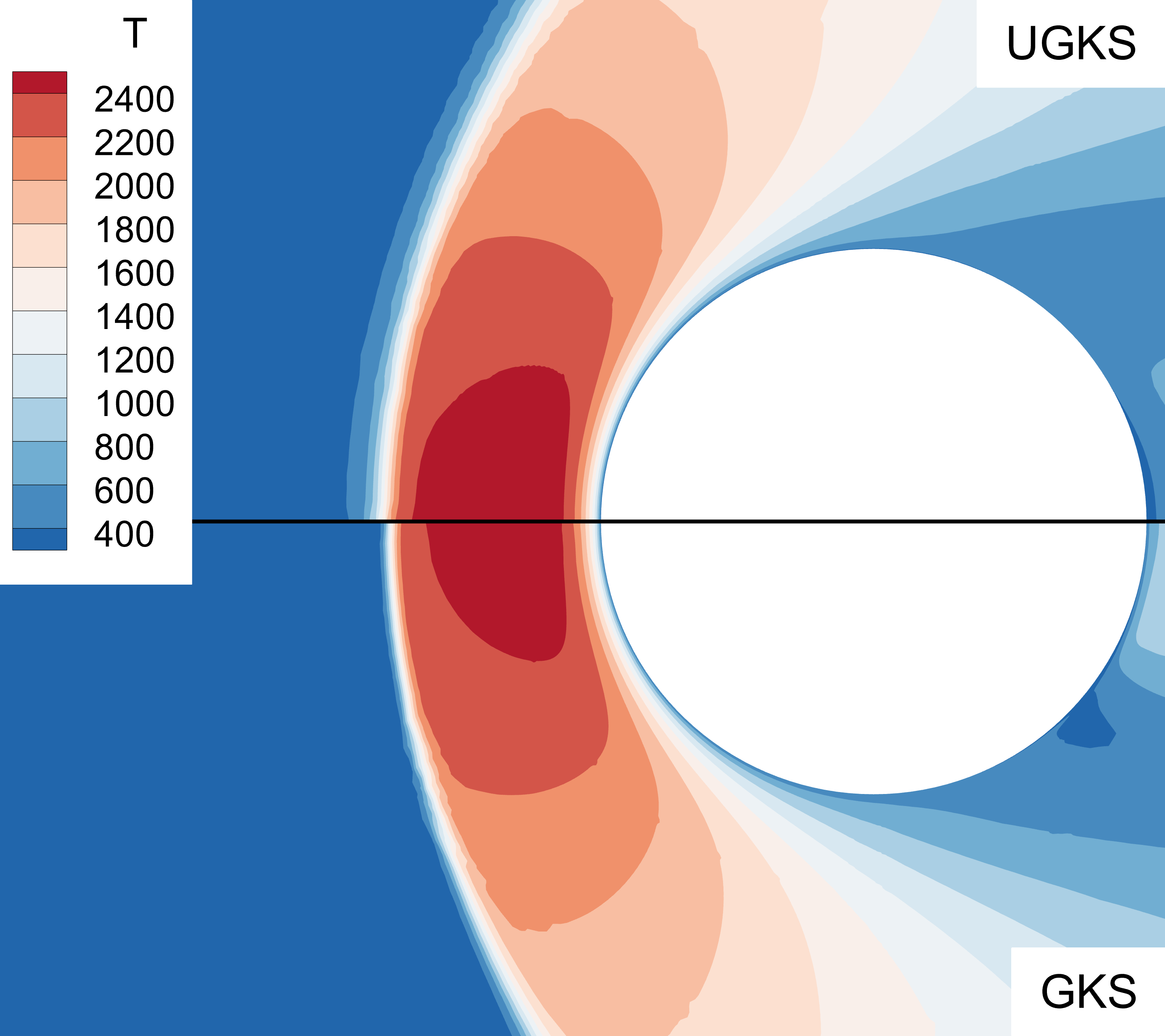}}~
	\subfloat[]{\includegraphics[width=0.33\textwidth]{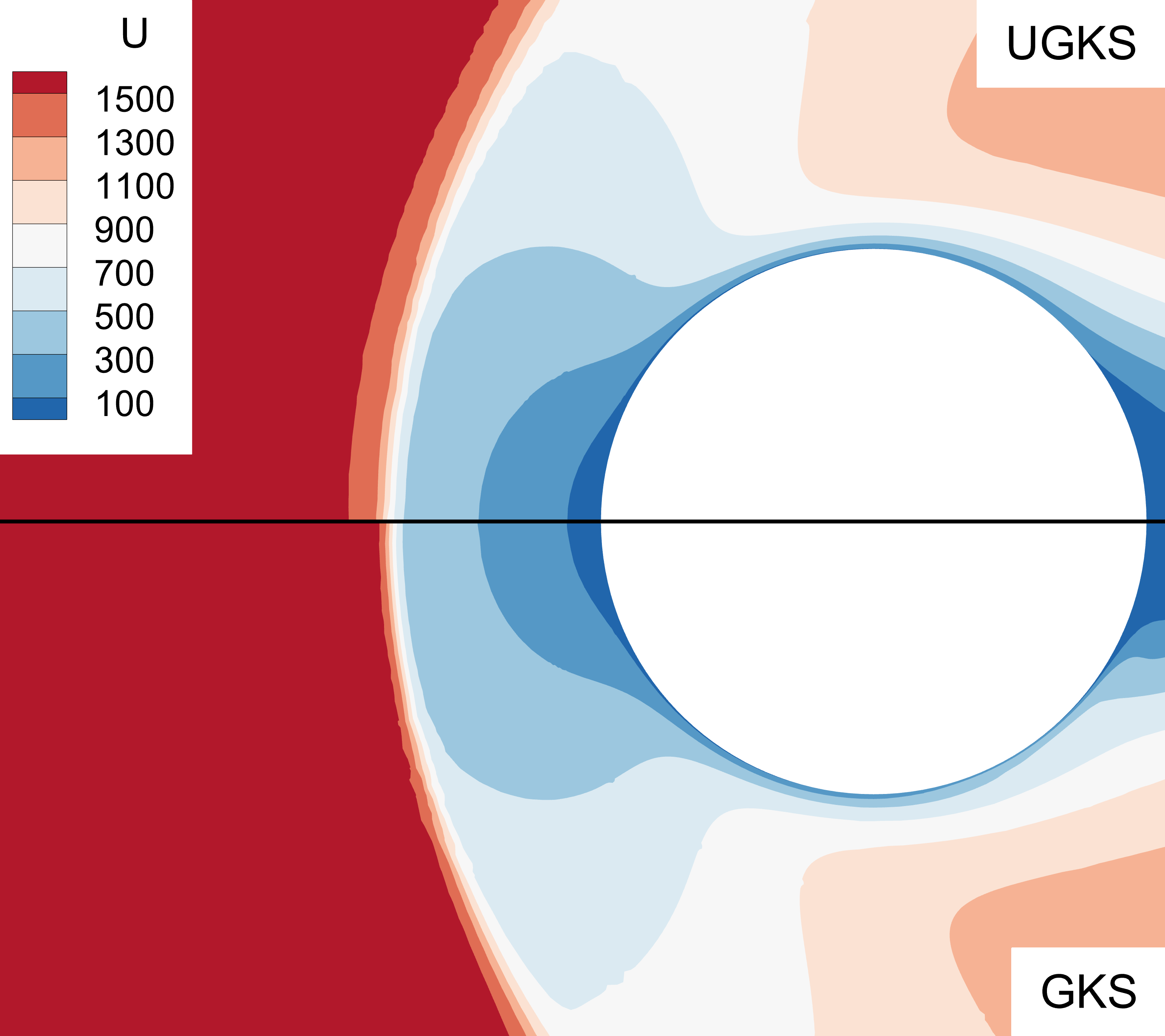}}\\
	\caption{Hypersonic flow at ${\rm Kn}_\infty = 0.02$ and ${\rm Ma}_\infty = 5$ passing over a circular cylinder by the GKS and UGKS. Distributions along the stagnation line: (a) pressure, (b) temperature, and (c) $x$-direction velocity $U$.}
	\label{fig:cylinder-Kn002-stag}
\end{figure}

\begin{figure}[H]
	\centering
	\subfloat[]{\includegraphics[width=0.33\textwidth]{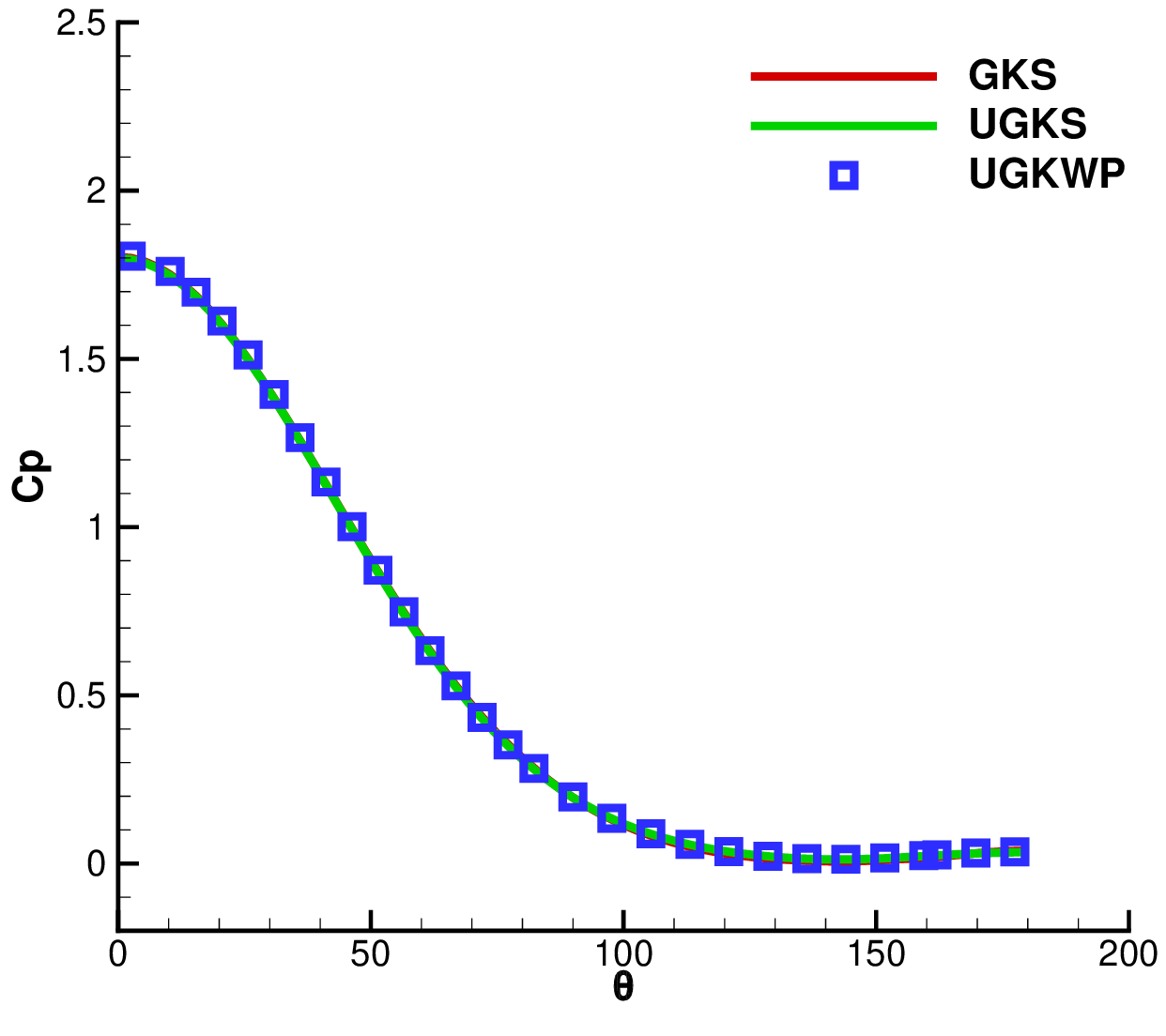}}
	\subfloat[]{\includegraphics[width=0.33\textwidth]{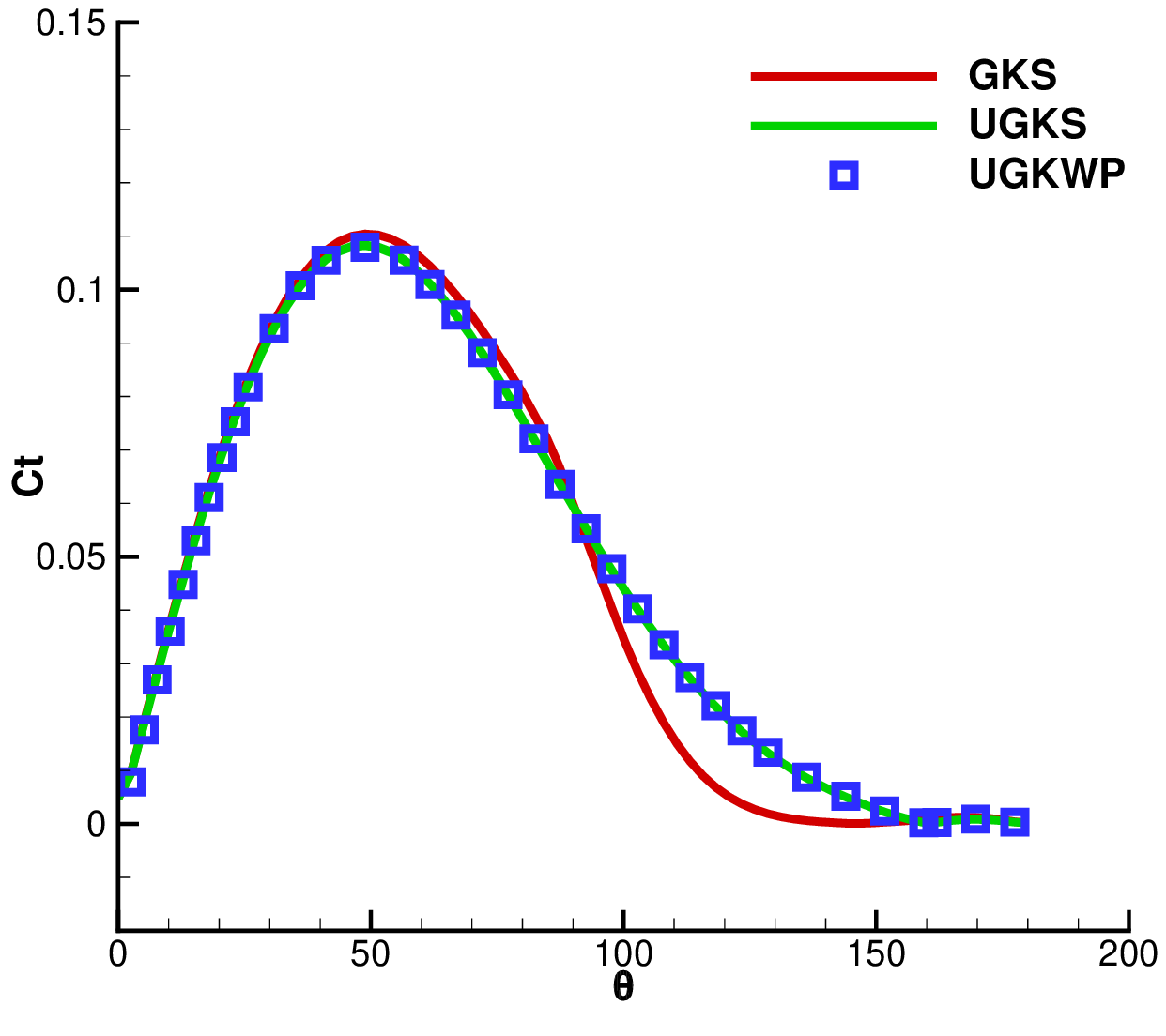}}
	\subfloat[]{\includegraphics[width=0.33\textwidth]{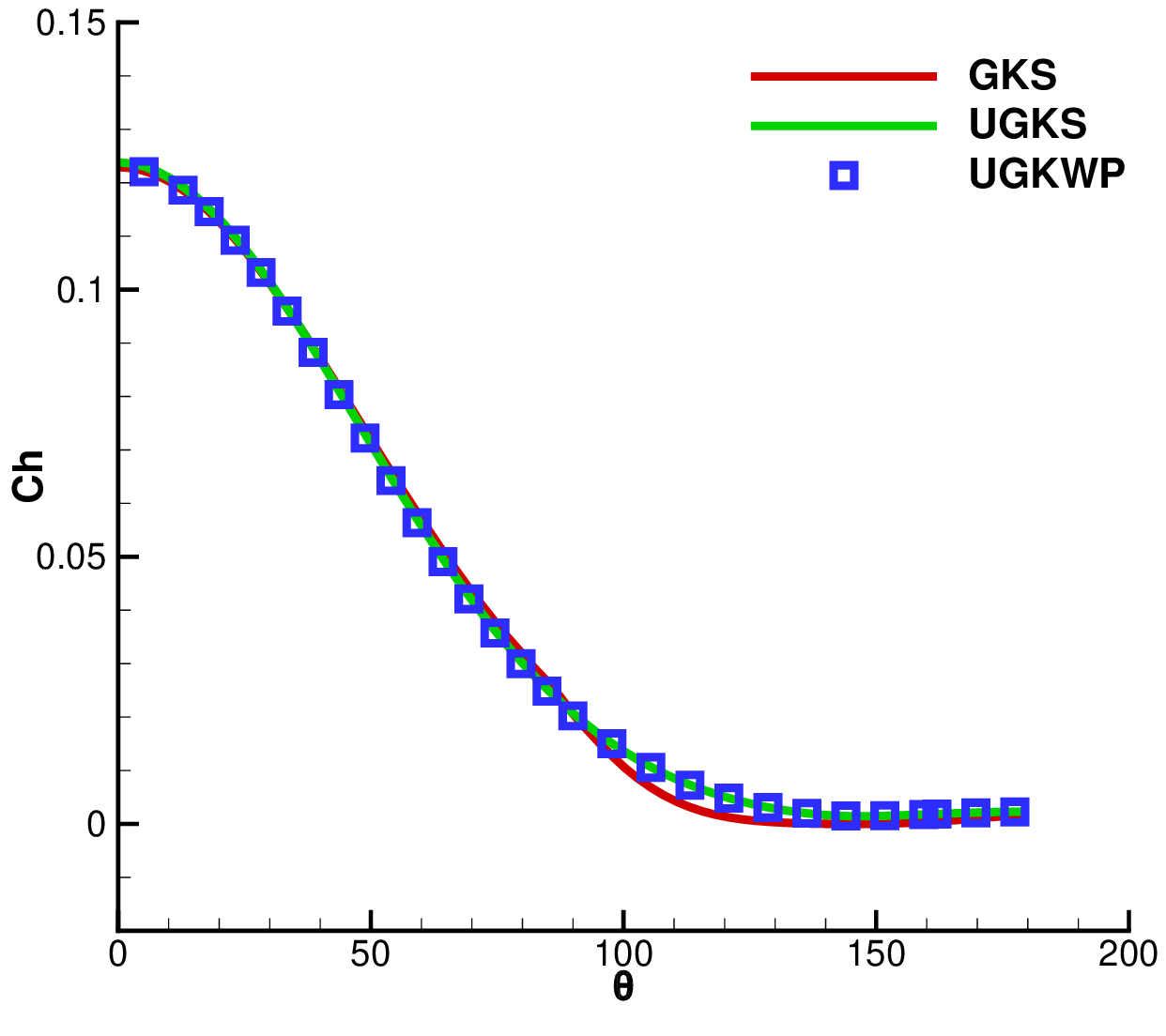}}\\
	\caption{Hypersonic flow at ${\rm Kn}_\infty = 0.02$ and ${\rm Ma}_\infty = 5$ passing over a circular cylinder by the GKS, UGKS, and UGKWP. Surface quantities distributions: (a) pressure coefficient, (b) shear stress coefficient, and (c) heat flux coefficient.}
	\label{fig:cylinder-surface-Kn002}
\end{figure}

For the less rarefied case with ${\rm Kn}_\infty = 0.02$, the GKS provides predictions with acceptable discrepancy in both
flow field and the surface quantities.
Figure~\ref{fig:cylinder-Kn002-stag} shows contours of the pressure, temperature and x-direction velocity simulated by the
GKS and the UGKS. For the GKS, since the nonlinear constitutive relations cannot be satisfied, the shock thickness is thinner
compared with the UGKS result. In the wake region where the relaxation limit is applied,
the flow structure shows noticeable difference, especially in the temperature and velocity contours.
The comparison of surface quantities with the UGKS and the UGKWP is depicted in
Fig.~\ref{fig:cylinder-surface-Kn002}. The GKS provides nearly the same $C_p$ result, while the heat flux coefficient and
friction coefficient have substantial error in the same region as the flow structure. Remarkably, the surface friction
dominated by viscosity shows relatively larger deviation.

\begin{figure}[H]
	\centering
	\subfloat[]{\includegraphics[width=0.33\textwidth]{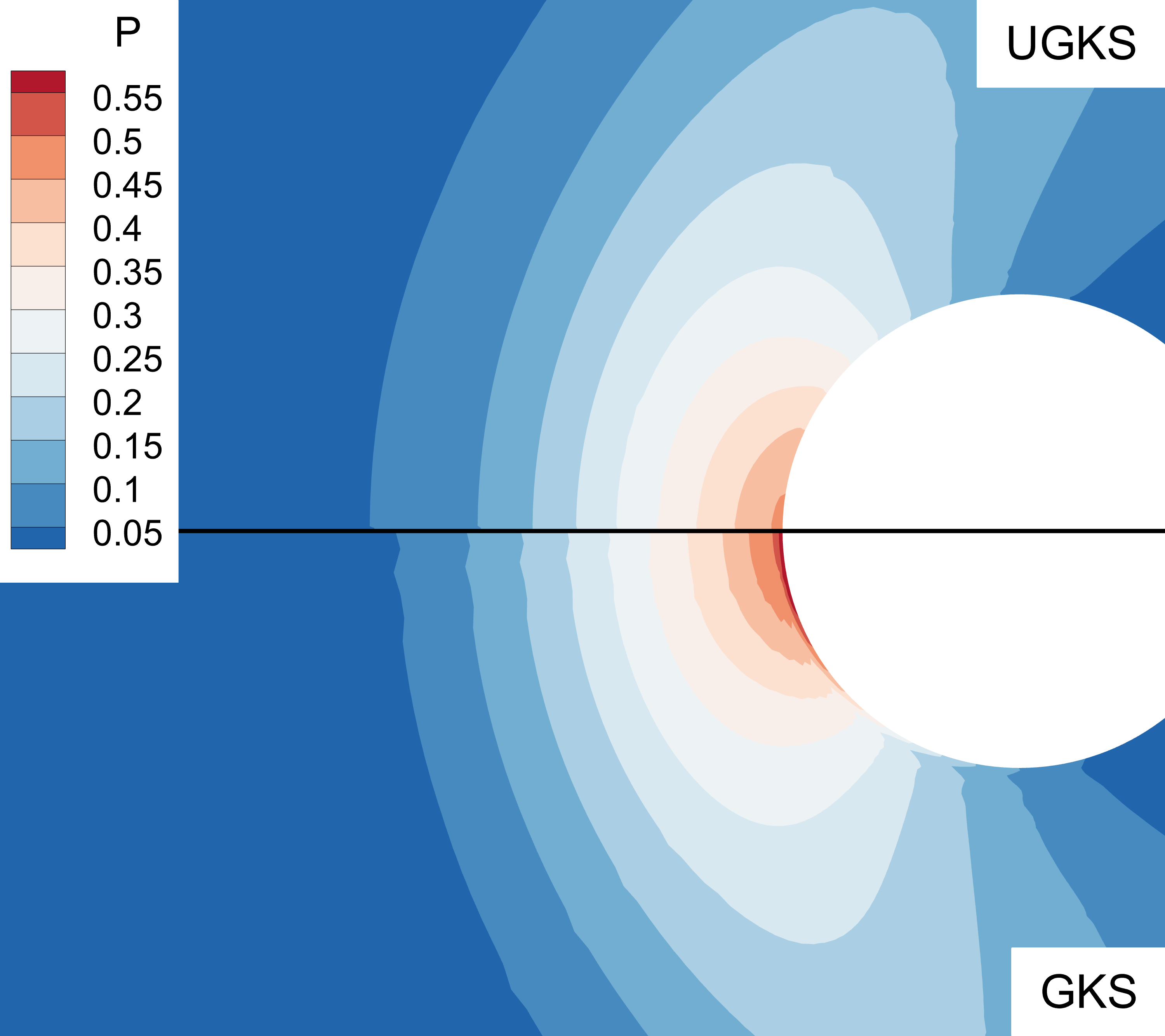}}~
	\subfloat[]{\includegraphics[width=0.33\textwidth]{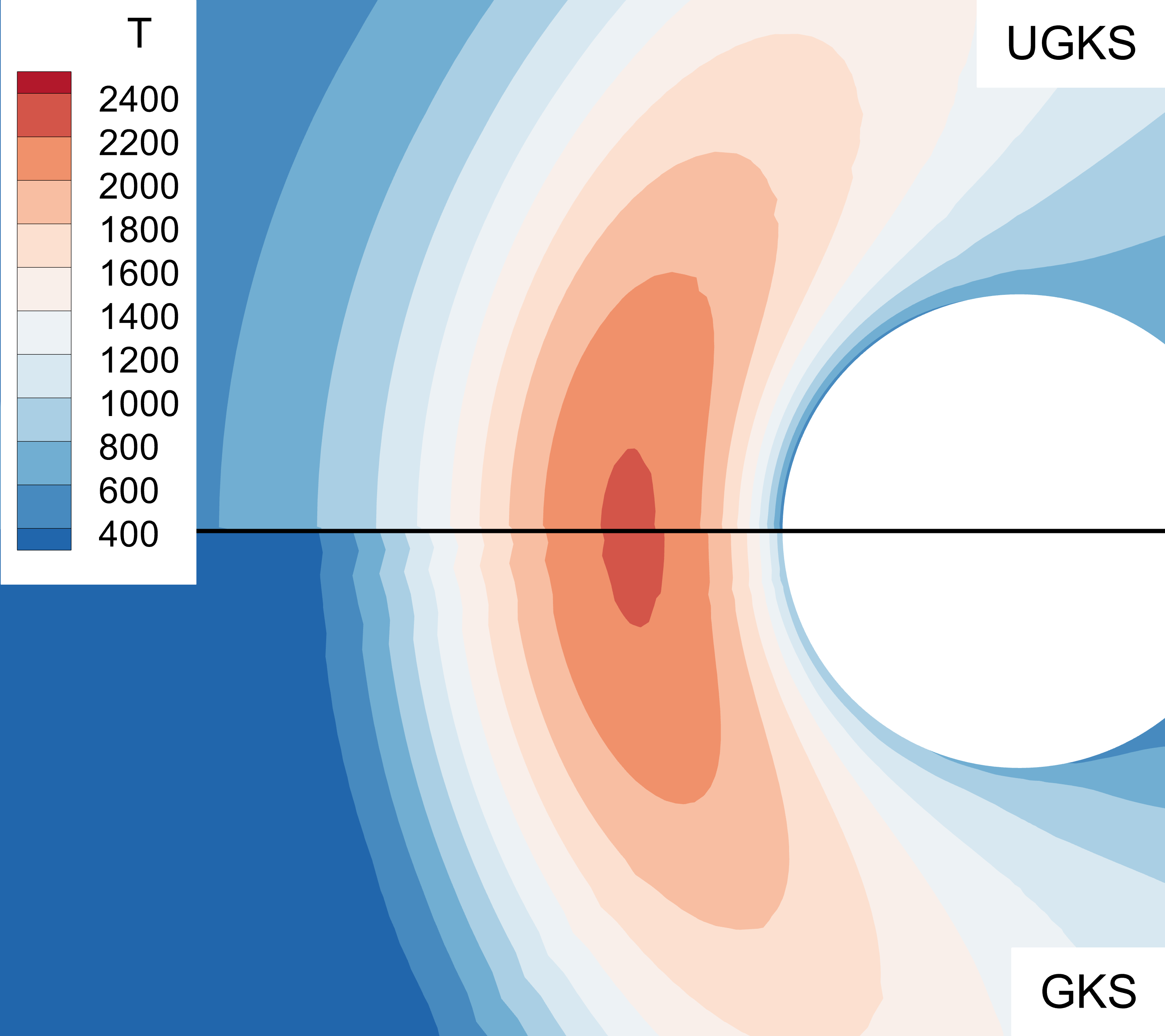}}~
	\subfloat[]{\includegraphics[width=0.33\textwidth]{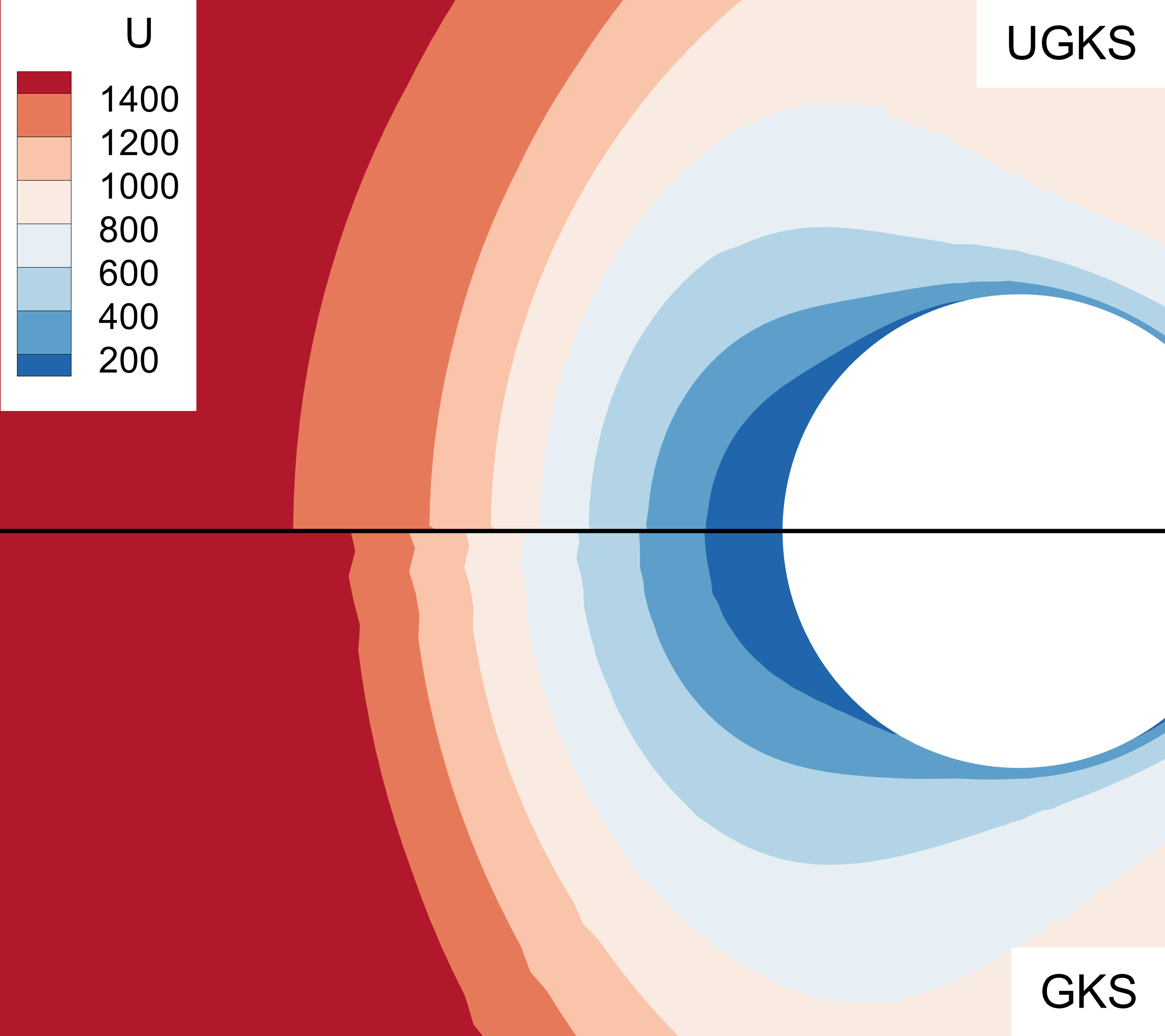}}\\
	\caption{Hypersonic flow at ${\rm Kn}_\infty = 0.4$ and ${\rm Ma}_\infty = 5$ passing over a circular cylinder by the GKS and UGKS. Distributions of (a) pressure, (b) temperature, and (c) $x$-direction velocity $U$.}
	\label{fig:cylinder-Kn04-field}
\end{figure}

\begin{figure}[H]
	\centering
	\subfloat[]{\includegraphics[width=0.33\textwidth]{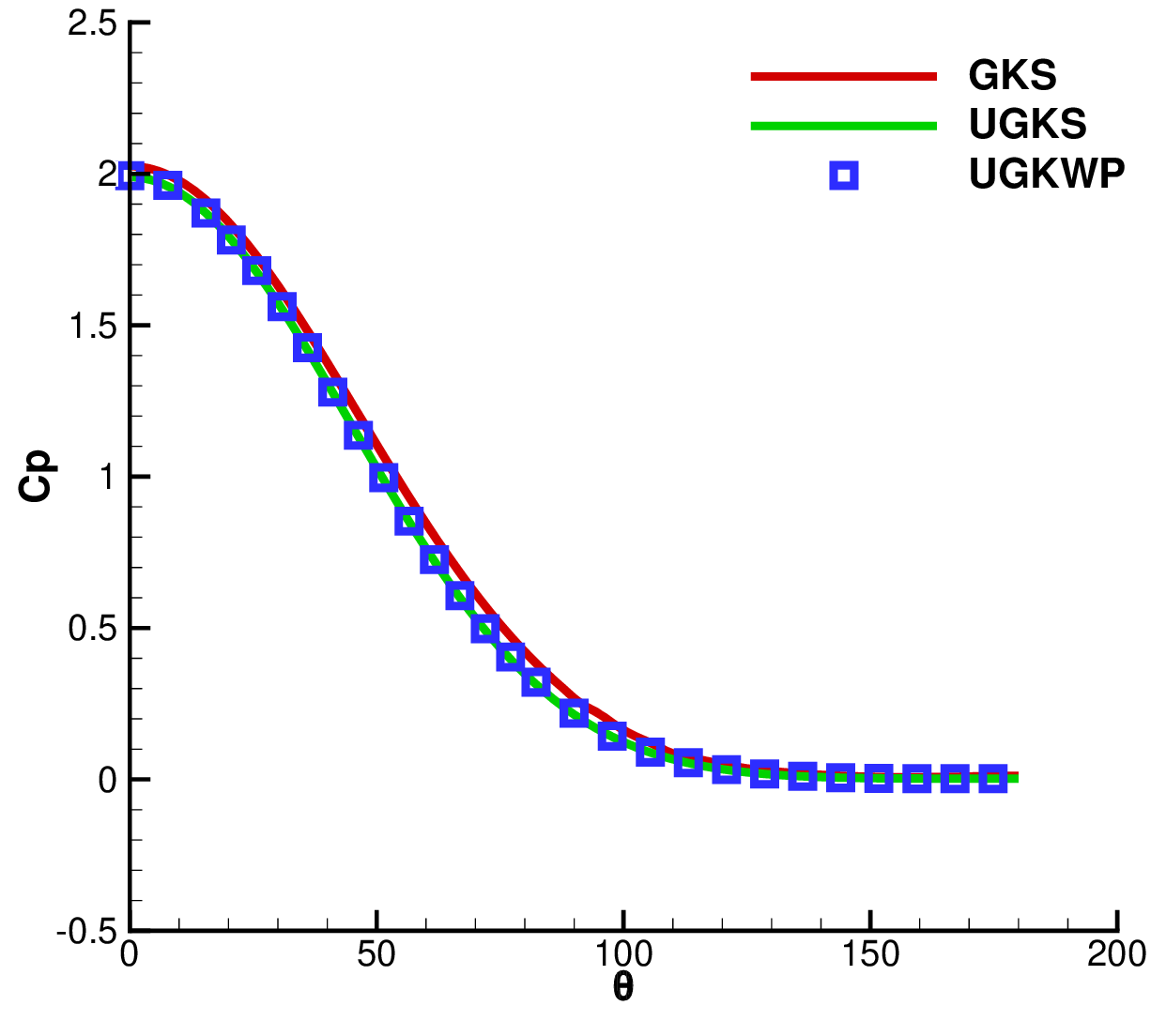}}
	\subfloat[]{\includegraphics[width=0.33\textwidth]{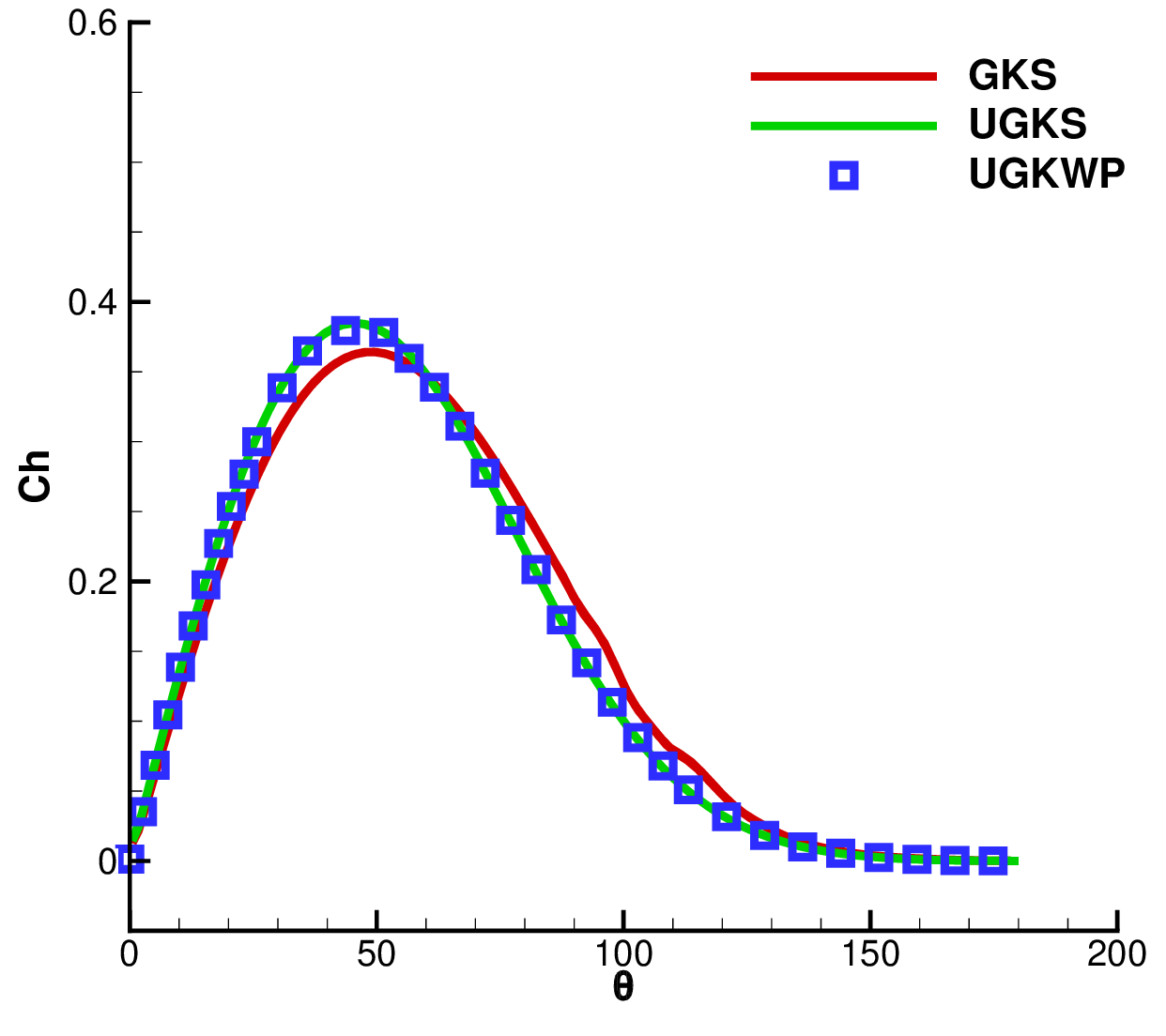}}
	\subfloat[]{\includegraphics[width=0.33\textwidth]{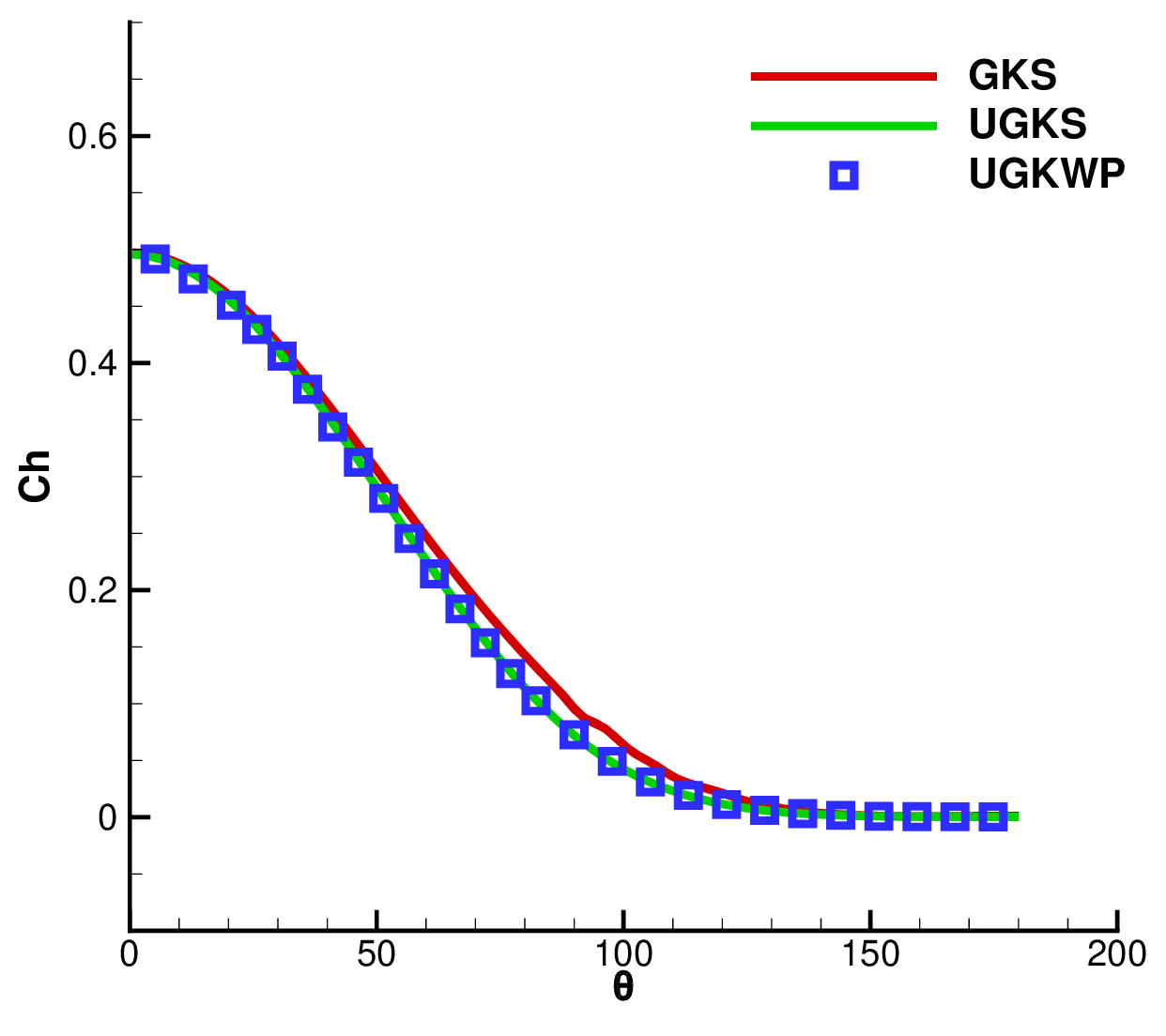}}\\
	\caption{Hypersonic flow at ${\rm Kn}_\infty = 0.4$ and ${\rm Ma}_\infty = 5$ passing over a circular cylinder by the GKS, UGKS, and UGKWP. Surface quantities distributions: (a) pressure coefficient, (b) shear stress coefficient, and (c) heat flux coefficient.}
	\label{fig:cylinder-surface-Kn04}
\end{figure}

For the rarefied case where ${\rm Kn}_\infty = 0.4$, with much stronger nonequilibrium,
figure~\ref{fig:cylinder-Kn04-field} shows significant difference in both the shock and wake regions.
The UGKS gives much thicker shock structure, as well as a lower pressure near the surface, which leads
to the departure of the GKS predicted $C_p$ with the multiscale methods in Fig.~\ref{fig:cylinder-surface-Kn04}.
Another noteworthy discrepancy lies in the downstream shift of the peak friction coefficient,
since the relaxation limit regime extends over a larger portion of the cylinder surface than ${\rm Kn}_\infty = 0.02$.

Surprisingly, heat flux predicted by the three methods shows excellent agreement. To investigate this
phenomenon with kinetic methodology, the distribution functions at the cylinder surface at $x = -1 $ in the
GKS and the UGKS simulation are compared. Based on the kinetic boundary, the incident distribution function
only adopts values when particle velocity $u > 0$, as a result, although the CE expansion distribution
function deviates greatly from the nonequilibrium distribution function, the incident portions are nearly identical.

\begin{figure}[H]
	\centering
	\includegraphics[width=0.6\textwidth]{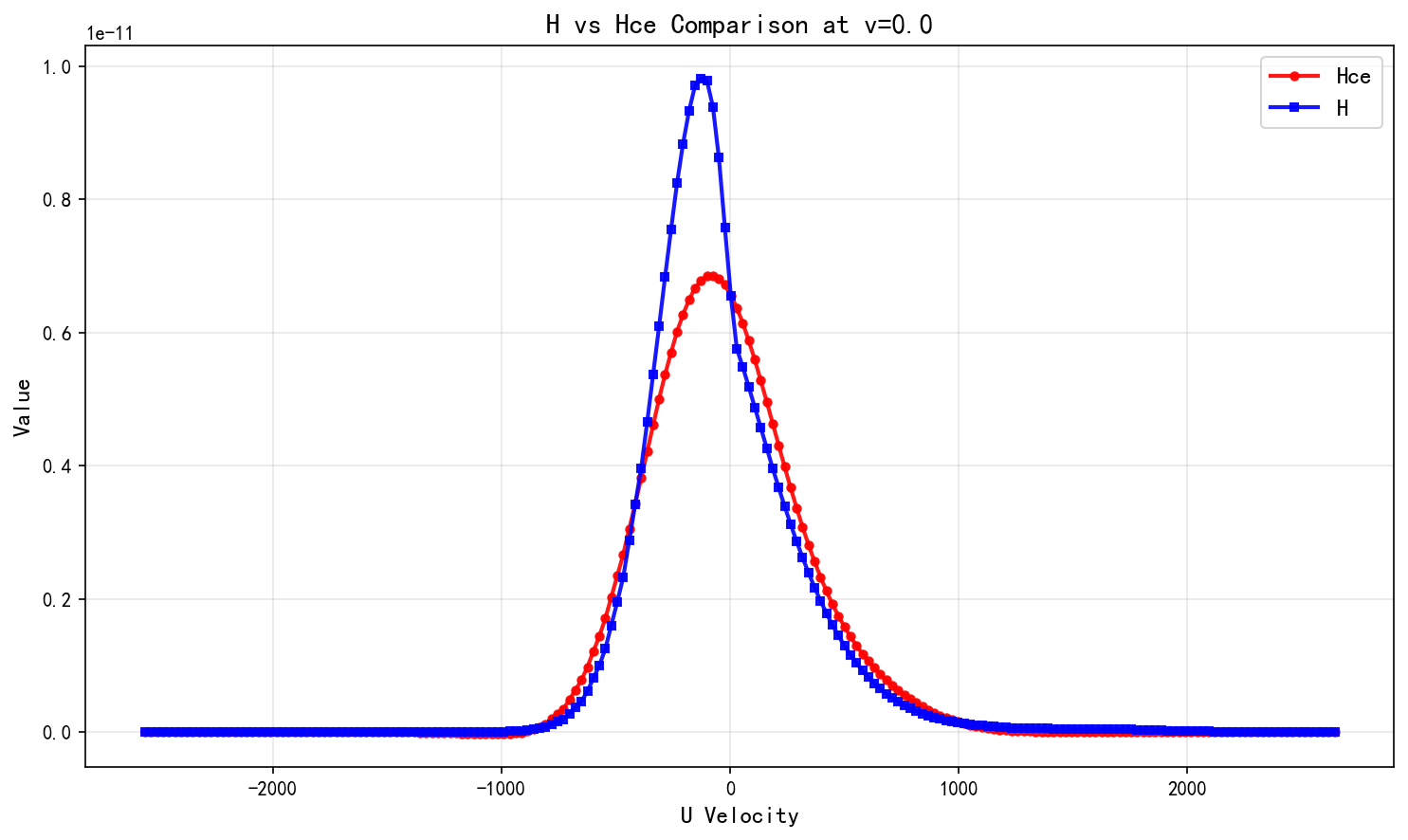}
	\caption{Hypersonic flow at ${\rm Kn}_\infty = 0.4$ and ${\rm Ma}_\infty = 5$ past a circular cylinder (GKS). Comparison of reduced nonequilibrium and CE expansion distribution functions.}
	\label{fig:cylinder-kn04-H-Hce}
\end{figure}

\subsection{Apollo~6 command module}\label{sec:case-apollo}

Simulations of hypersonic flow around an Apollo~6 command module are conducted for nitrogen
over a range of freestream conditions to represent three altitudes in the re-entry process, at $H = 85$, $100$, and $120$~km.
All the initial and boundary conditions as well as the gas properties are given in Table~\ref{tab:apollo-altitude}.
Following the reference conditions of the following papers~\cite{moss2006dsmc,zhang2024conservative}, the angle of attack is fixed at $-25^\circ$
and the freestream velocity is set as $U_\infty = 9.6$~km/s.
The reference length $L_{ref}$ is taken as the base diameter of the capsule $d = 3.91$~m,
and the reference area for normalizing the aerodynamic coefficients is $A_{ref} = \pi d^2/4 = 12.0$~m$^2$.
Note that the thermal nonequilibrium is not considered in this study.
\begin{table}[htb]
	\centering
	\caption{Freestream and wall parameters for nitrogen in the Apollo~6 command module case at selected altitudes.}
	\begin{tabular}{ccccccccc}
		\hline
		Altitude (km) & Ma & Kn & $\rho_\infty$ (kg/m$^3$) & $T_\infty$ (K) & $T_w$ (K) & $R$ (J/(kg$\cdot$K)) & $\omega$ \\
		\hline
		120 & 23.72 & 0.773 & $2.26 \times 10^{-8}$ & 368 & 675 & 317.85 & 0.7535 \\
		100 & 33.96 & 0.0338 & $5.58 \times 10^{-7}$ & 194 & 1146 & 294.24 & 0.7476 \\
		85 & 35.59 & 0.0024 & $7.96 \times 10^{-6}$ & 181 & 1598 & 287.10 & 0.7471 \\
		\hline
	\end{tabular}
	\label{tab:apollo-altitude}
\end{table}

\begin{figure}[H]
	\centering
	\subfloat[$H=85$~km]{\includegraphics[width=0.32\textwidth]{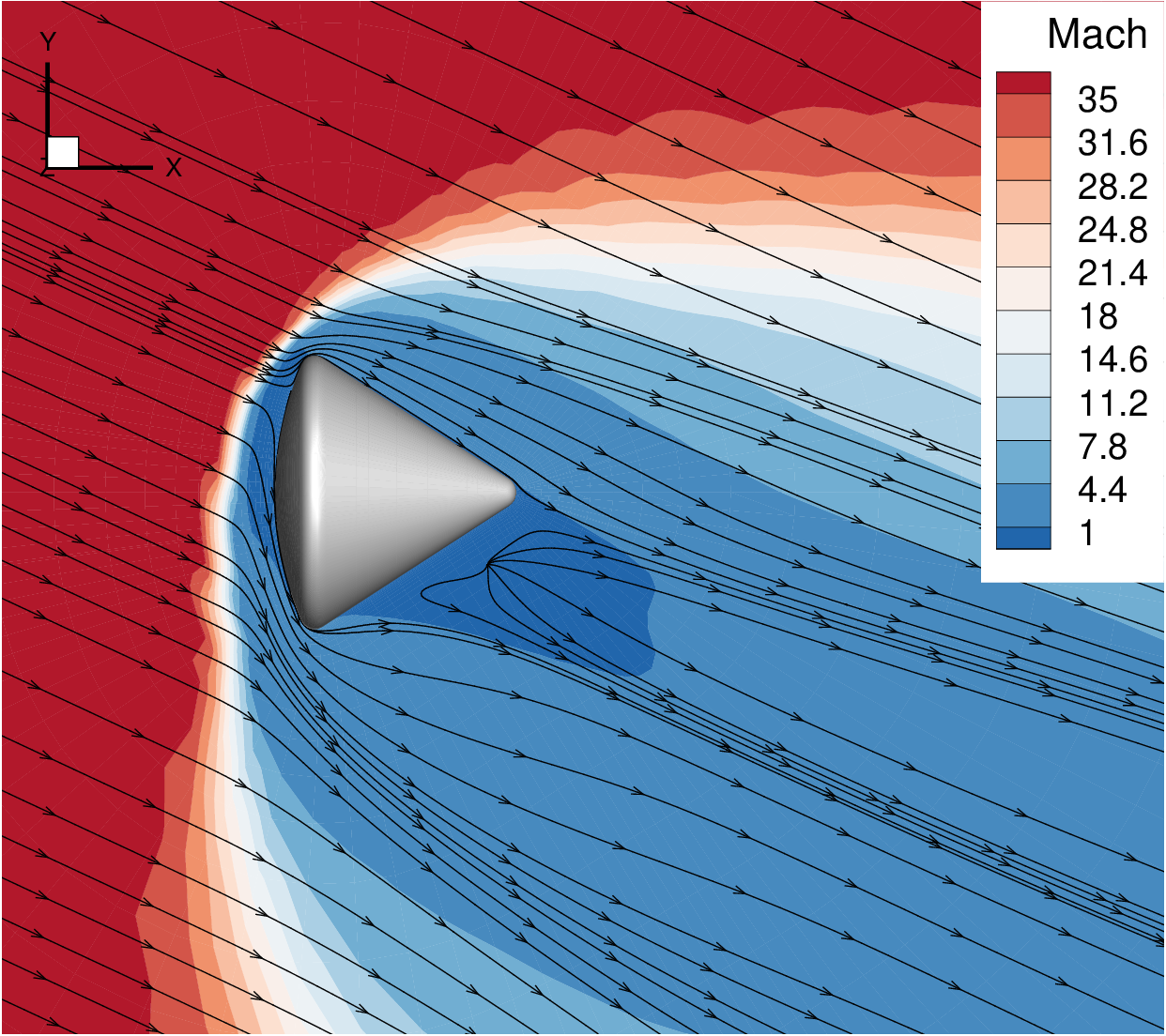}}~
	\subfloat[$H=100$~km]{\includegraphics[width=0.32\textwidth]{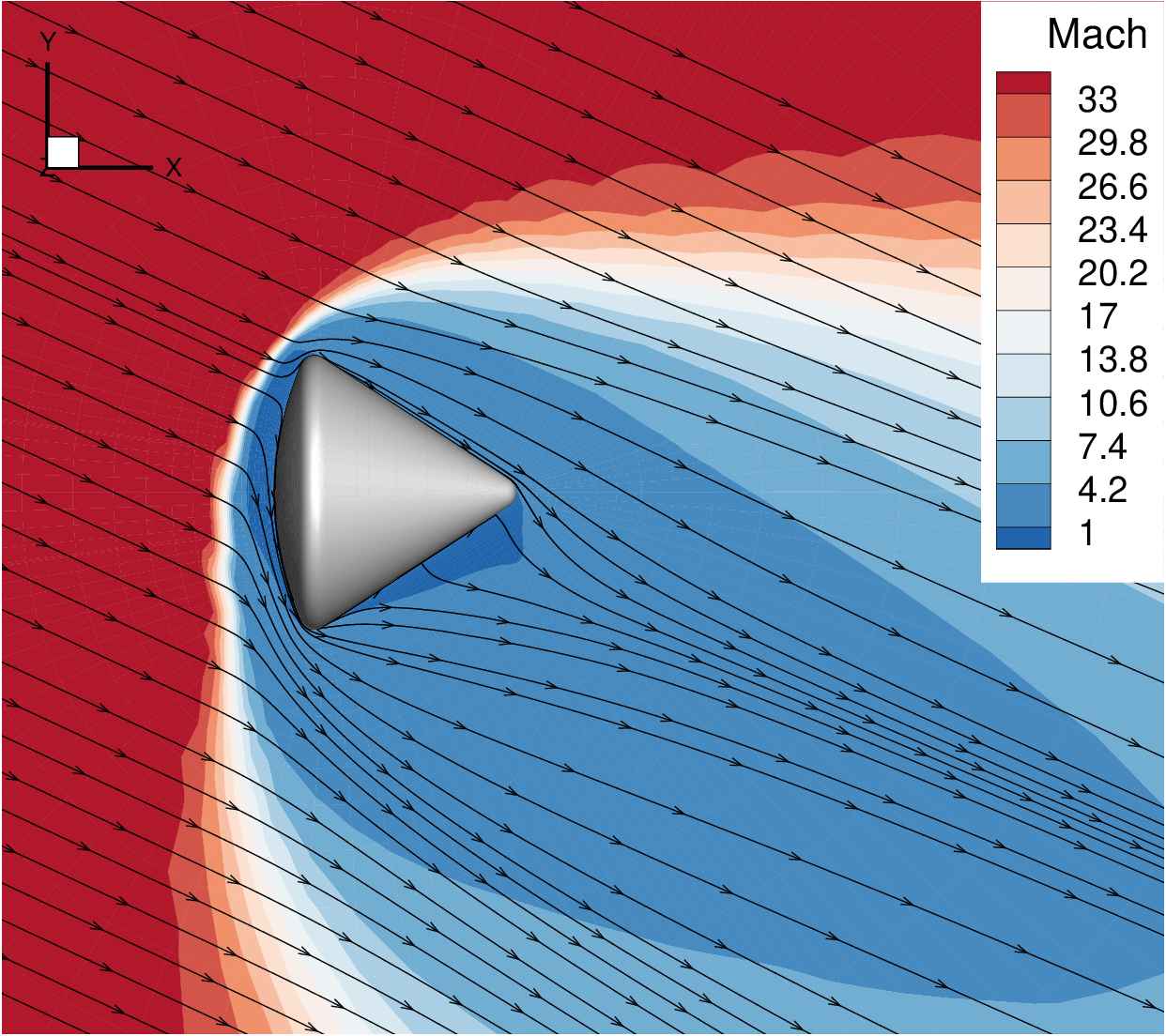}}~
	\subfloat[$H=120$~km]{\includegraphics[width=0.32\textwidth]{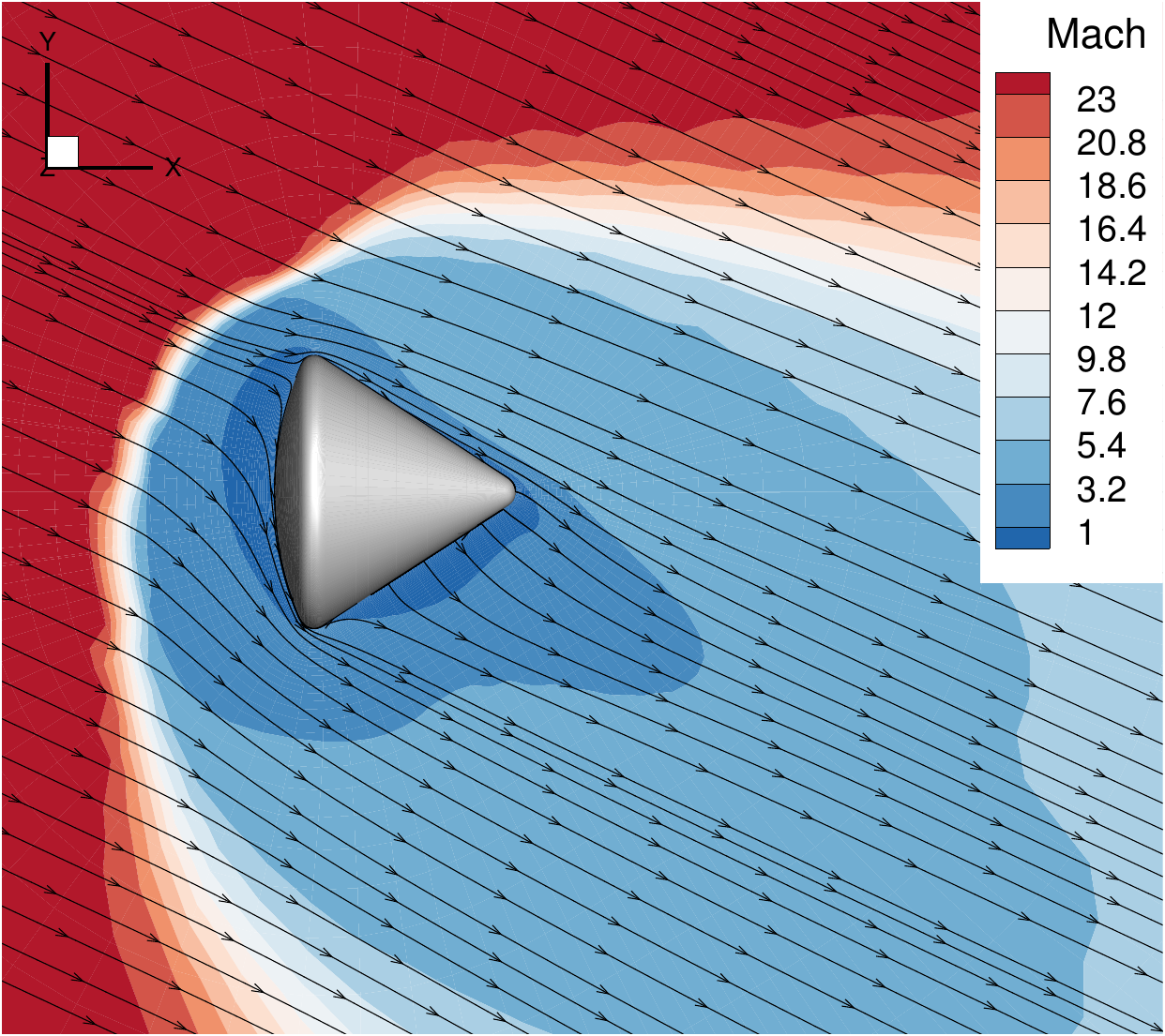}}\\
	\caption{Mach number contours around the Apollo~6 command module at three altitudes, with streamlines on the symmetry ($Z$) plane.}
	\label{fig:apollo-contour}
\end{figure}

Figure~\ref{fig:apollo-contour} shows Mach number contours with streamlines on the symmetry plane around the capsule
at different altitudes computed by the GKS. Both the shock thickness and the bow-shock distance from the capsule increase
with altitude. For the case with $H = 120$~km, the GKS provides false shock prediction, due to the strong nonequilibrium effect
under the Kn number of 0.77. The streamlines reveal a well-defined near-wake recirculation on the capsule afterbody at $H = 85$~km,
where the shear layer separating from the shoulder reattaches downstream and drives reverse flow toward the base surface.
As the altitude increases to $H = 100$~km and above, increased rarefaction smears the shear layer and suppresses reattachment.

\begin{figure}[H]
	\centering
	\subfloat[]{\includegraphics[width=0.32\textwidth]{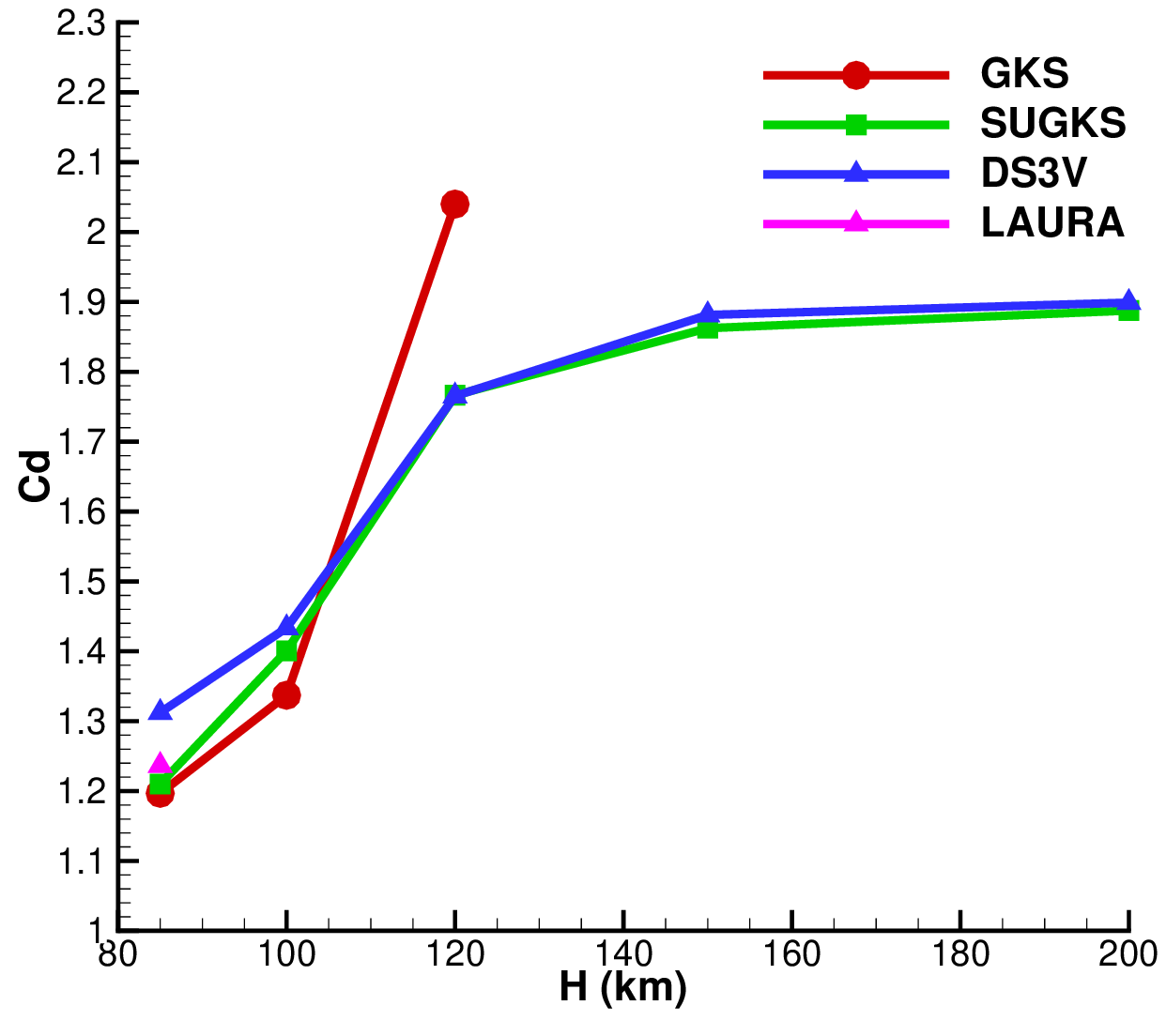}}~
	\subfloat[]{\includegraphics[width=0.32\textwidth]{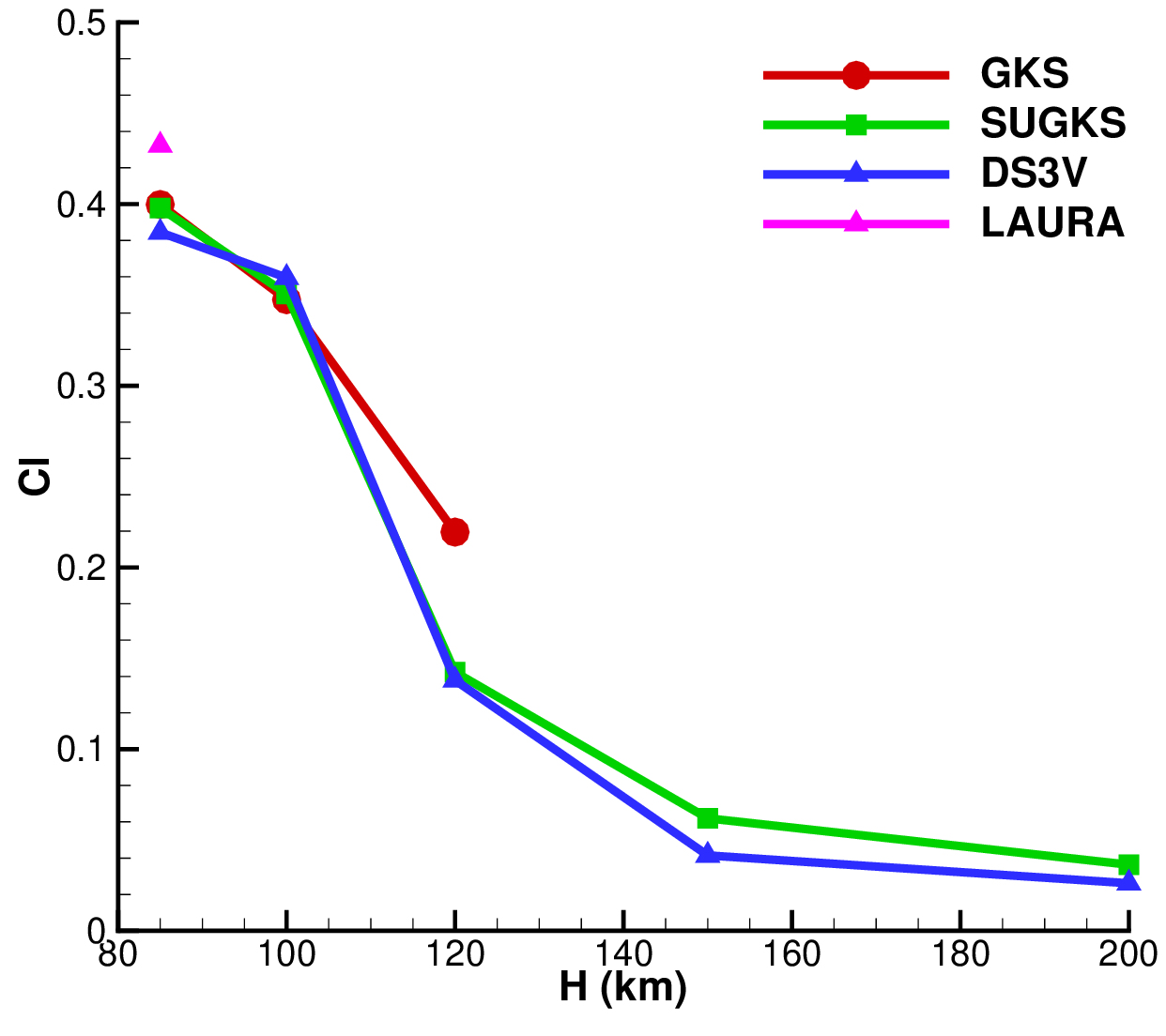}}~
	\subfloat[]{\includegraphics[width=0.32\textwidth]{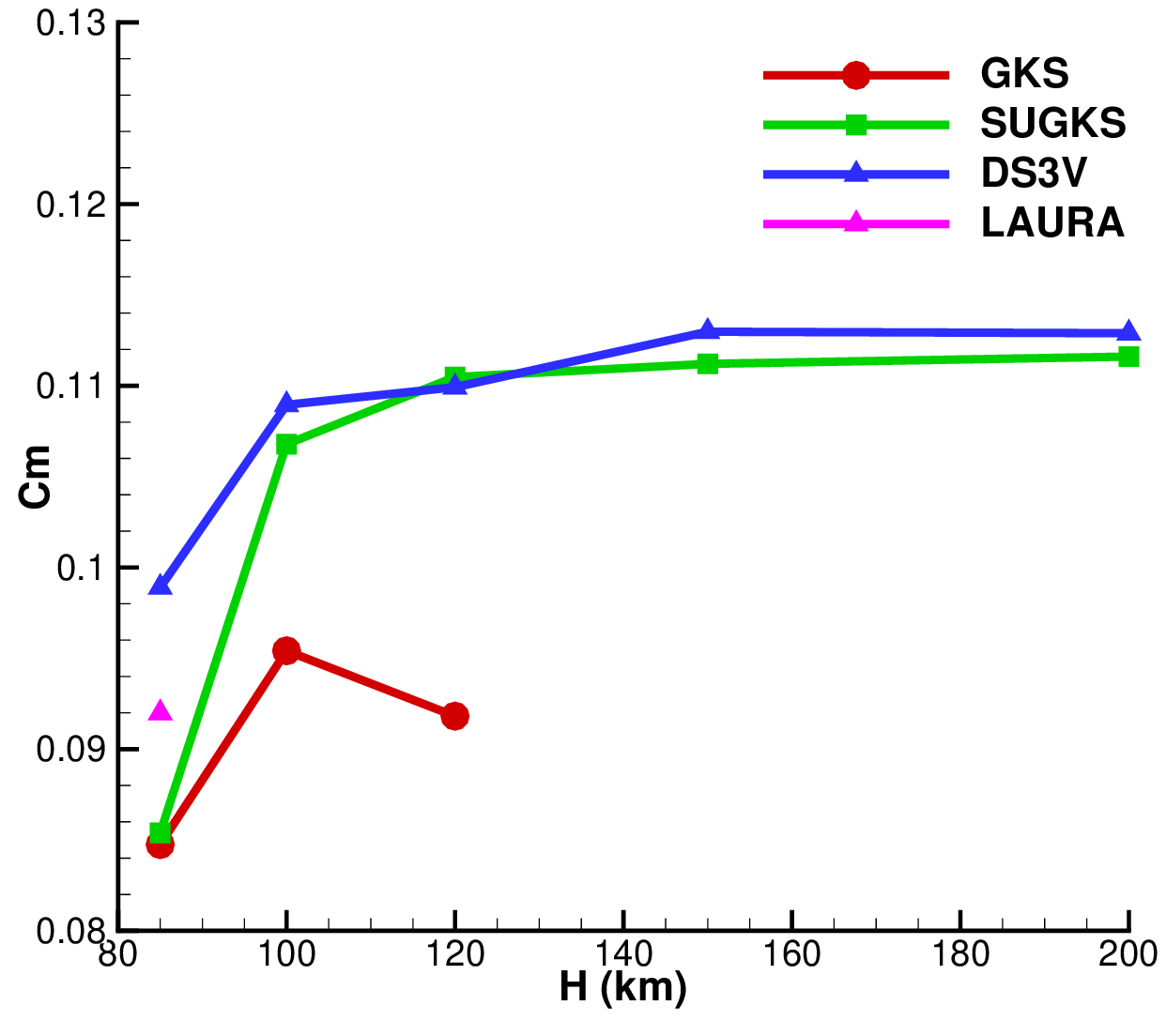}}\\
	\caption{Apollo~6 command module: integrated aerodynamic coefficients versus altitude $H$ for the GKS, SUGKS, and DSMC. (a) Drag coefficient $C_d$, (b) lift coefficient $C_l$, and (c) pitching moment coefficient $C_m$.}
	\label{fig:apollo-coeffs}
\end{figure}
For integrated aerodynamic quantities versus altitude, the GKS is further
compared with a conservative implicit simplified unified gas-kinetic scheme (SUGKS)~\cite{zhang2024conservative},
and with the DSMC data obtained from two different DSMC codes of LAURA and DS3V~\cite{moss2006dsmc}.
Figure~\ref{fig:apollo-coeffs} summarizes $C_d$, $C_l$, and $C_m$ as functions of altitude in one figure.
For the case with $H = 85$~km, the GKS provides nearly identical results with the SUGKS, while they all deviate from the DSMC.
Considering the difference in the data given by different DSMC codes, the multiscale method of SUGKS is more reliable.
When the altitude increases to $H = 100$~km, the GKS can still work well and give acceptable results in both drag and lift coefficients,
but the pitching moment coefficient shows a deviation from the DSMC and the SUGKS.
Comparing with the drag or lift coefficients, the pitching moment is more sensitive to the nonequilibrium effect.
The drag and lift forces are dominated by the windward stagnation pressure and the forebody, while the pitching moment is
affected by the shear stress distribution, especially the leeward afterbody.
The GKS predicts all the aerodynamic coefficients with large discrepancy at $H = 120$~km, where the nonequilibrium effect
dominates most regions.

In this case, we observe that the GKS with the kinetic boundary condition gives good predictions below $H = 100$~km.
The large blunt forebody of the Apollo capsule creates a windward region where the strong compression keeps the flow
more continuous than the incoming flow and the leeward region, so that the drag and lift are well captured.
The nonequilibrium effect first appears in the leeward region, causing the failure in pitching moment prediction
while having limited impact on the forebody-dominated forces.

\subsection[Hypersonic flow over a 70-degree blunted cone with a cylindrical sting]{Hypersonic flow over a \texorpdfstring{$70^\circ$}{70 degree} blunted cone with a cylindrical sting}\label{sec:case-70cone}

Hypersonic flow over a $70^\circ$ blunted cone with a cylindrical sting is a well-established benchmark for hypersonic flows.
The experimental data were reported in~\cite{allegre1997forces} from tests in the SR3 wind tunnel at CNRS Meudon.
The model, whose forebody configuration matches that of the Mars Pathfinder probe, was instrumented with an
external balance that provided direct measurements of drag, lift, and pitching moment.
This test case has since been widely adopted for validation of numerical methods~\cite{palharini2015benchmark,moss1995dsmc,schouler2020survey,cao2026adaptive,cao2024nccr}.
Two experimental conditions from the reference are adopted in the
present study, corresponding to ${\rm Kn}_\infty = 0.013$ (Case~1) and ${\rm Kn}_\infty = 5.4\times 10^{-4}$ (Case~3).
Scaling the model to the full-scale Mars Pathfinder vehicle (base diameter $2.65$~m)
using the US Standard Atmosphere 1976, these freestream Knudsen numbers correspond to
$H \sim 80$~km and $H \sim 57$~km, respectively.
The flow parameters, together with the freestream mean free path $\lambda_\infty$ compiled by
Schouler et al.~\cite{schouler2020survey}, are summarized in Table~\ref{tab:cone-conditions}.
In the experiment, the dynamic viscosity coefficient is calculated by a combination of the Sutherland viscosity law
(for $T \ge 100$~K) and a linear viscosity law (for $T < 100$~K), from which $\lambda_\infty$ is derived.
The working gas is nitrogen (see Table~\ref{tab:gas}),
with the base diameter of $50$~mm taken as the reference length $L_{ref}$ and the base cross-sectional area $\pi d^2/4$ used for normalizing the aerodynamic coefficients.
Simulations are performed at four angles of attack
$0^\circ$, $10^\circ$, $20^\circ$, and $30^\circ$, with the height of the first cell layer at the wall set to $0.075$~mm.

\begin{table}[htb]
	\centering
	\caption{Flow conditions for the two experimental cases of the $70^\circ$ blunted cone with sting~\cite{allegre1997forces}.}
	\begin{tabular}{@{}lcc@{}}
		\hline
		& Case 1 & Case 3 \\
		\hline
		Kn$_\infty$ & 0.013 & $5.4\times 10^{-4}$ \\
		Ma$_\infty$ & 20.2 & 20.5 \\
		$U_\infty$ (m/s) & 1503 & 1634 \\
		$T_\infty$ (K) & 13.3 & 15.3 \\
		$\rho_\infty$ (kg/m$^3$) & $1.73\times 10^{-5}$ & $4.66\times 10^{-4}$ \\
		$\lambda_\infty$ (mm) & 0.671 & 0.027 \\
		$T_w$ (K) & 350 & 300 \\
		\hline
	\end{tabular}
	\label{tab:cone-conditions}
\end{table}

\begin{figure}[H]
	\centering
	\subfloat[$U$]{\includegraphics[width=0.32\textwidth]{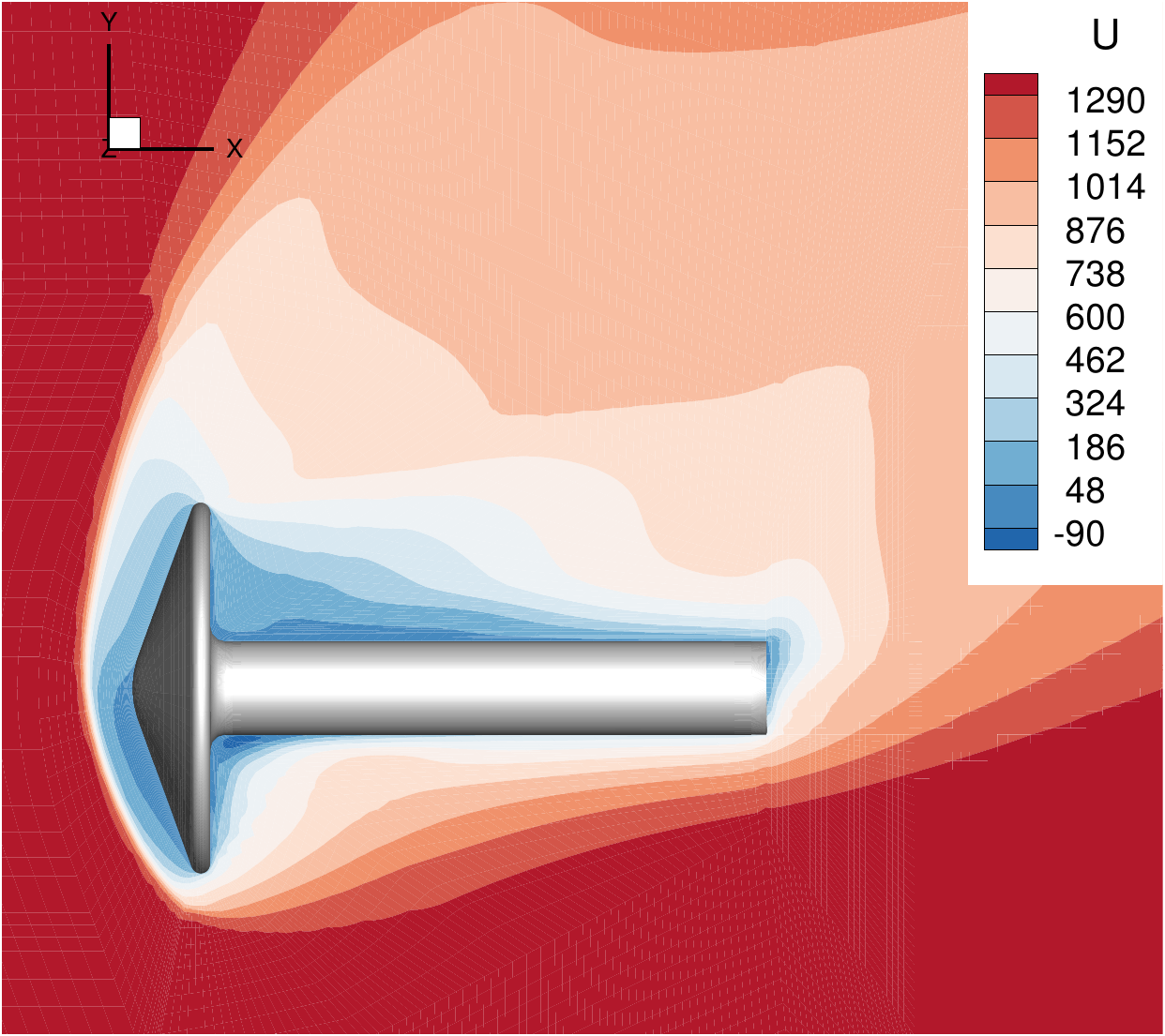}}~
	\subfloat[$\rho$]{\includegraphics[width=0.32\textwidth]{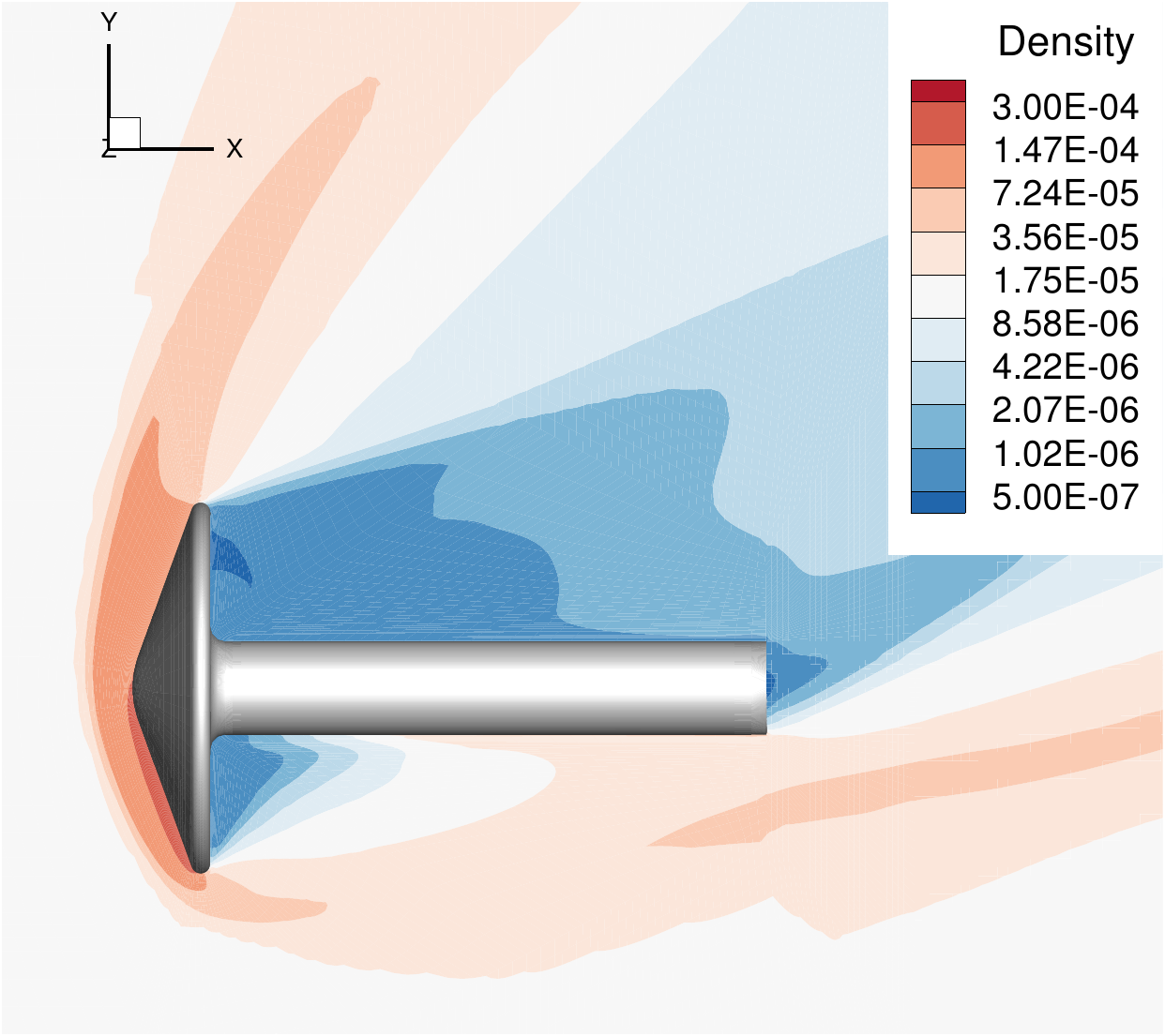}}~
	\subfloat[$T$]{\includegraphics[width=0.32\textwidth]{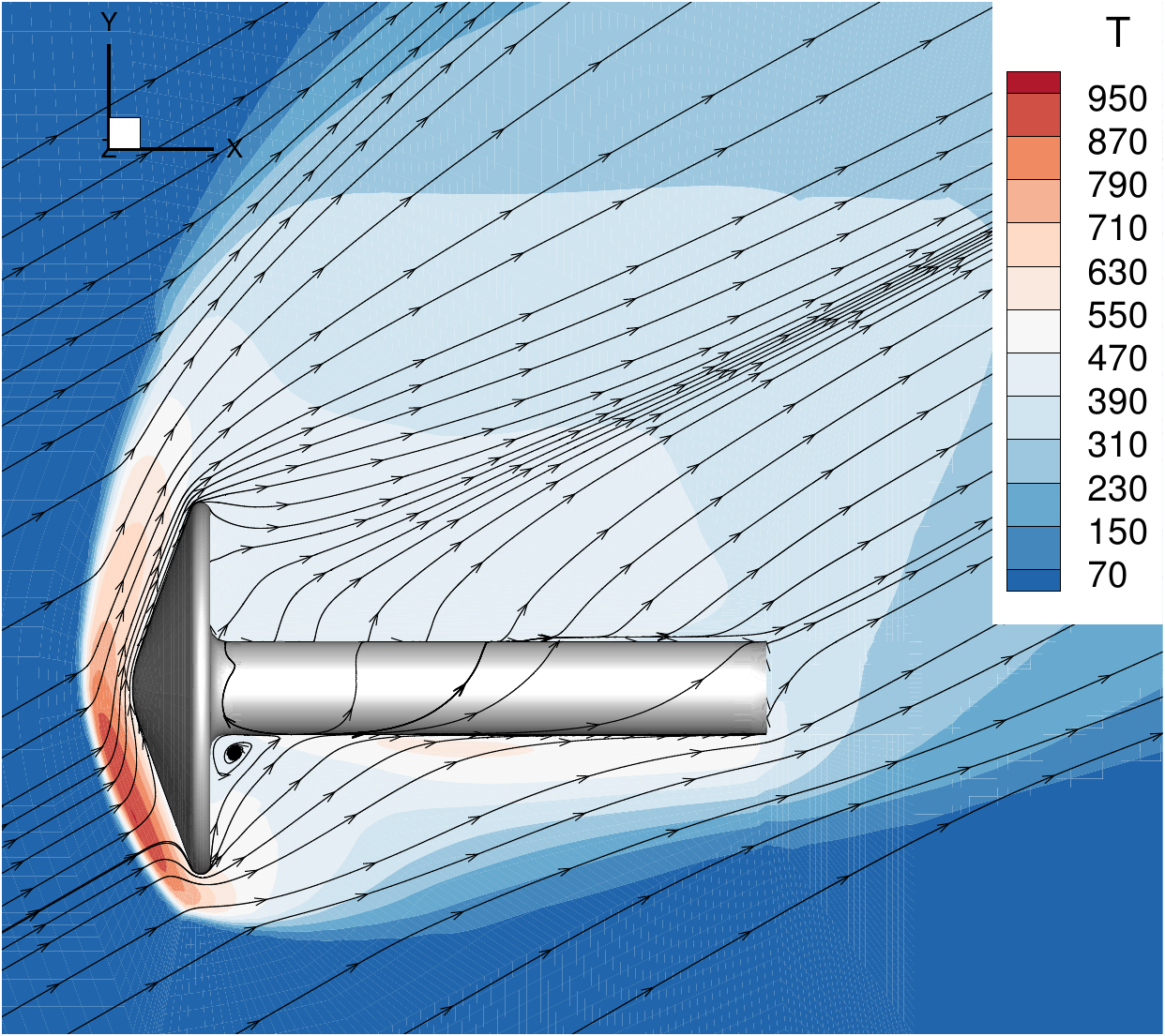}}\\
	\caption{Hypersonic flow over a $70^\circ$ blunted cone with sting at $\alpha = 30^\circ$, ${\rm Ma}_\infty = 20.2$, and ${\rm Kn}_\infty = 0.013$ ($H \sim 80$~km) by the GKS. Contours on the symmetry plane: (a) $x$-direction velocity $U$, (b) density, and (c) temperature with streamlines.}
	\label{fig:cone-U-case1}
\end{figure}

\begin{figure}[H]
	\centering
	\subfloat[$U$]{\includegraphics[width=0.32\textwidth]{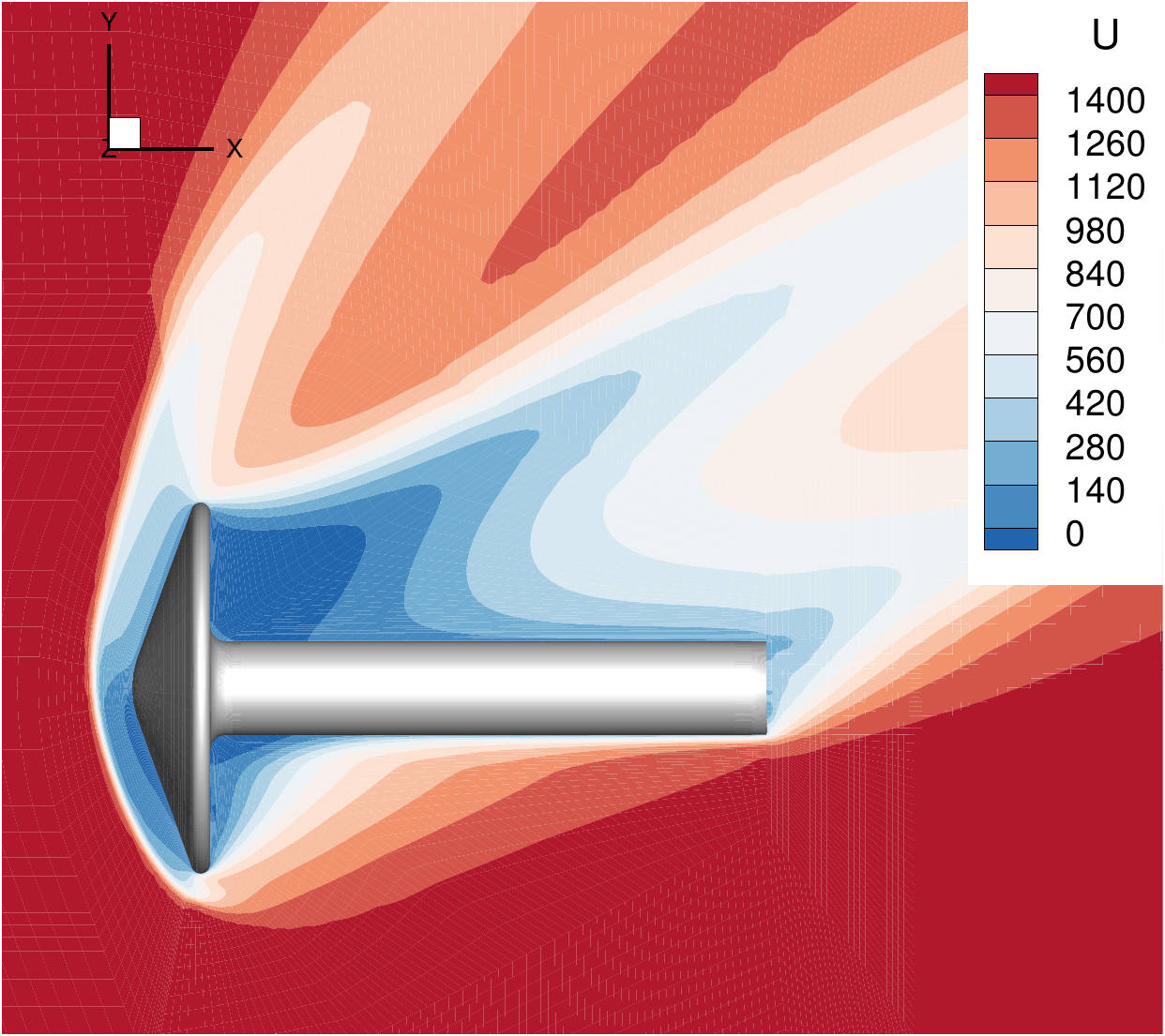}}~
	\subfloat[$\rho$]{\includegraphics[width=0.32\textwidth]{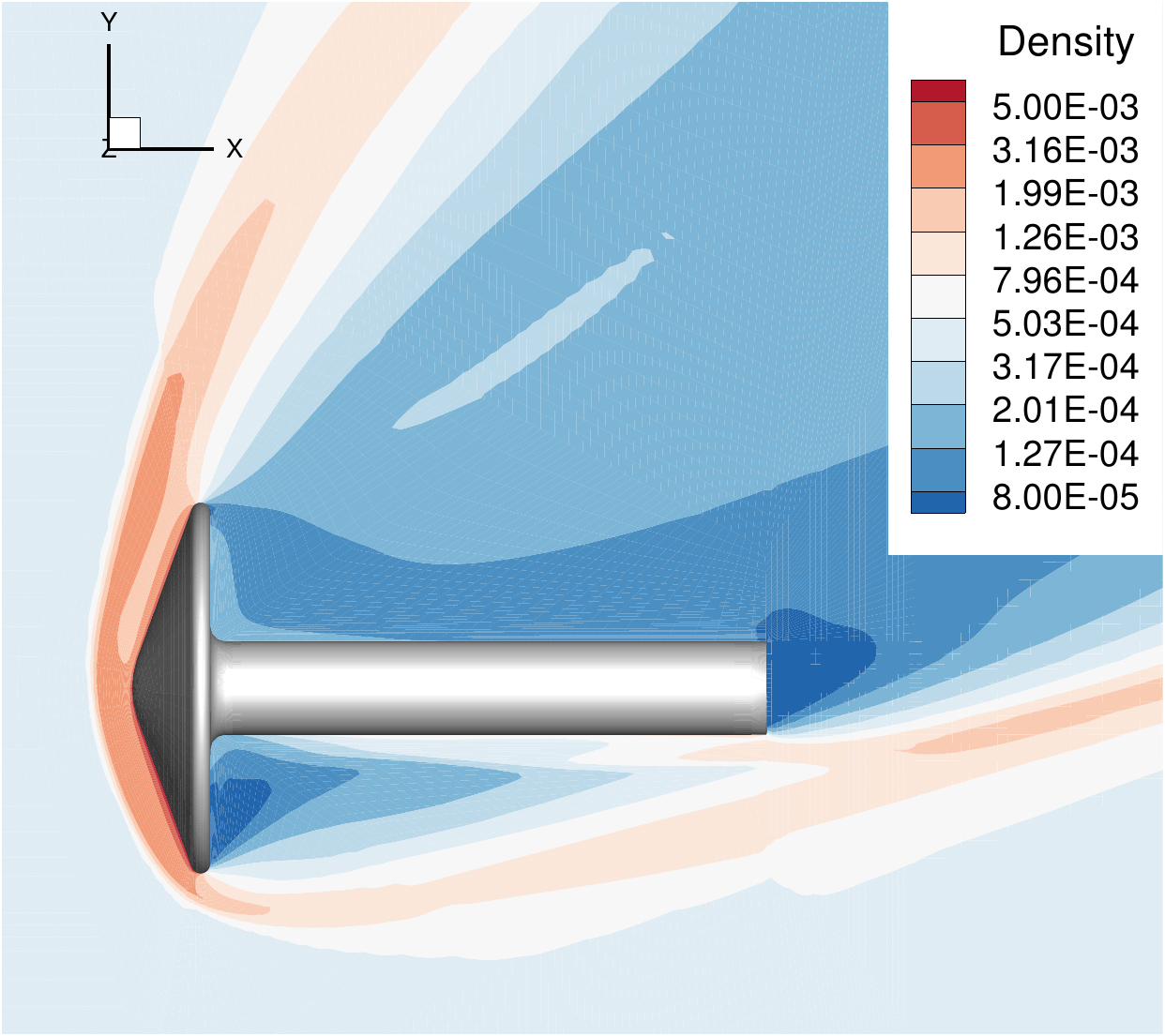}}~
	\subfloat[$T$]{\includegraphics[width=0.32\textwidth]{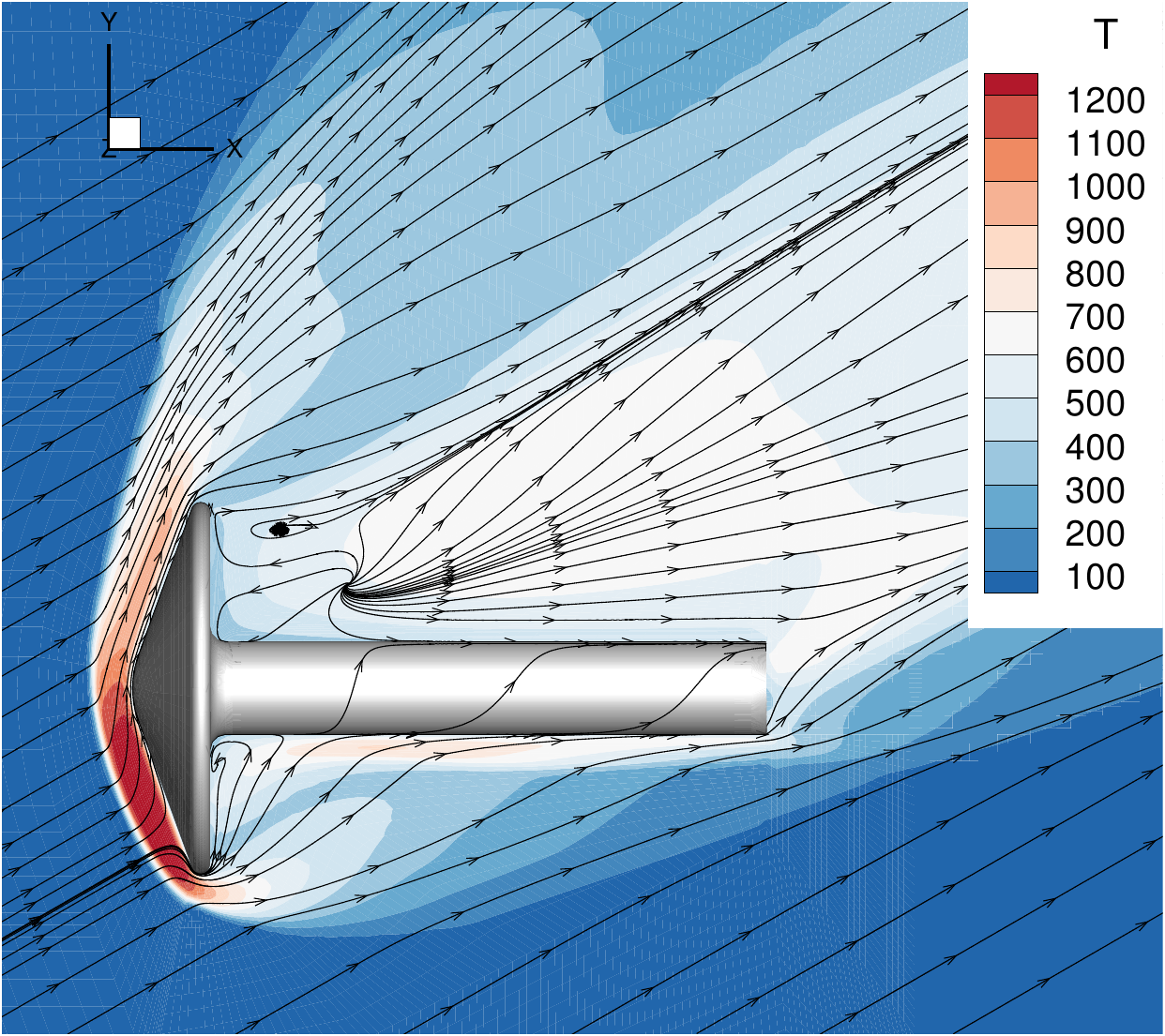}}\\
	\caption{Hypersonic flow over a $70^\circ$ blunted cone with sting at $\alpha = 30^\circ$, ${\rm Ma}_\infty = 20.5$, and ${\rm Kn}_\infty = 5.4\times 10^{-4}$ ($H \sim 57$~km) by the GKS. Contours on the symmetry plane: (a) $x$-direction velocity $U$, (b) density, and (c) temperature with streamlines.}
	\label{fig:cone-U-case3}
\end{figure}
Figures~\ref{fig:cone-U-case1} and \ref{fig:cone-U-case3} show $x$-direction velocity $U$, density, and temperature contours
with streamlines on the symmetry plane at $\alpha = 30^\circ$ for ${\rm Kn}_\infty = 0.013$ ($H \sim 80$~km) and $5.4\times 10^{-4}$ ($H \sim 57$~km), respectively.
For the case with ${\rm Kn}_\infty = 0.013$, a small circulation can be observed in the leeward region of the forebody,
below the cylindrical sting. This pattern is due to the
flow that passes around the lower edge of the forebody and recirculates after impinging on the cylindrical sting.
For the case with ${\rm Kn}_\infty = 5.4\times 10^{-4}$, a distinct flow feature
appears on the upper side
of the cylindrical sting just behind the forebody. The symmetry-plane streamlines in this region
exhibit a source-like pattern, diverging outward from a localized zone. This apparent source is a
three-dimensional effect: the flow diverges laterally around the forebody and converges back
toward the symmetry plane in the near wake, producing the outwardly diverging streamlines.
This is similar to the reverse flow structure observed in the Apollo~6 command module.

\begin{figure}[H]
	\centering
	\subfloat[${\rm Kn}_\infty = 0.013$ ($H \sim 80$~km)]{\includegraphics[width=0.48\textwidth]{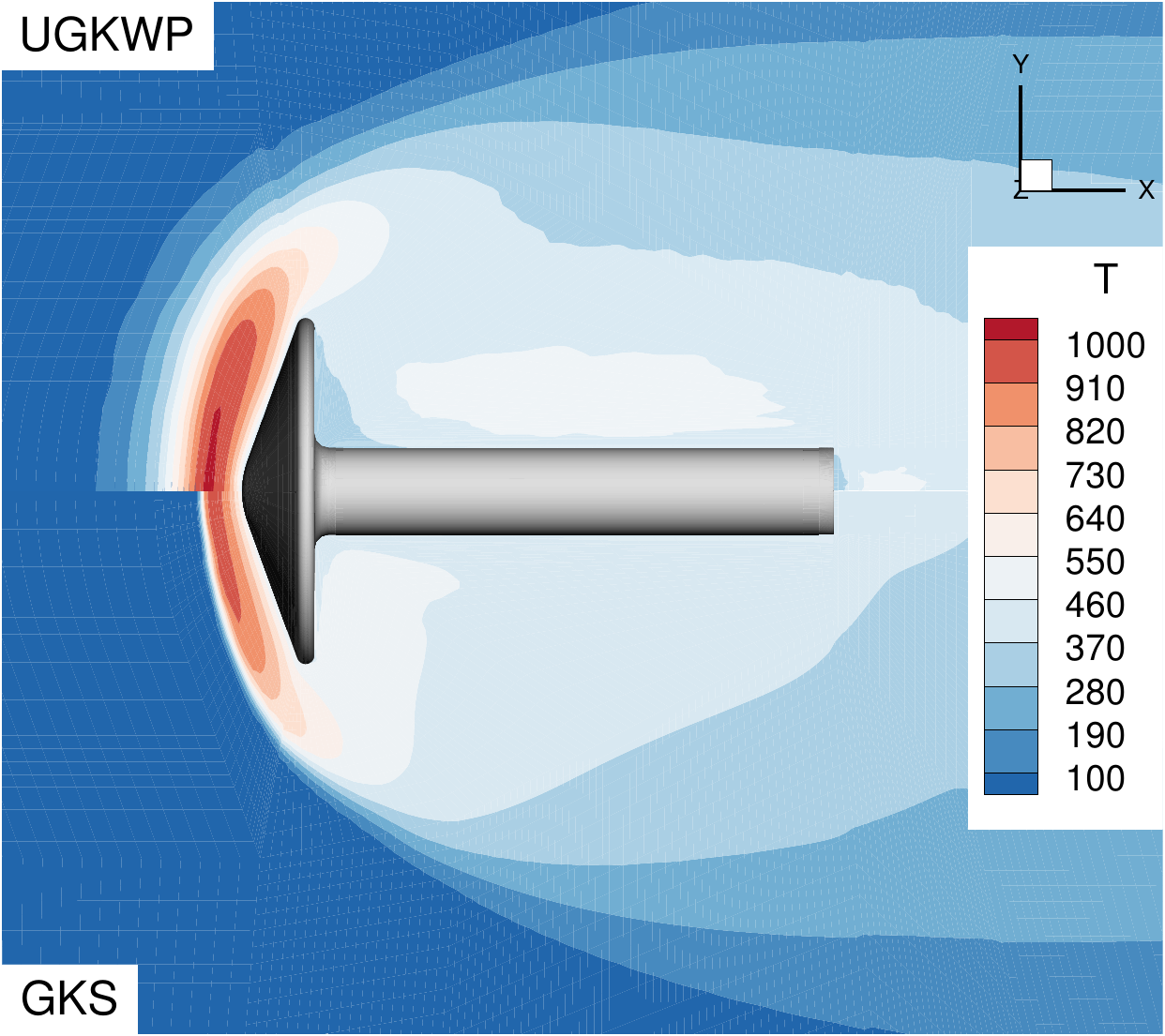}}~
	\subfloat[${\rm Kn}_\infty = 5.4\times 10^{-4}$ ($H \sim 57$~km)]{\includegraphics[width=0.48\textwidth]{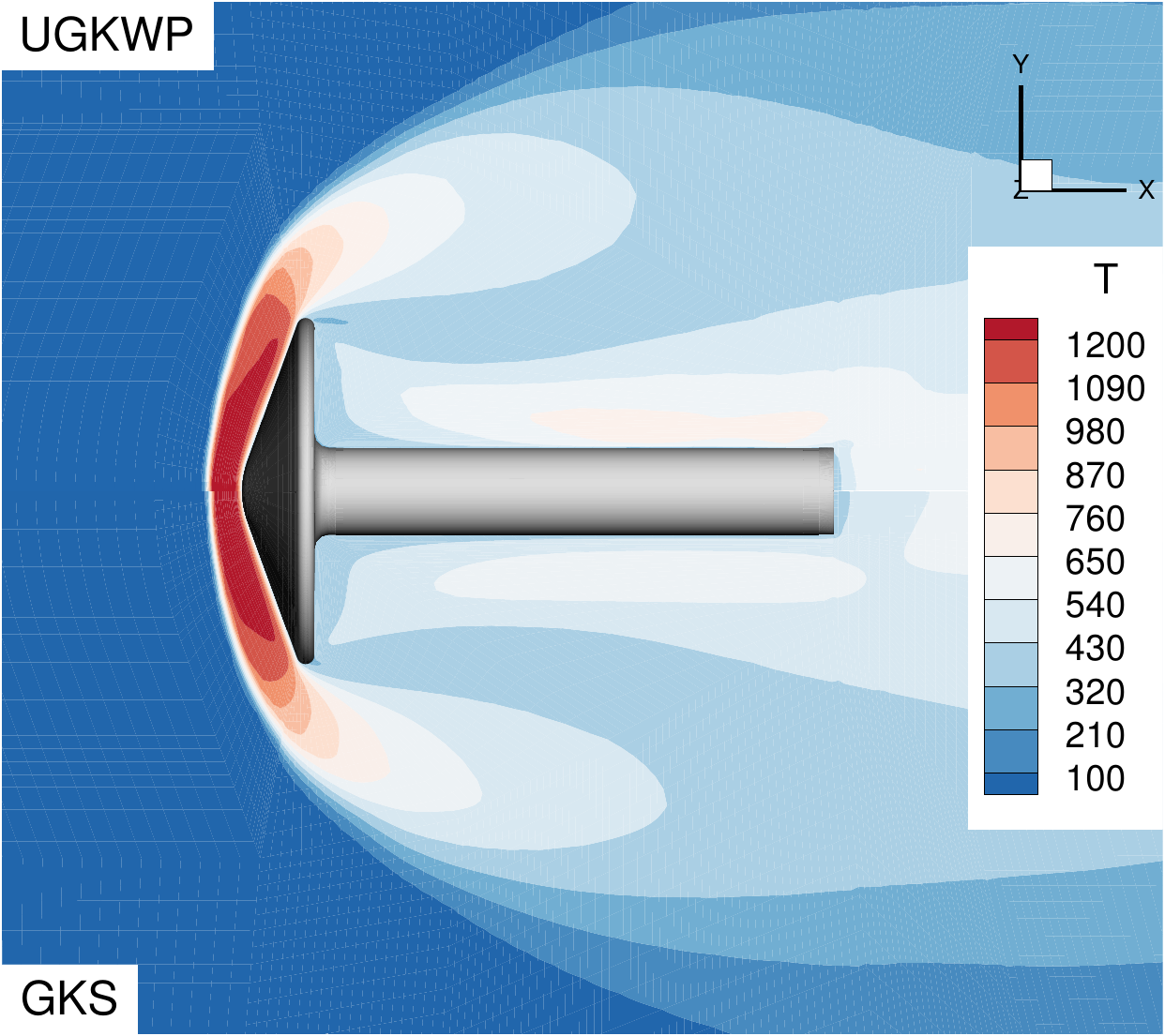}}
	\caption{Comparison of the time-averaged temperature by the UGKWP and the temperature by the GKS on the symmetry plane of the $70^\circ$ blunted cone with sting at $\alpha = 0^\circ$. (a) ${\rm Ma}_\infty = 20.2$, ${\rm Kn}_\infty = 0.013$ ($H \sim 80$~km); (b) ${\rm Ma}_\infty = 20.5$, ${\rm Kn}_\infty = 5.4\times 10^{-4}$ ($H \sim 57$~km).}
	\label{fig:cone-T-compare}
\end{figure}

Figure~\ref{fig:cone-T-compare} compares the time-averaged temperature by the UGKWP and the
temperature by the GKS along the symmetry plane at $\alpha = 0^\circ$ for both cases.
For case at ${\rm Kn}_\infty = 5.4\times 10^{-4}$ ($H \sim 57$~km), the GKS predicts almost identical results
with the UGKWP at the forebody, while exhibits a lower temperature around the cylindrical sting at
the leeward region, showing that the nonequilibrium effect
can still be observed in such regions while the incoming flow is near-continuum.
At ${\rm Kn}_\infty = 0.013$ ($H \sim 80$~km), more significant deviations exist in both the leeward region
and the shock region ahead of the forebody, as indicated in the cylinder flow section~\ref{sec:case-cylinder}.

\begin{figure}[H]
	\centering
	\subfloat[]{\includegraphics[width=0.33\textwidth]{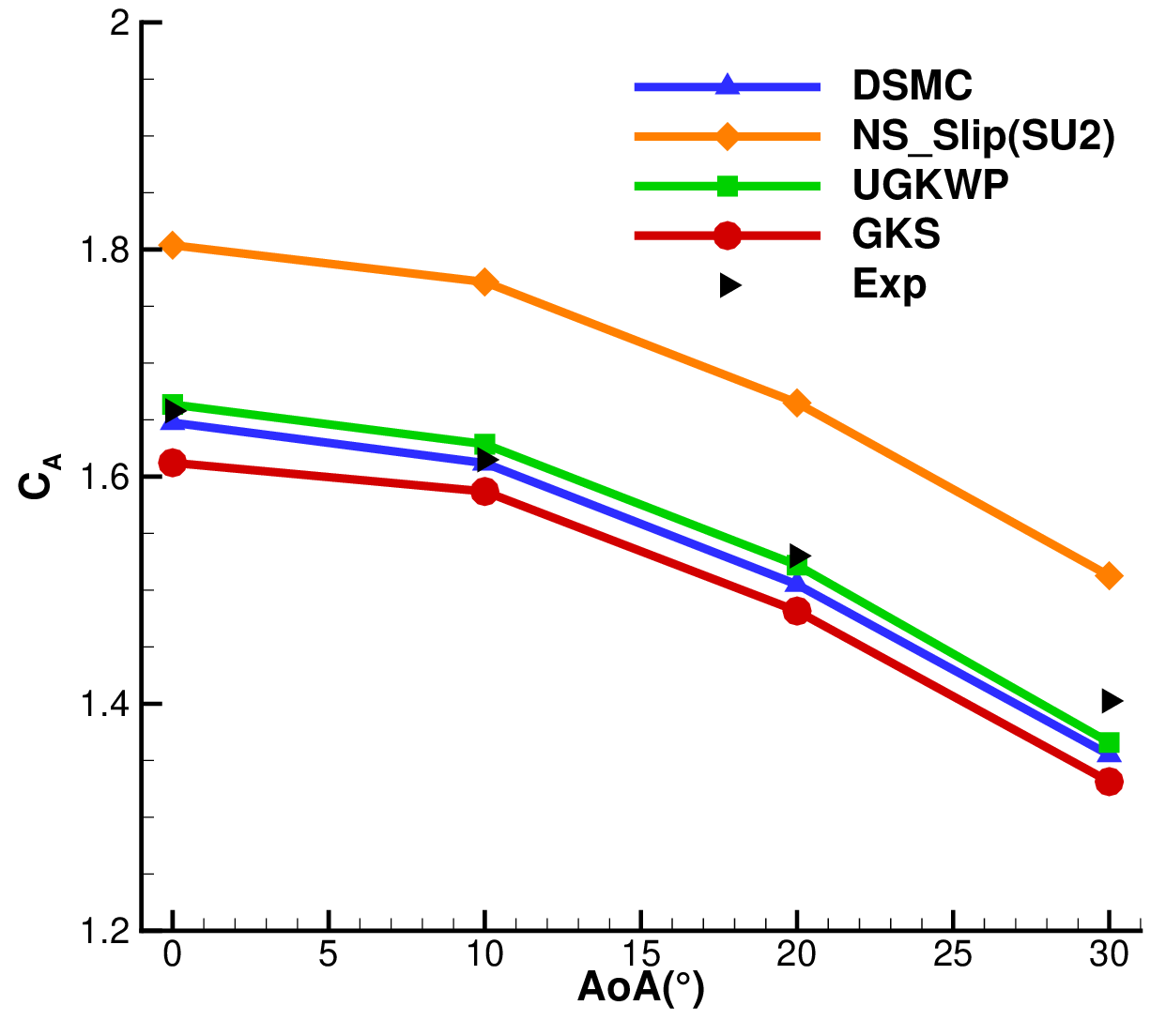}}
	\subfloat[]{\includegraphics[width=0.33\textwidth]{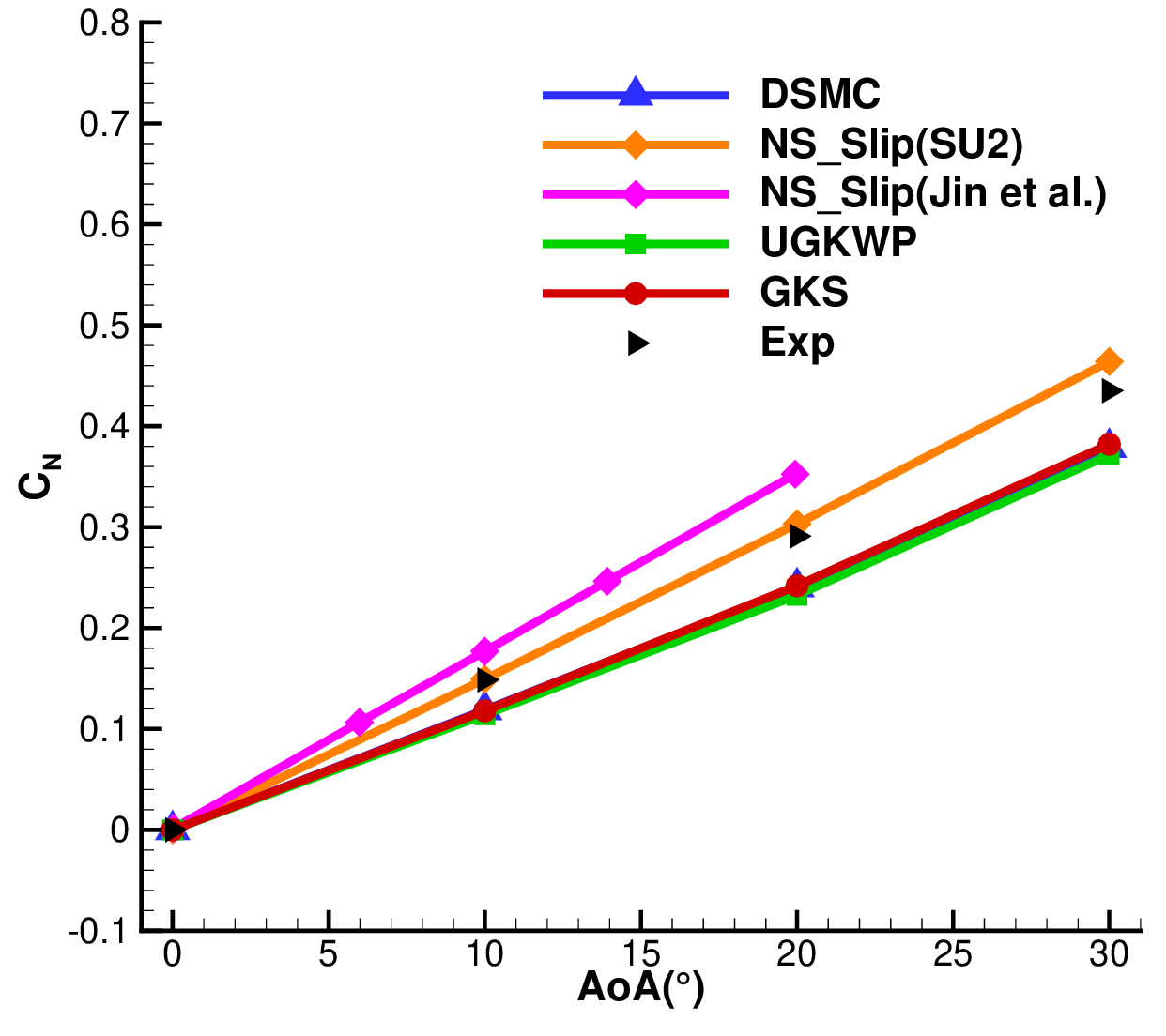}}
	\subfloat[]{\includegraphics[width=0.33\textwidth]{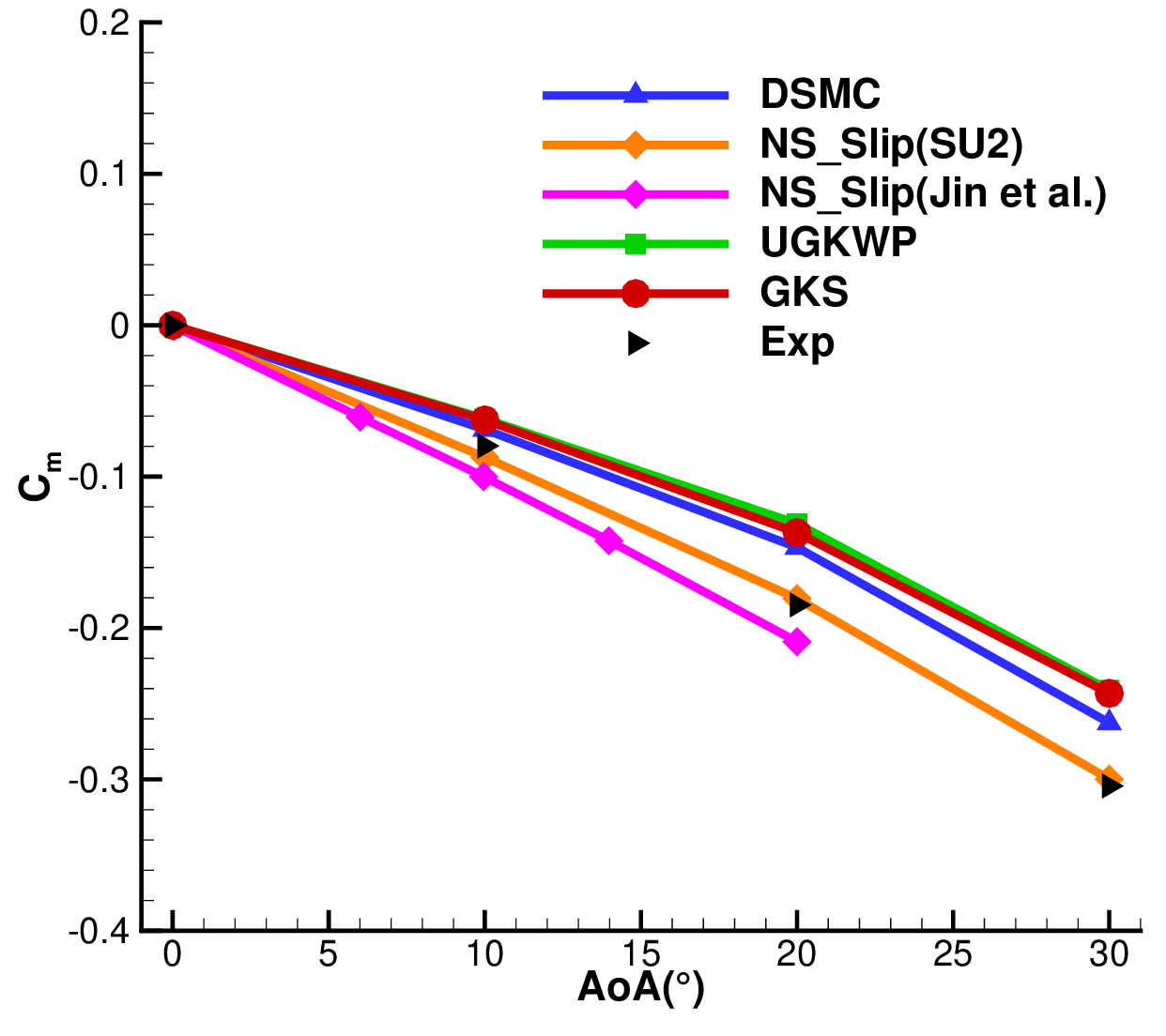}}\\
	\caption{Hypersonic flow over a $70^\circ$ blunted cone with sting at ${\rm Kn}_\infty = 0.013$ ($H \sim 80$~km) and ${\rm Ma}_\infty = 20.2$. Aerodynamic coefficients versus angle of attack: (a) axial force $C_A$, (b) normal force $C_N$, and (c) pitching moment coefficient $C_m$. Comparisons include experiment, GKS, UGKWP, DSMC, NS with slip wall~\cite{jin2026computational}.}
	\label{fig:cone-case1-coef}
\end{figure}

\begin{figure}[H]
	\centering
	\subfloat[]{\includegraphics[width=0.33\textwidth]{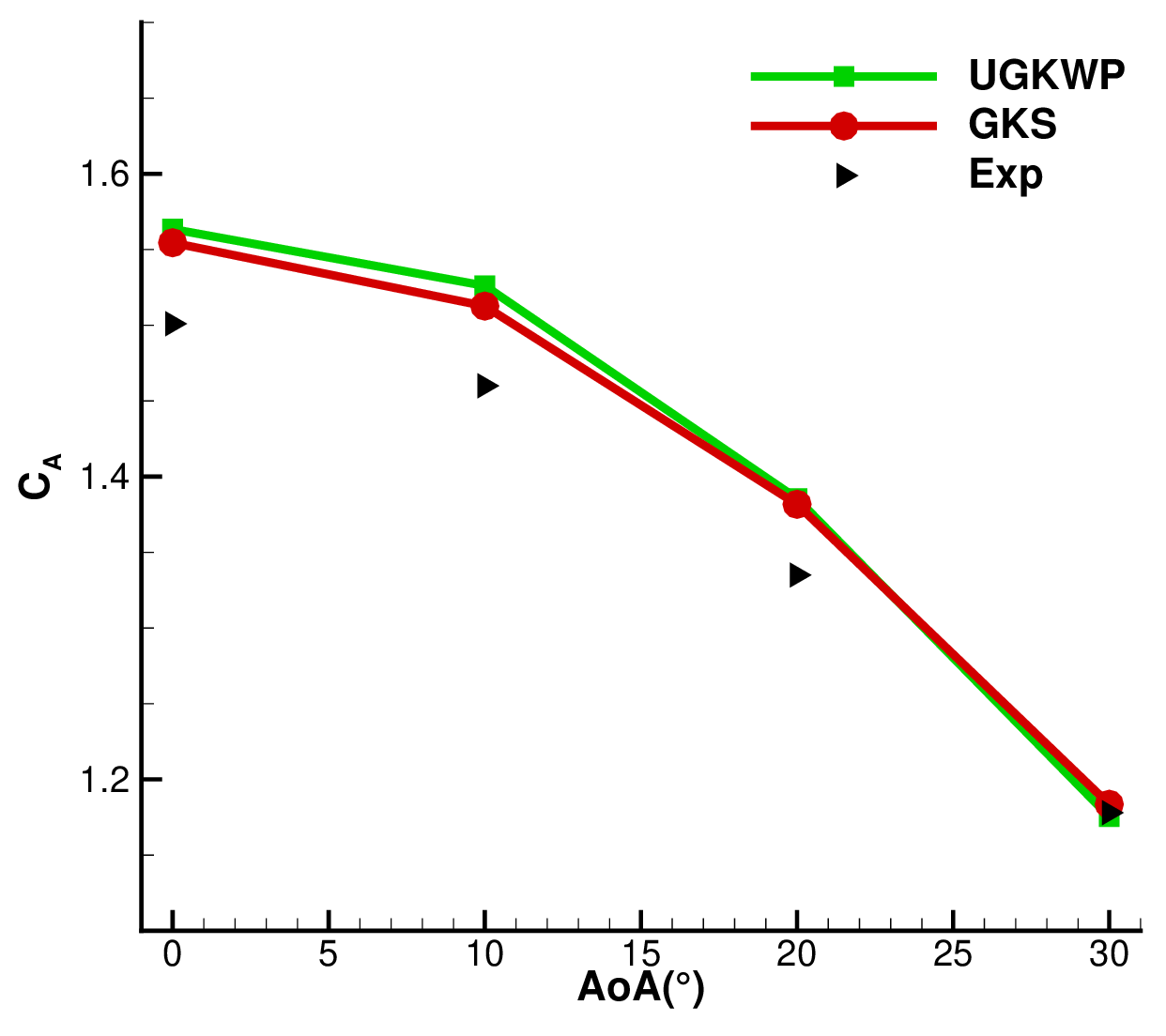}}
	\subfloat[]{\includegraphics[width=0.33\textwidth]{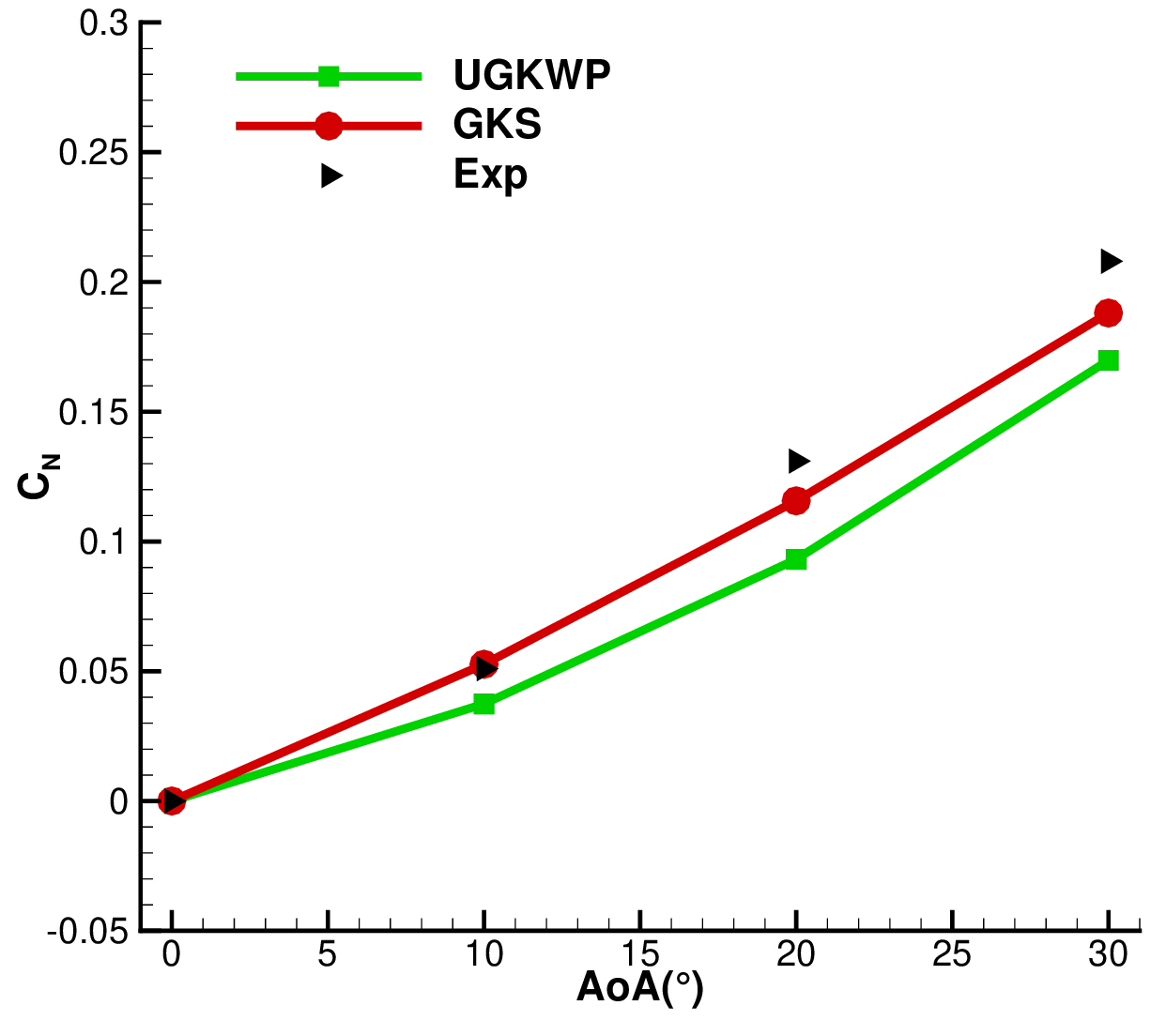}}
	\subfloat[]{\includegraphics[width=0.33\textwidth]{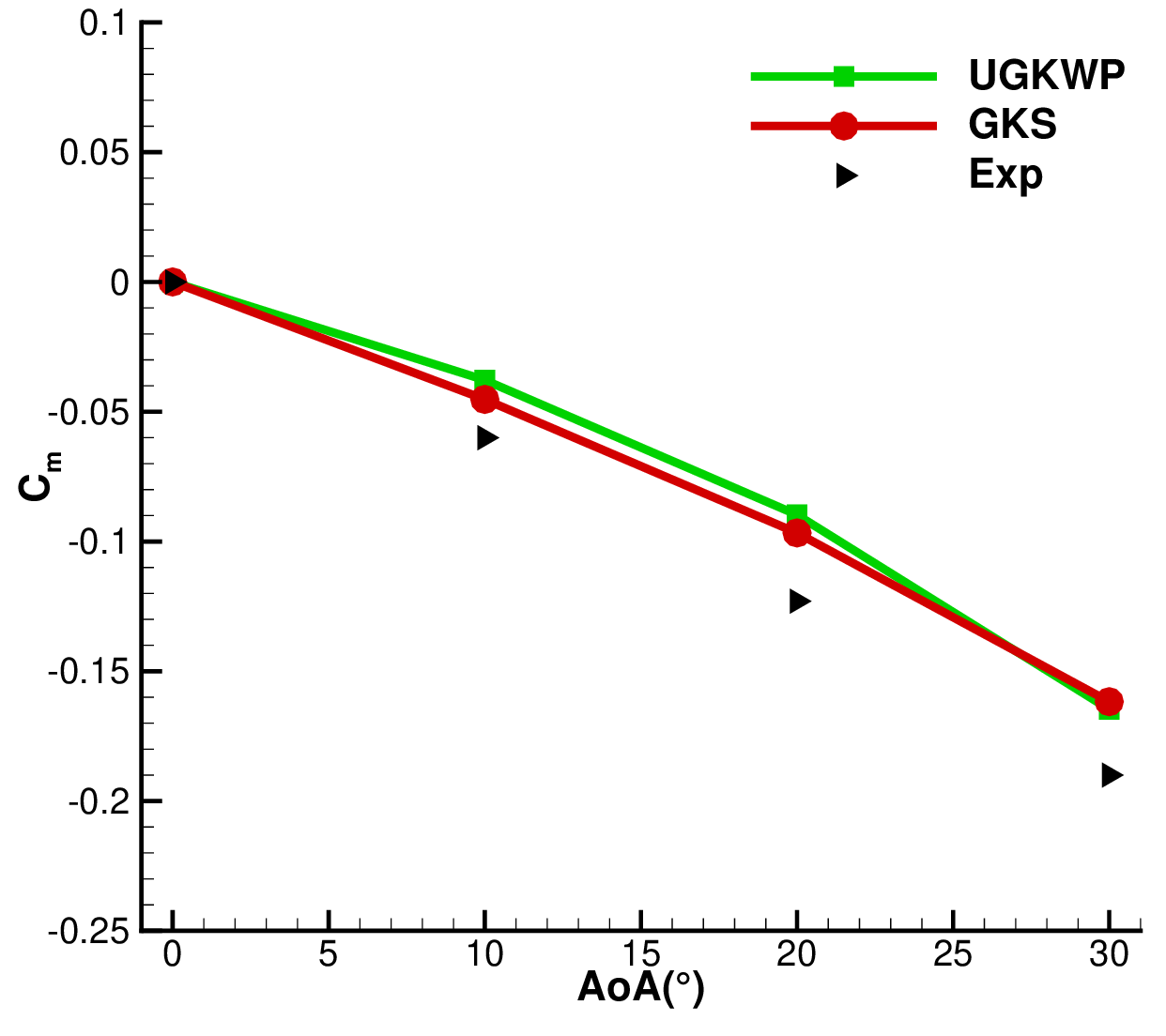}}\\
	\caption{Hypersonic flow over a $70^\circ$ blunted cone with sting at ${\rm Kn}_\infty = 5.4\times 10^{-4}$ ($H \sim 57$~km) and ${\rm Ma}_\infty = 20.5$. Aerodynamic coefficients versus angle of attack: (a) axial force $C_A$, (b) normal force $C_N$, and (c) pitching moment coefficient $C_m$. Comparisons include the experiment, the UGKWP, and the GKS.}
	\label{fig:cone-case3-coef}
\end{figure}

Figures~\ref{fig:cone-case1-coef} and \ref{fig:cone-case3-coef} present the axial force, normal force,
and pitching moment coefficients ($C_A$, $C_N$, $C_m$) versus angle of attack, normalized by the base
cross-sectional area with the pitching moment referred to the nose of the blunted cone.
Figure~\ref{fig:cone-case1-coef} corresponds to the more rarefied condition
${\rm Kn}_\infty = 0.013$ ($H \sim 80$~km), where the GKS is compared with the UGKWP, the experimental data~\cite{allegre1997forces}, DSMC results~\cite{schouler2020survey}, and a
conventional CFD solver with slip boundary condition, solved via the SU2 framework~\cite{palacios2014su2,cao2024nccr}.
The NS slip results obtained here agree with those reported in Ref.~\cite{jin2026computational}, and both confirm the same trend.
The GKS agrees closely with the UGKWP, the DSMC and the experiment across all three
coefficients, with only minor deviations from the experimental data in $C_N$
and $C_m$. The NS slip solver, however, deviates markedly from all the computational approaches
, with the largest discrepancy appearing in $C_A$.
Figure~\ref{fig:cone-case3-coef} shows the same coefficients at the more continuum condition
${\rm Kn}_\infty = 5.4\times 10^{-4}$ ($H \sim 57$~km), where the comparison focuses on the GKS, the UGKWP,
and the experiment.
Here the GKS remains consistent with the UGKWP, and both are close to the experimental data.
At $H \sim 80$~km and $H \sim 57$~km, the GKS performs well, consistent with the Apollo~6 findings in
Section~\ref{sec:case-apollo}. This is because the large-angle blunt forebody creates a
windward compression zone where the flow remains closer to continuum, as in the Apollo~6 case.
In the present configuration, although the afterbody includes a slender cylindrical sting, it lies
within the low-density leeward wake and therefore contributes little to the integrated
aerodynamic loads.

\subsection{Hypersonic flow around a $9^\circ$ Blunted Cone}\label{sec:case-bluntcone}

The hypersonic flow over a $9^\circ$ half-angle blunted cone was tested by Boylan and
Potter~\cite{boylan1967aerodynamics} in the AEDC-VKF Tunnel~L at ${\rm Ma}_\infty = 10.15$ and
${\rm Kn}_\infty = 0.065$ ($H \sim 80$~km), providing measurements of lift, drag, and pitching moment.
This case has since been studied by several computational approaches~\cite{padilla2009assessment,zhang2024conservative}
.
The present work uses the GKS and the UGKWP with the same kinetic boundary condition to simulate this
benchmark as Case~A. A more rarefied condition, Case~B, is constructed by reducing the
freestream density by a factor of ten while keeping all other parameters unchanged;
here the GKS is also compared with the UGKWP.
Assuming a full-scale vehicle with a base diameter of $0.55$~m,
and using the US Standard Atmosphere 1976, the freestream Knudsen numbers
of Case~A and Case~B correspond to
$H \sim 80$~km and $H \sim 94$~km, respectively. The flow conditions are summarized in
Table~\ref{tab:bluntcone-conditions}.

\begin{table}[H]
	\centering
	\caption{Freestream conditions for the $9^\circ$ blunted cone cases.}
	\begin{tabular}{lcccccc}
		\hline
		Case & Kn$_\infty$ & Ma$_\infty$ & $U_\infty$ (m/s) & $T_\infty$ (K) & $\rho_\infty$ (kg/m$^3$) & $T_w$ (K) \\
		\hline
		A & 0.065 & 10.15 & 2478 & 143.5 & $6.15\times 10^{-5}$ & 600 \\
		B & 0.65  & 10.15 & 2478 & 143.5 & $6.15\times 10^{-6}$ & 600 \\
		\hline
	\end{tabular}
	\label{tab:bluntcone-conditions}
\end{table}

All simulations are performed at $\alpha = 0^\circ, 10^\circ, 20^\circ$, and $25^\circ$.
The working gas is nitrogen, and the dynamic viscosity is obtained from the Sutherland law
with $\mu_0 = 1.6579\times 10^{-5}$~Pa$\cdot$s, $T_0 = 273$~K, and $S = 111$~K~\cite{zhang2024conservative}.
The base diameter $d = 15.24$~mm is taken as the reference length.
\begin{figure}[H]
	\centering
	\subfloat[$U$]{\includegraphics[width=0.32\textwidth]{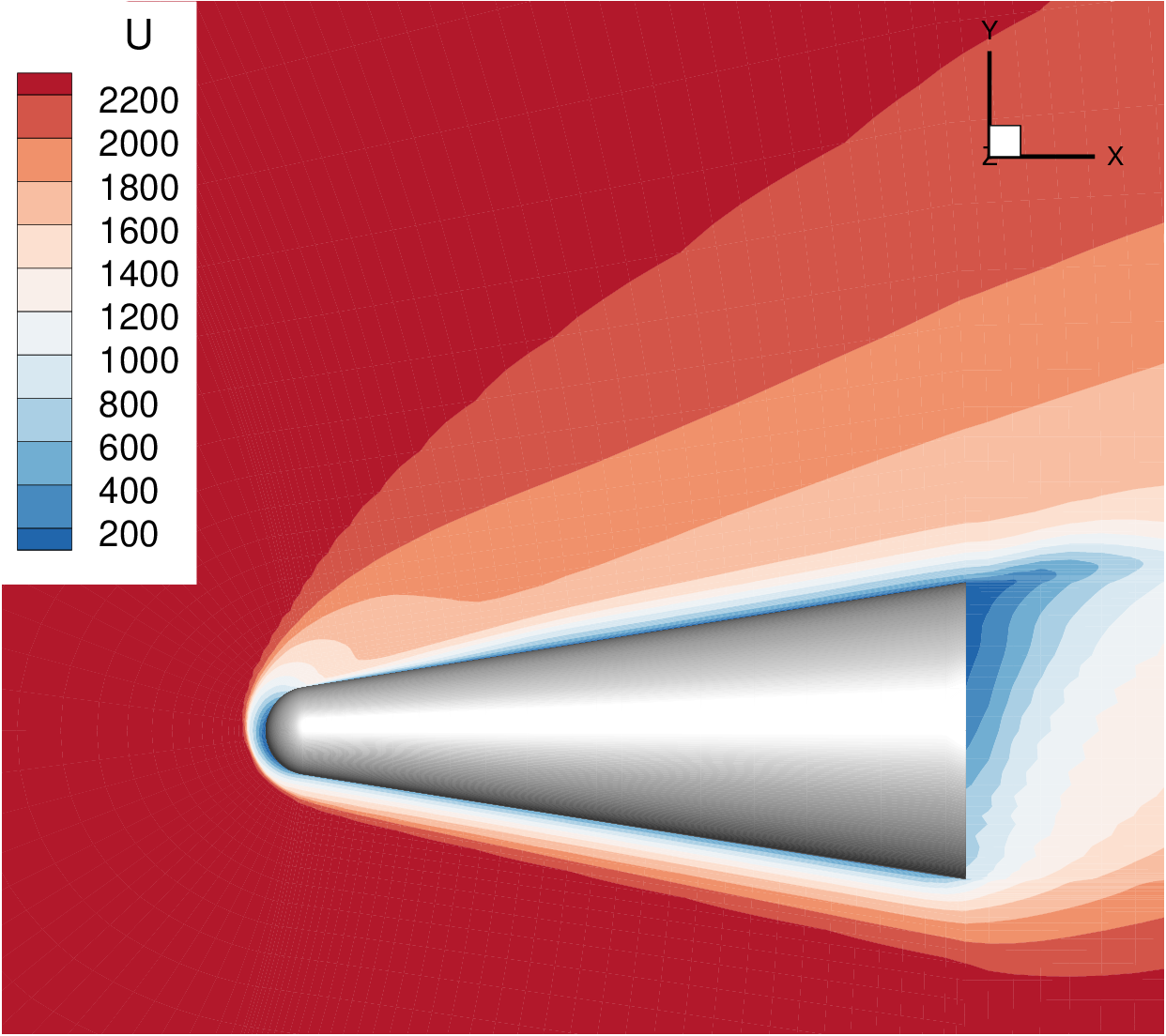}}~
	\subfloat[$\rho$]{\includegraphics[width=0.32\textwidth]{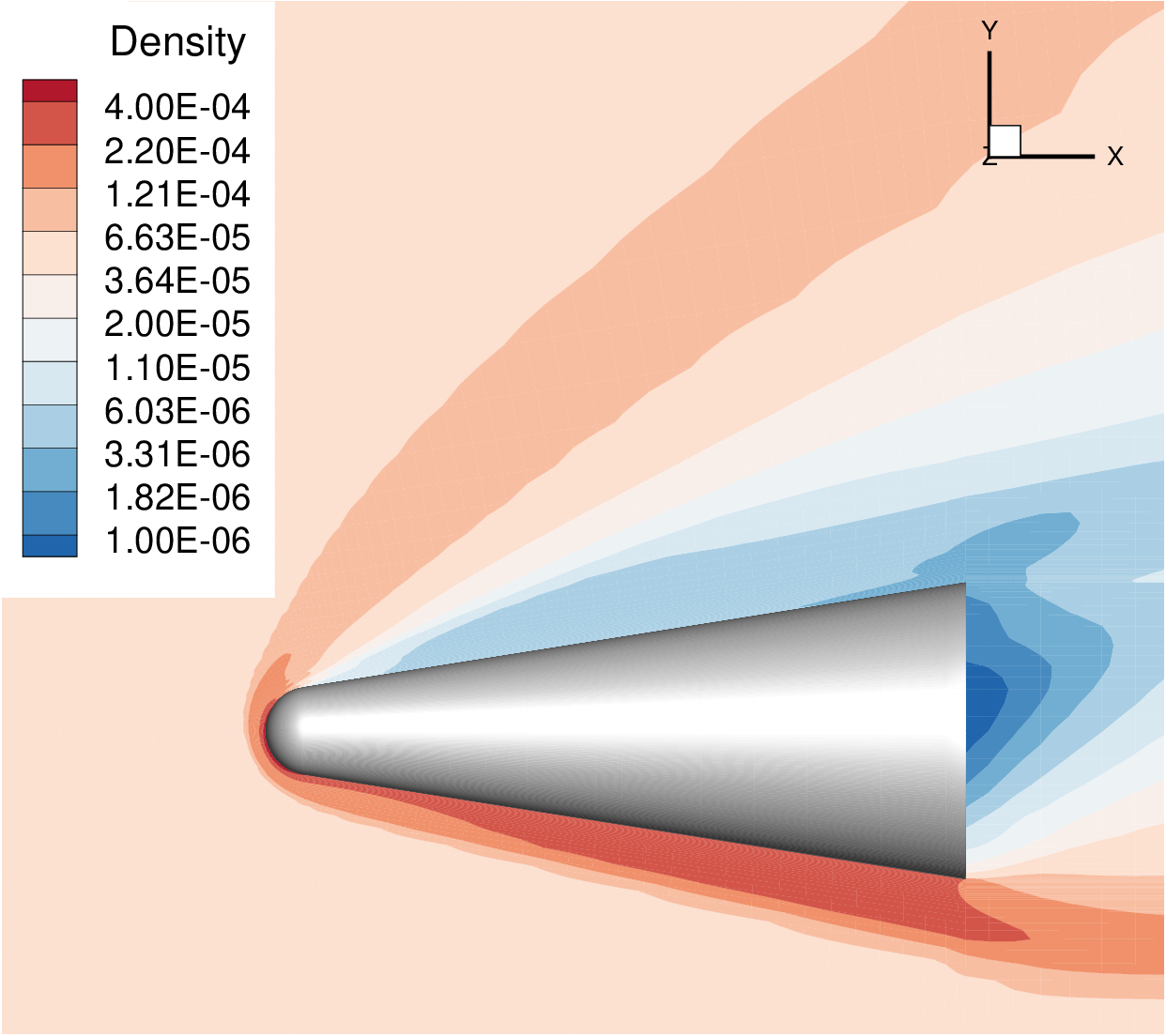}}~
	\subfloat[$T$]{\includegraphics[width=0.32\textwidth]{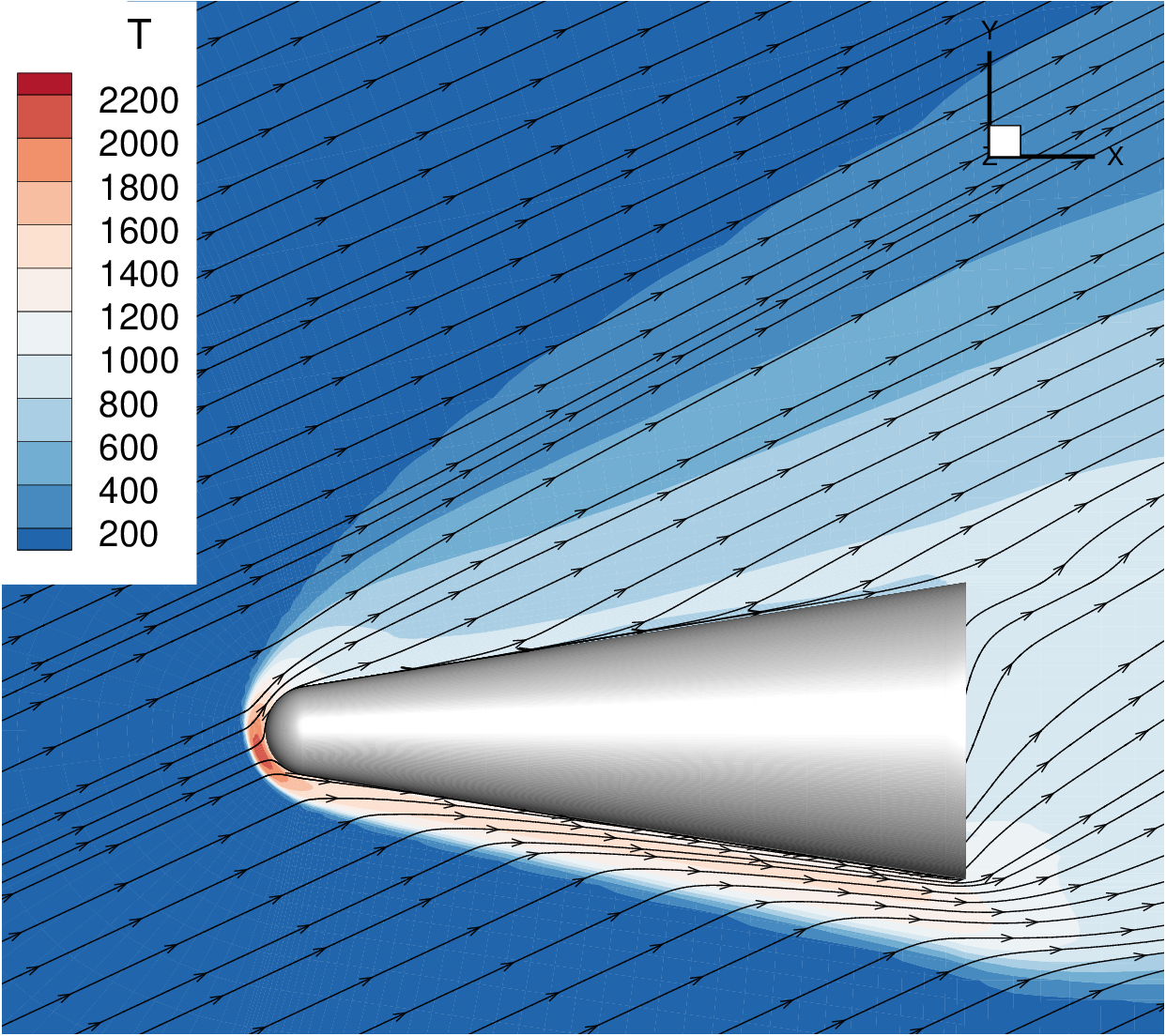}}\\
	\caption{Hypersonic flow over $9^\circ$ blunted cone at ${\rm Kn}_\infty = 0.065$ ($H \sim 80$~km), ${\rm Ma}_\infty = 10.15$, and $\alpha = 25^\circ$ by the GKS. Contours on the symmetry plane: (a) $x$-direction velocity $U$, (b) density, and (c) temperature with streamlines.}
	\label{fig:bluntcone-caseA-contour}
\end{figure}

\begin{figure}[H]
	\centering
	\subfloat[$U$]{\includegraphics[width=0.32\textwidth]{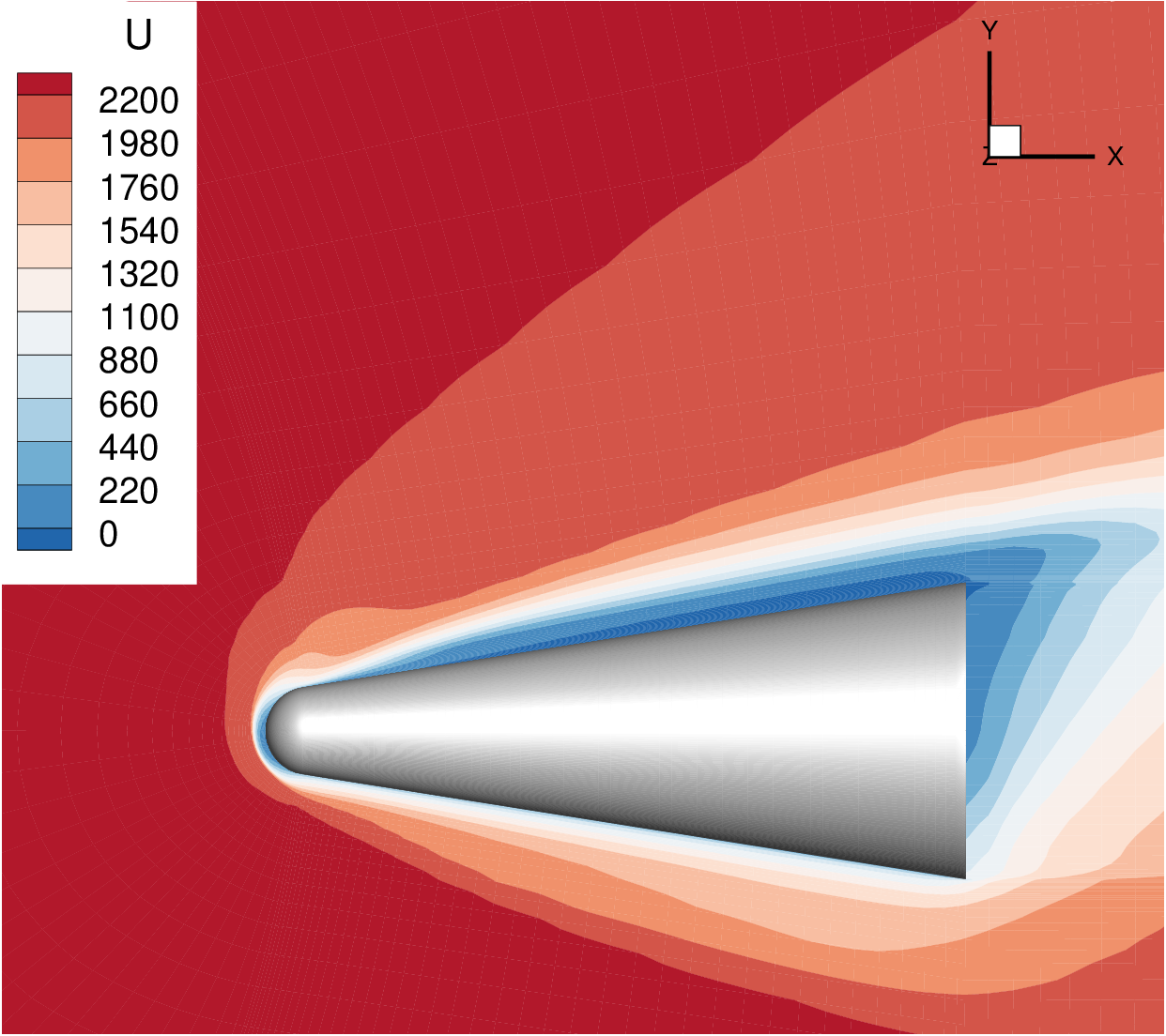}}~
	\subfloat[$\rho$]{\includegraphics[width=0.32\textwidth]{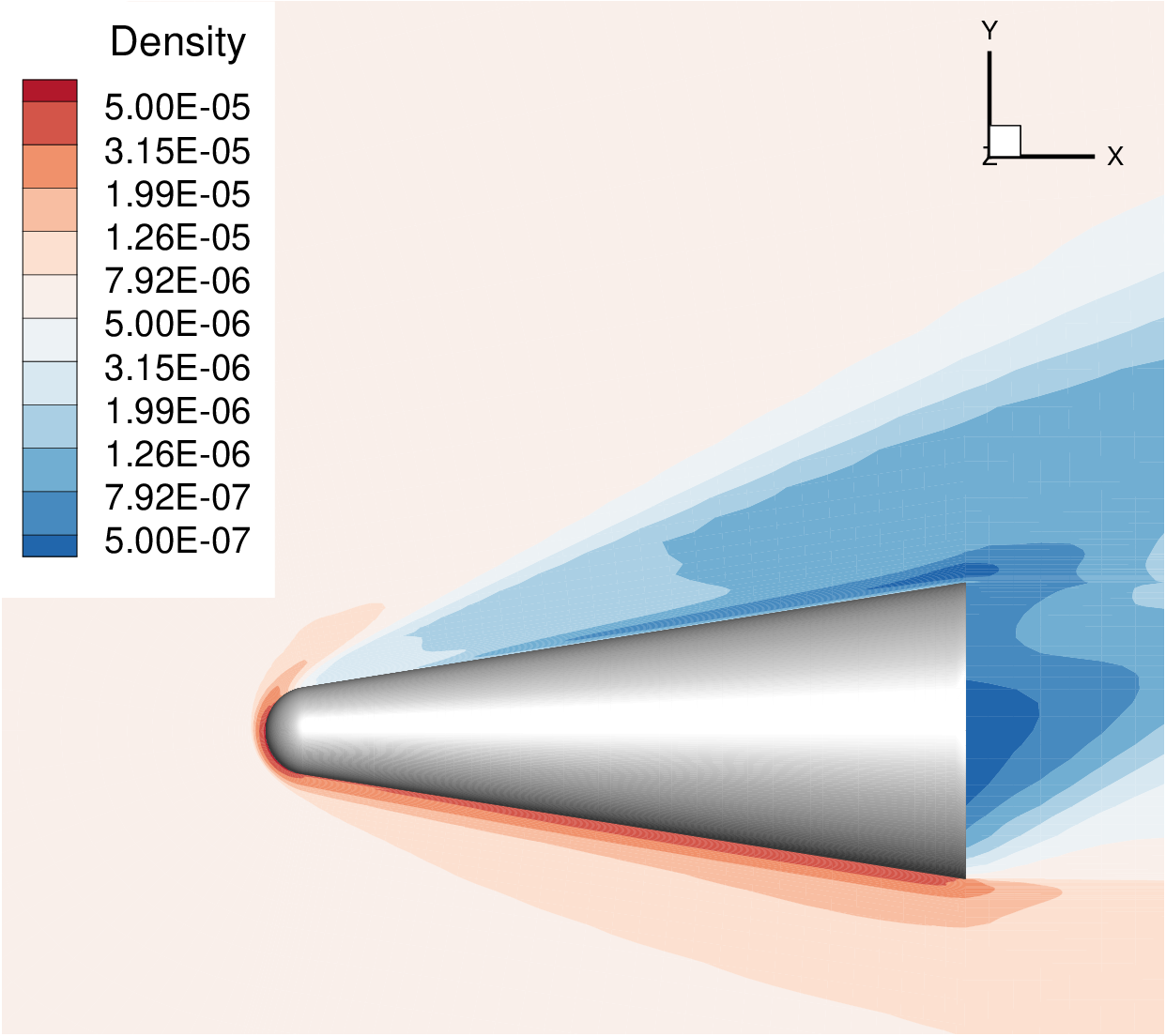}}~
	\subfloat[$T$]{\includegraphics[width=0.32\textwidth]{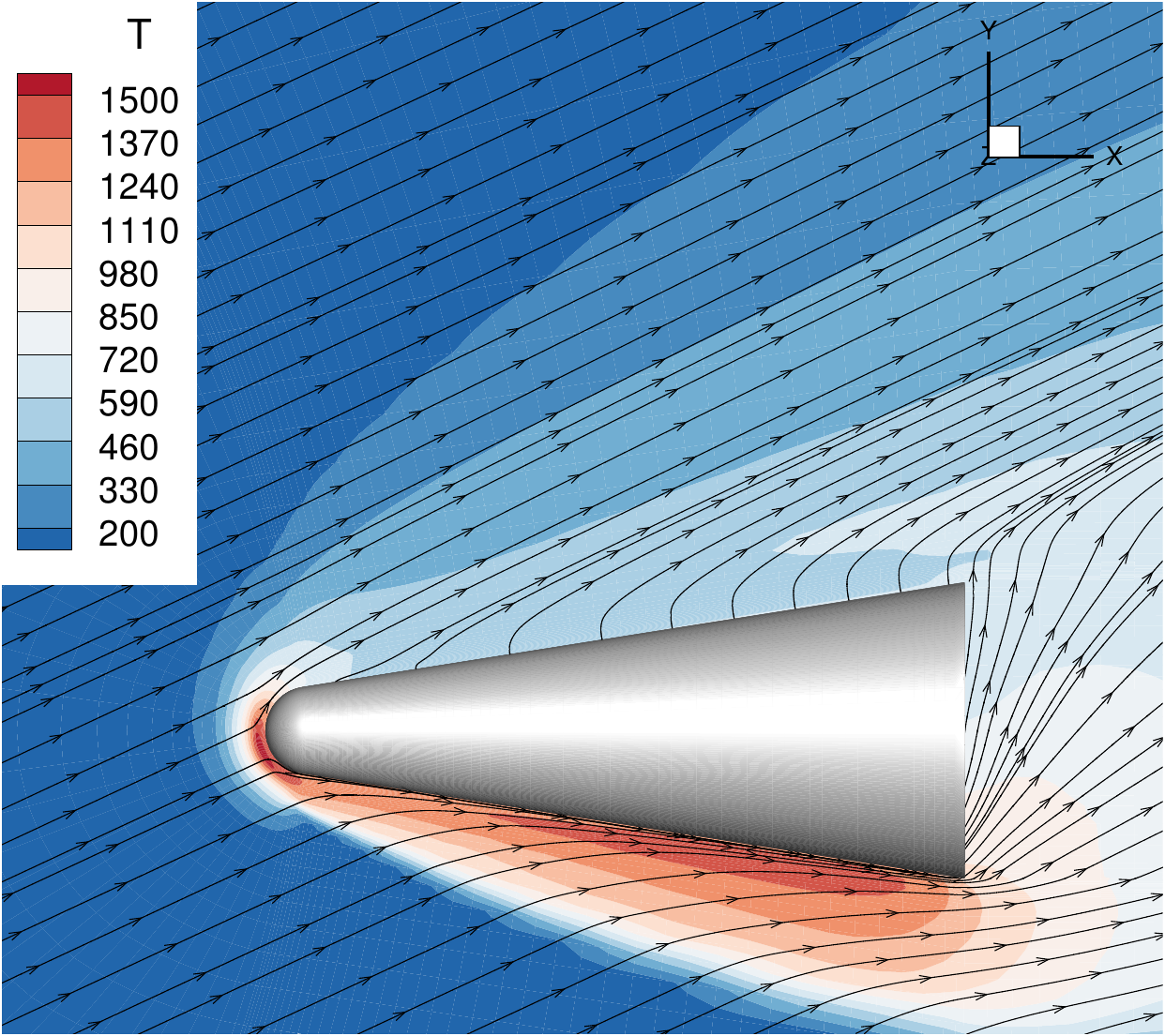}}\\
	\caption{Hypersonic flow over $9^\circ$ blunted cone at ${\rm Kn}_\infty = 0.65$ ($H \sim 94$~km), ${\rm Ma}_\infty = 10.15$, and $\alpha = 25^\circ$ by the GKS. Contours on the symmetry plane: (a) $x$-direction velocity $U$, (b) density, and (c) temperature with streamlines.}
	\label{fig:bluntcone-caseB-contour}
\end{figure}

Figures~\ref{fig:bluntcone-caseA-contour} and \ref{fig:bluntcone-caseB-contour} show the
$x$-direction velocity, density, and temperature contours on the symmetry plane at
$\alpha = 25^\circ$ for both cases.
From the streamlines, both cases show few recirculation features in the leeward region,
in contrast to the Apollo~6 command module and the $70^\circ$ blunted cone.

\begin{figure}[H]
	\centering
	\subfloat[${\rm Kn}_\infty = 0.065$ ($H \sim 80$~km)]{\includegraphics[width=0.48\textwidth]{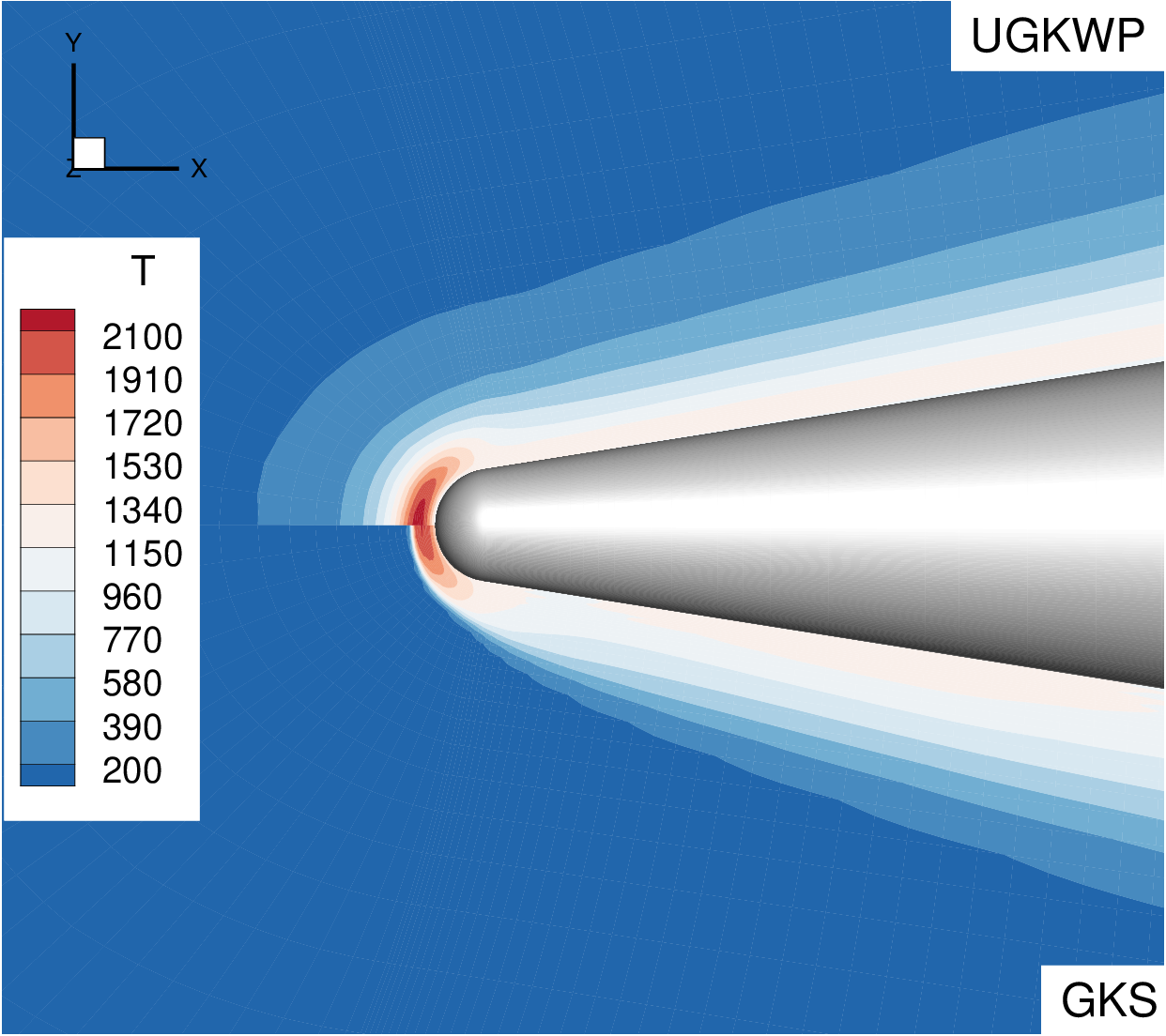}}~
	\subfloat[${\rm Kn}_\infty = 0.65$ ($H \sim 94$~km)]{\includegraphics[width=0.48\textwidth]{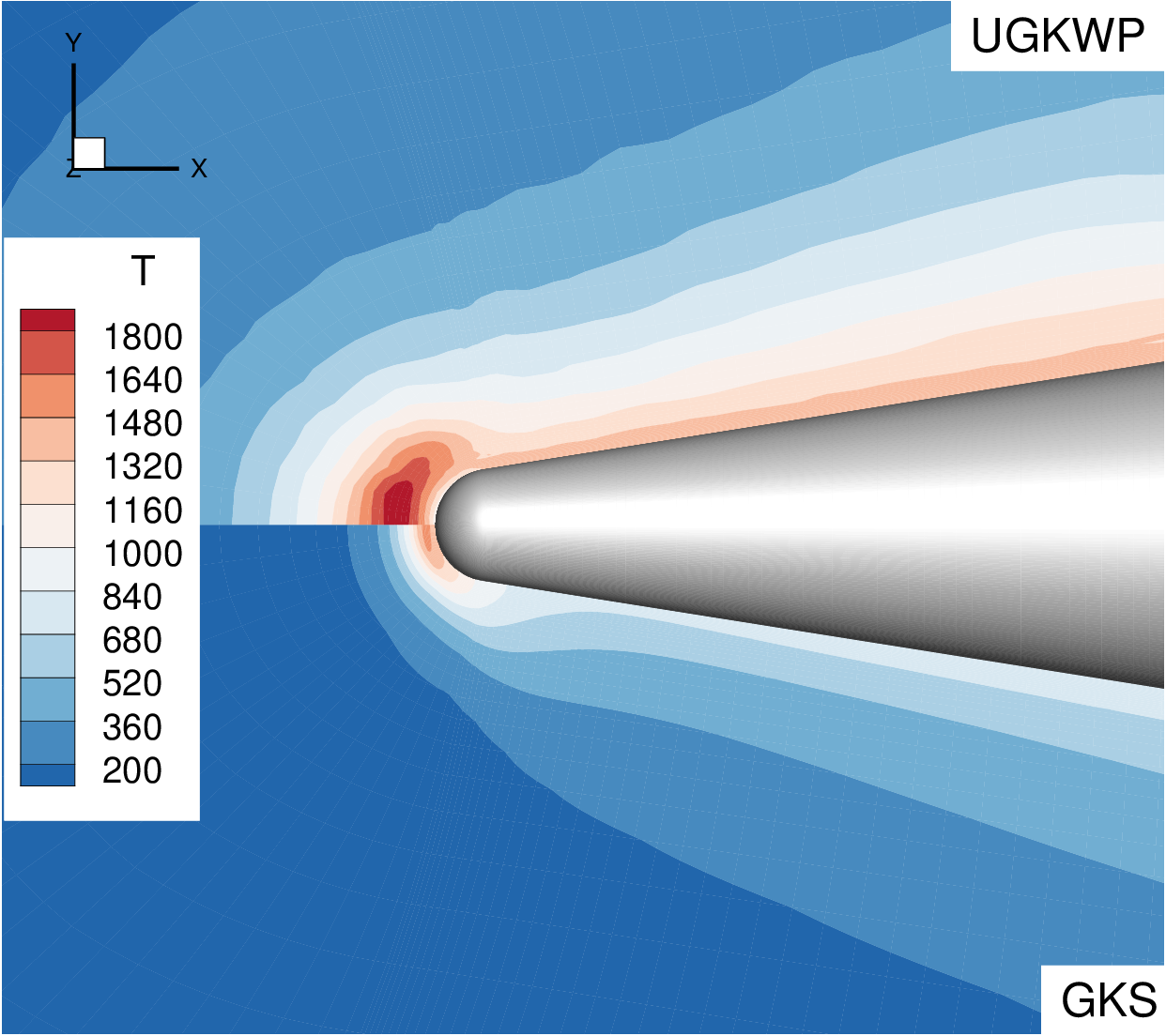}}
	\caption{Comparison of the time-averaged UGKWP temperature and the GKS temperature for the $9^\circ$ blunted cone at ${\rm Ma}_\infty = 10.15$ and $\alpha = 0^\circ$. (a) ${\rm Kn}_\infty = 0.065$ ($H \sim 80$~km); (b) ${\rm Kn}_\infty = 0.65$ ($H \sim 94$~km).}
	\label{fig:bluntcone-T-compare}
\end{figure}

Figure~\ref{fig:bluntcone-T-compare} compares the
temperature fields by the GKS and the UGKWP for both cases.
These comparisons indicate that above $H \sim 80$~km, the GKS exhibits a
markedly weaker ability to resolve the nonequilibrium effect of the flow field
than the UGKWP: the shock thickness predicted by the GKS is substantially smaller,
the near-wall temperature is higher, and the peak temperature is reduced.
Although the GKS does produce a thicker shock at the more rarefied condition ($H \sim 94$~km) than at $H \sim 80$~km, at this altitude the UGKWP gives the result where the shock and the near-wall
boundary layer have largely merged. This indicates a fundamentally different flow
mechanism in which the true gas distribution function departs significantly from the
CE expansion.

\begin{figure}[H]
	\centering
	\subfloat[]{\includegraphics[width=0.33\textwidth]{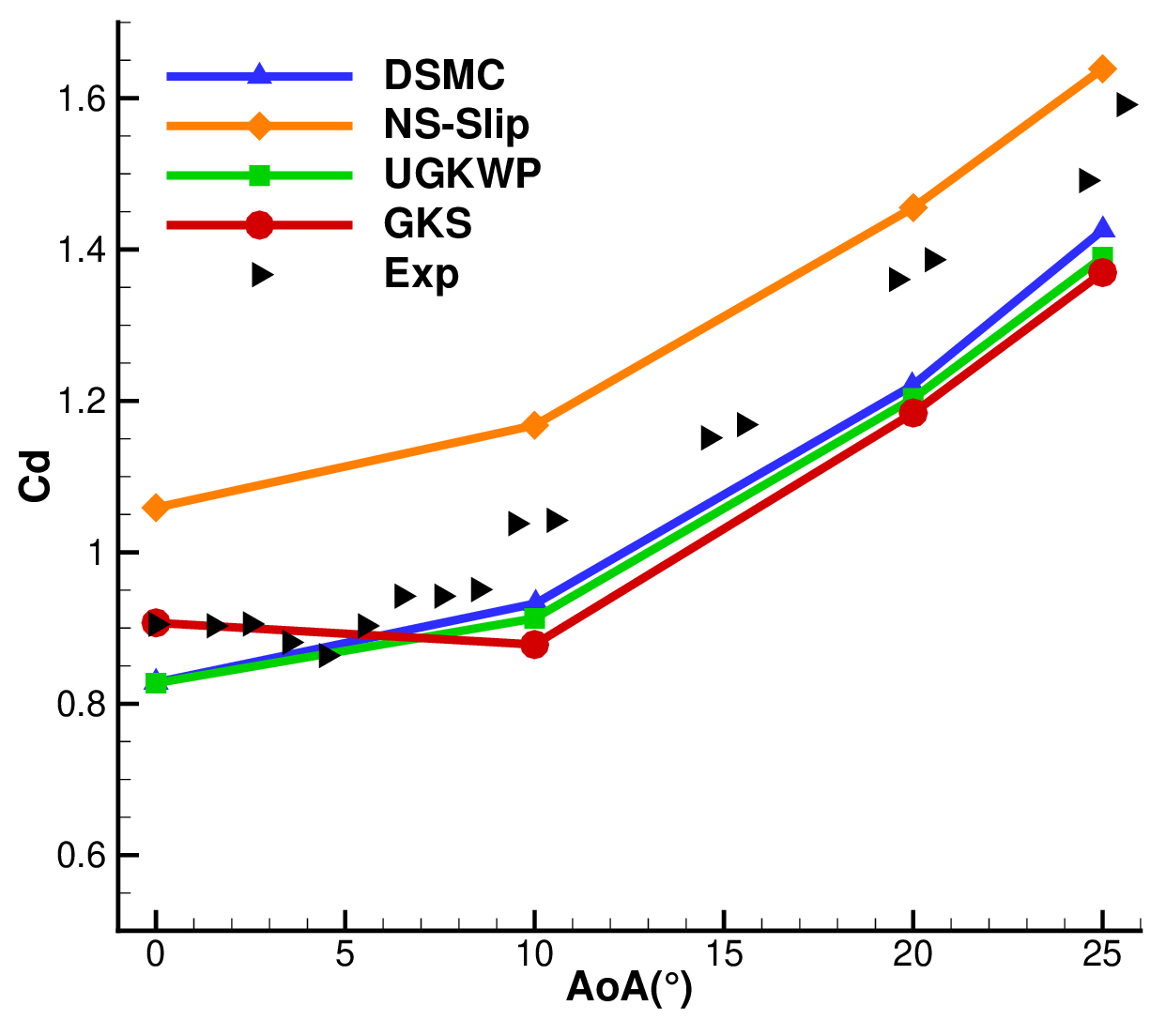}}~
	\subfloat[]{\includegraphics[width=0.33\textwidth]{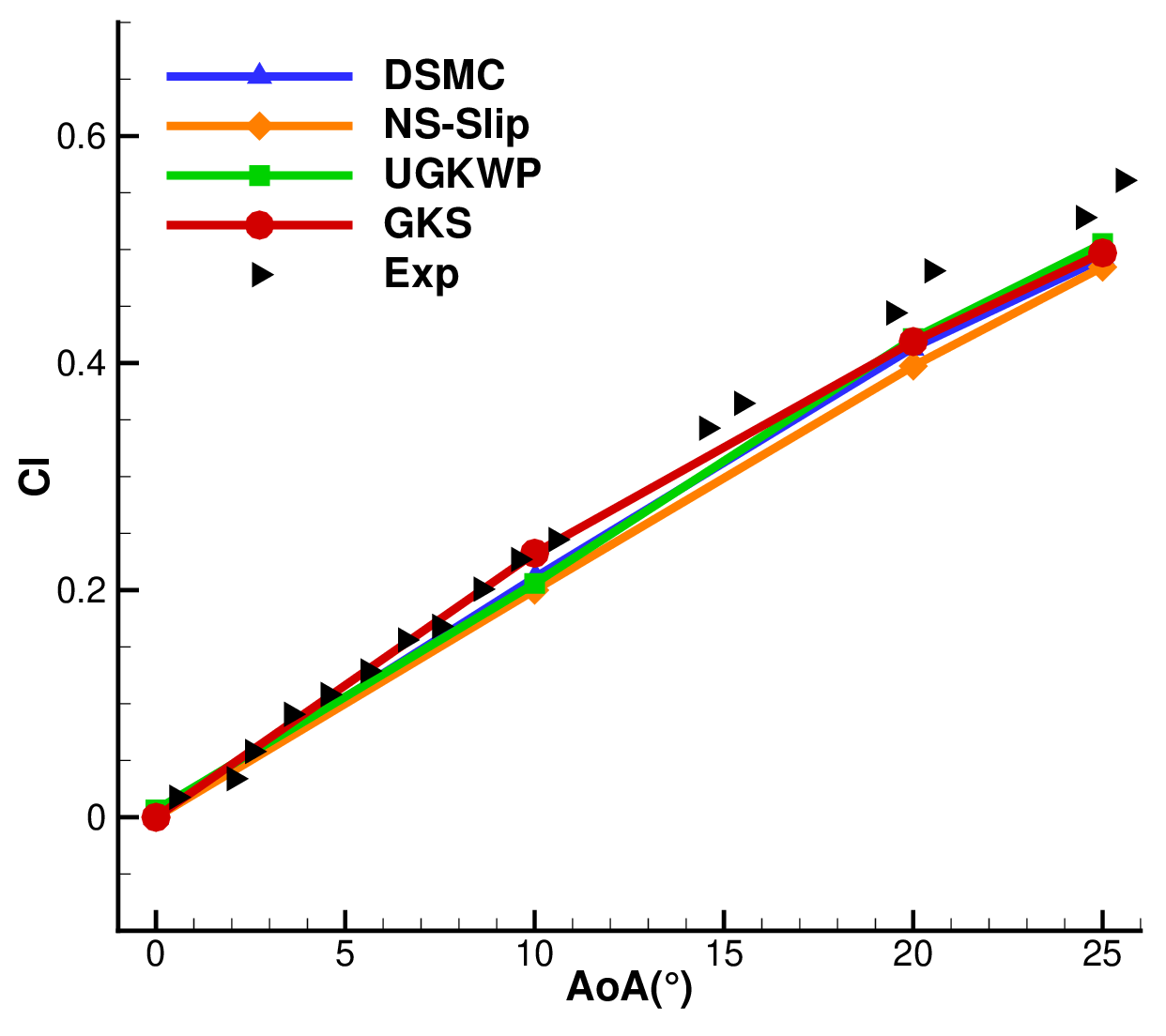}}~
	\subfloat[]{\includegraphics[width=0.33\textwidth]{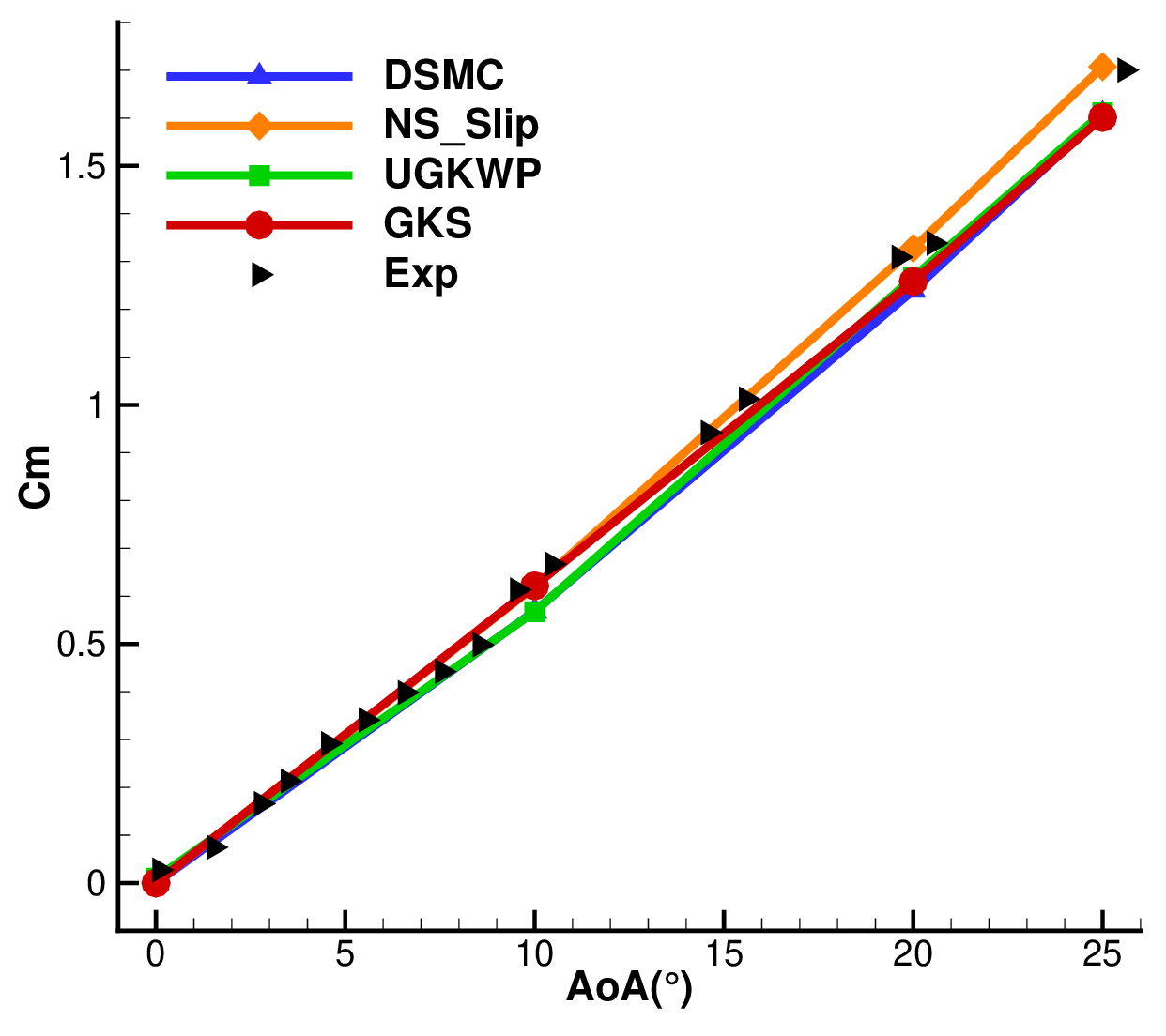}}\\
	\caption{Hypersonic flow over $9^\circ$ blunted cone at ${\rm Kn}_\infty = 0.065$ ($H \sim 80$~km), ${\rm Ma}_\infty = 10.15$. Aerodynamic coefficients versus angle of attack: (a) drag coefficient $C_d$, (b) lift coefficient $C_l$, and (c) pitching moment coefficient $C_m$. Comparisons include experiment~\cite{boylan1967aerodynamics}, DSMC~\cite{padilla2009assessment}, NS with slip boundary, and the GKS.}
	\label{fig:bluntcone-caseA}
\end{figure}

\begin{figure}[H]
	\centering
	\subfloat[]{\includegraphics[width=0.33\textwidth]{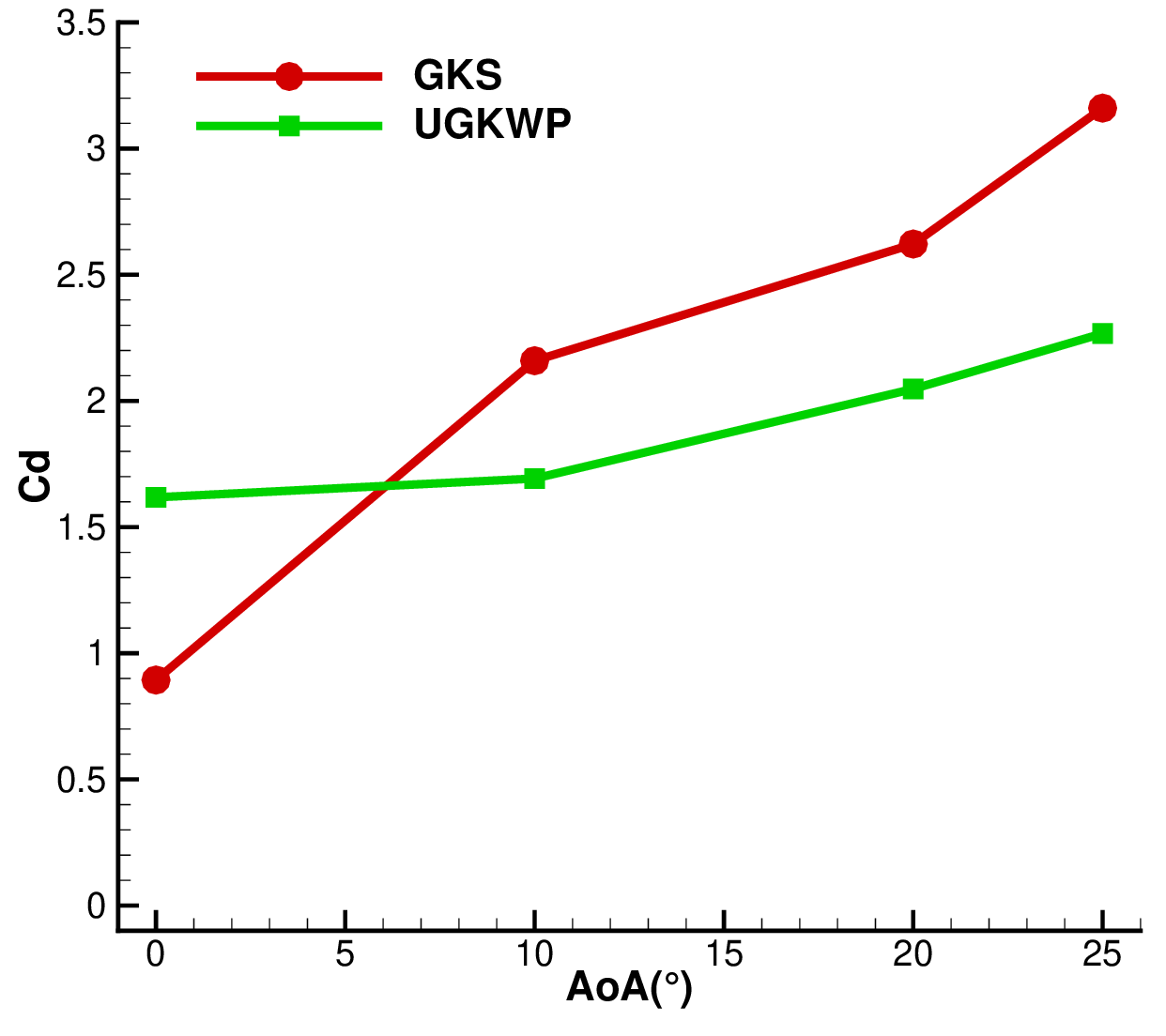}}~
	\subfloat[]{\includegraphics[width=0.33\textwidth]{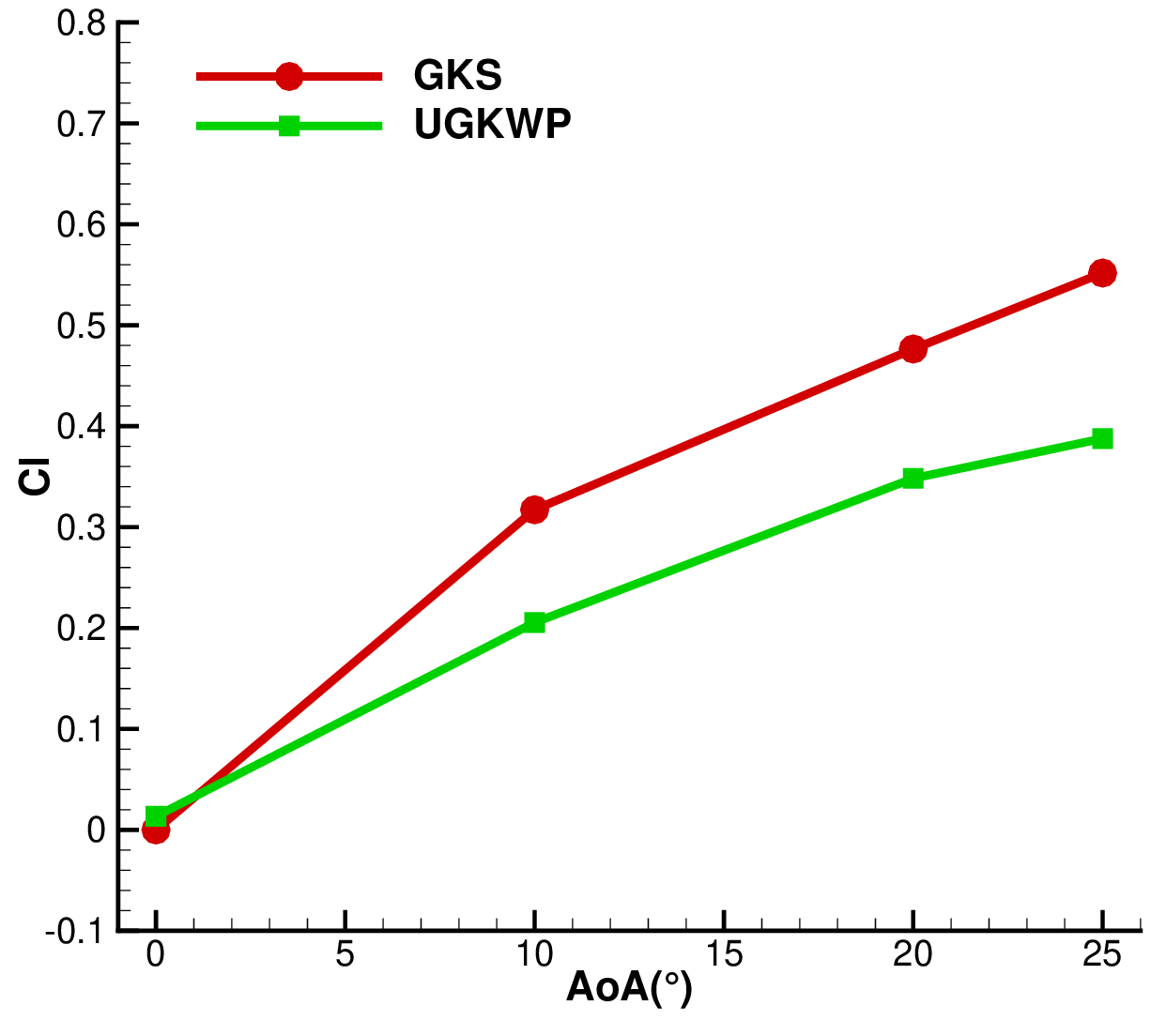}}~
	\subfloat[]{\includegraphics[width=0.33\textwidth]{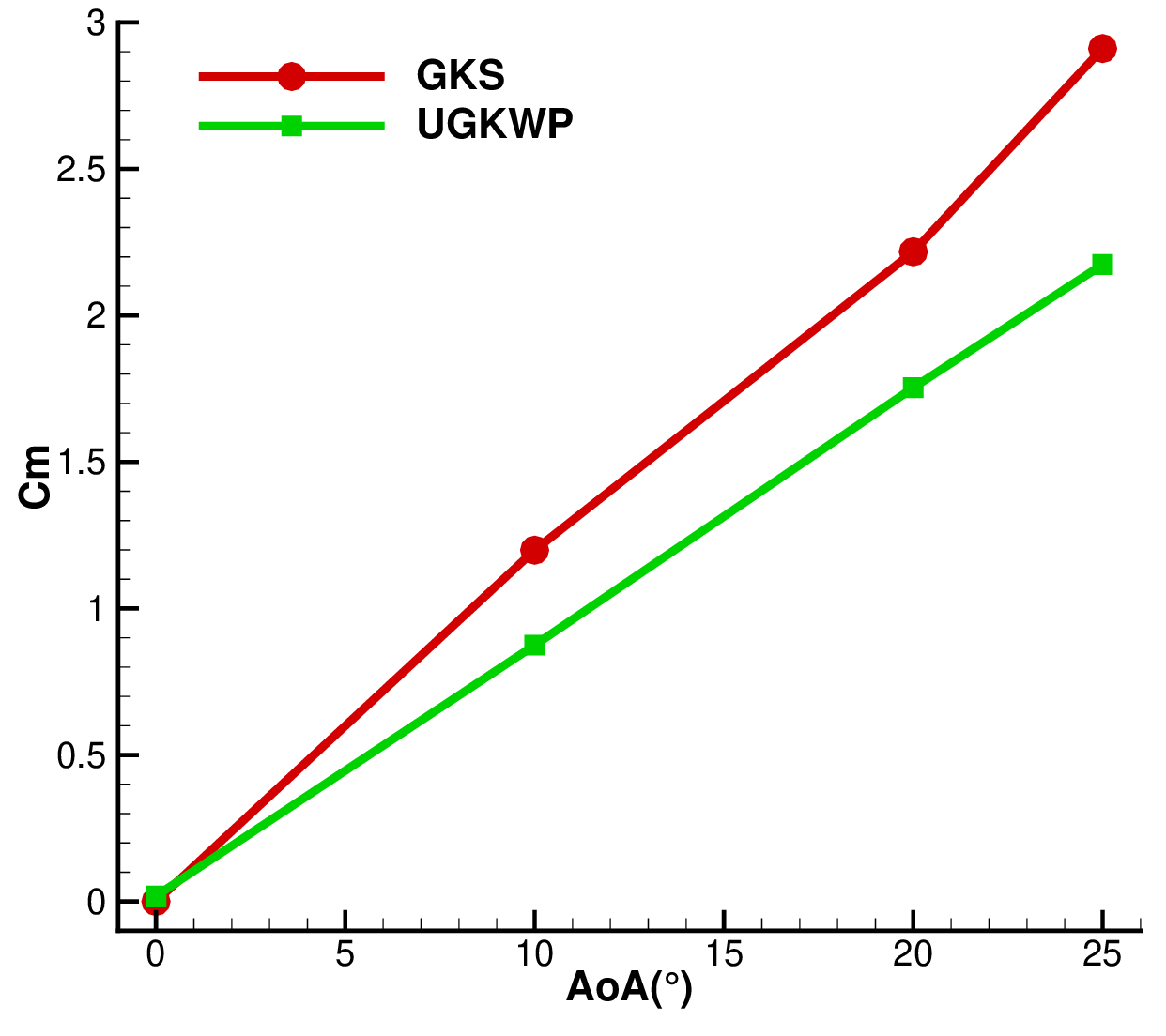}}\\
	\caption{Hypersonic flow over $9^\circ$ blunted cone at ${\rm Kn}_\infty = 0.65$ ($H \sim 94$~km), ${\rm Ma}_\infty = 10.15$. Aerodynamic coefficients versus angle of attack: (a) drag coefficient $C_d$, (b) lift coefficient $C_l$, and (c) pitching moment coefficient $C_m$. Comparisons include the UGKWP and the GKS.}
	\label{fig:bluntcone-caseB}
\end{figure}

Figure~\ref{fig:bluntcone-caseA} compares the GKS with the experiment~\cite{boylan1967aerodynamics},
the DSMC~\cite{padilla2009assessment}, and the same in-house NS solver
with slip boundary condition used in Section~\ref{sec:case-70cone}
at ${\rm Kn}_\infty = 0.065$ ($H \sim 80$~km) for the drag, lift, and pitching moment coefficients
($C_d$, $C_l$, $C_m$).
Although the GKS shows notable differences in the predicted flow-field structure
relative to the UGKWP, the aerodynamic coefficients remain in good
agreement with the reference data.
A slight deviation in $C_d$ at $\alpha = 0^\circ$ between the GKS and the UGKWP
is observed; at this angle of attack the windward projected area is smallest, so
the flow is locally more rarefied and therefore more sensitive to the differences
in the underlying gas-distribution modeling.
Most importantly, the GKS result lies consistently closer to the other
reference results than the in-house NS with slip boundary does for all three
coefficients, especially in the drag coefficient.
Figure~\ref{fig:bluntcone-caseB} shows the same coefficients at the more rarefied
condition ${\rm Kn}_\infty = 0.65$ ($H \sim 94$~km), where the GKS and the UGKWP are compared.
Here the GKS has largely lost the ability to predict the aerodynamic coefficients,
in contrast to the Apollo~6 case in Section~\ref{sec:case-apollo}, where the GKS
retains partial accuracy up to $100$~km.
A contributing factor is the slenderness of the $9^\circ$ cone: even on the
windward side, the shallower shock angle produces weaker compression, so that
the flow remains farther from equilibrium at a given altitude than for a
large-angle blunt forebody such as the Apollo command module.

\section{Conclusion}\label{sec:conclusion}

In this work, the applicability limits of the gas-kinetic scheme (GKS) equipped with kinetic boundary conditions have been systematically evaluated for hypersonic flows around near-space vehicles. 
Below an altitude of $80$~km, the GKS predicts aerodynamic coefficients with an accuracy comparable to the UGKWP and DSMC methods, maintaining close agreement with available experimental data. For the Apollo 6 command module, which features a large blunt forebody, partial predictive capability is retained even at $100$~km.
However, at such high altitudes, the flow-field contours differ visibly from the reference solutions. These discrepancies are primarily concentrated in the leeward region and across the shock wave, where strong non-equilibrium effects invalidate the Chapman–Enskog (CE) expansion. This apparent paradox, where aerodynamic forces remain reliable despite inaccurate flow-field contours, can be explained by observations from the cylinder case. At ${\rm Kn}_\infty = 0.4$, the contour mismatch is similarly pronounced, yet the surface pressure, shear stress, and heat flux remain remarkably close to the UGKS results. Examining the distribution function at the stagnation point (spatially at $x = -1$, and on the $v = 0$ plane in velocity space) reveals that while the non-equilibrium distribution and the CE expansion differ in the $u < 0$ (outgoing) regime, they show close agreement in the incident portion ($u > 0$). Under identical kinetic boundary conditions, the GKS wall fluxes consequently remain close to those of the UGKS and UGKWP. This mechanism extends to three-dimensional cases, explaining the highly accurate aerodynamic force predictions observed in the present results.

Furthermore, the GKS with kinetic boundary conditions consistently outperforms conventional Navier–Stokes (NS) solvers utilizing Maxwell-type slip conditions. In both the $70^\circ$ and $9^\circ$ blunted cone cases, the force coefficients predicted by the GKS align more closely with the UGKWP and DSMC results than those of the NS slip approach; in particular, the NS slip axial force coefficient $C_A$ exhibits a large departure from the reference data. This advantage stems directly from the kinetic-level description of the GKS. The kinetic boundary condition preserves the local non-equilibrium distribution induced by wall interactions, thereby yielding wall fluxes that are closer to the reference kinetic solutions. Concurrently, the $\tau_{\mathrm{eff}}$ limiter suppresses unphysical incident distributions in regions of strong non-equilibrium, further contributing to the enhanced accuracy.

In summary, the present approach offers a clear improvement over conventional CFD with slip boundary conditions for predicting aerodynamic forces on hypersonic vehicles in near space at altitudes up to $80$~km, demonstrating considerable engineering potential. However, its inability to accurately predict temperature fields may limit its applicability in thermal non-equilibrium scenarios, such as reactive flows, dissociation, and plasma phenomena, where truly multiscale methods like the UGKWP and UGKS remain indispensable.

\section*{Acknowledgment}
This work was supported by National Key R$\&$D program of China 2022YFA1004500,
National Natural Science Foundation of China (Grant Nos. 92371107),
and Hong Kong research grant council (16208324).

\bibliographystyle{elsarticle-num}
\bibliography{main}

@article{bird1970direct,
	title = {Direct simulation and the {Boltzmann} equation},
	author = {Bird, Graeme A.},
	journal = {Physics of Fluids},
	volume = {13},
	number = {11},
	pages = {2676--2681},
	year = {1970},
	doi = {10.1063/1.1692849},
}

@article{ivanov1998hypersonic,
	title = {Computational hypersonic rarefied flows},
	author = {Ivanov, M. S. and Gimelshein, S. F.},
	journal = {Annual Review of Fluid Mechanics},
	volume = {30},
	number = {1},
	pages = {469--505},
	year = {1998},
	doi = {10.1146/annurev.fluid.30.1.469},
}

@book{boyd2017nonequilibrium,
	title = {Nonequilibrium Gas Dynamics and Molecular Simulation},
	author = {Boyd, Iain D. and Schwartzentruber, Thomas E.},
	year = {2017},
	publisher = {Cambridge University Press},
	doi = {10.1017/9781139683494},
}

@article{boyd2007hypersonic_continuum,
	title = {Effects of continuum breakdown on hypersonic aerothermodynamics},
	author = {Lofthouse, Andrew J. and Boyd, Iain D. and Wright, Michael J.},
	journal = {Physics of Fluids},
	volume = {19},
	number = {2},
	pages = {027105},
	year = {2007},
	doi = {10.1063/1.2710289},
}

@book{bird1994molecular,
	title = {Molecular Gas Dynamics and the Direct Simulation of Gas Flows},
	author = {Bird, Graeme A.},
	year = {1994},
	publisher = {Oxford University Press},
	address = {Oxford},
	doi = {10.1093/oso/9780198561958.001.0001},
}

@article{broadwell1964study,
	title = {Study of rarefied shear flow by the discrete velocity method},
	author = {Broadwell, James E.},
	journal = {Journal of Fluid Mechanics},
	volume = {19},
	number = {3},
	pages = {401--414},
	year = {1964},
	doi = {10.1017/S0022112064000817},
}

@article{shakhov1968,
	author = {Shakhov, E. M.},
	title = {Generalization of the {Krook} kinetic relaxation equation},
	journal = {Fluid Dynamics},
	volume = {3},
	number = {5},
	pages = {95--96},
	year = {1968},
	doi = {10.1007/BF01029546},
}

@article{zhu_implicit_2019,
	title = {An implicit unified gas-kinetic scheme for unsteady flow in all {Knudsen} regimes},
	volume = {386},
	journal = {Journal of Computational Physics},
	author = {Zhu, Yajun and Zhong, Chengwen and Xu, Kun},
	year = {2019},
	pages = {190--217},
	doi = {10.1016/j.jcp.2019.01.033},
}

@article{zhu_implicit_2016,
	title = {Implicit unified gas-kinetic scheme for steady state solutions in all flow regimes},
	volume = {315},
	journal = {Journal of Computational Physics},
	author = {Zhu, Yajun and Zhong, Chengwen and Xu, Kun},
	year = {2016},
	pages = {16--38},
	doi = {10.1016/j.jcp.2016.03.038},
}

@article{maxwell_1879,
	title = {On stresses in rarified gases arising from inequalities of temperature},
	volume = {170},
	journal = {Philosophical Transactions of the Royal Society of London},
	author = {Maxwell, James Clerk},
	year = {1879},
	pages = {231--256},
	doi = {10.1098/rstl.1879.0067},
}

@article{xu2001gas,
	title = {A gas-kinetic {BGK} scheme for the {Navier--Stokes} equations and its connection with artificial dissipation and {Godunov} method},
	author = {Xu, Kun},
	journal = {Journal of Computational Physics},
	volume = {171},
	number = {1},
	pages = {289--335},
	year = {2001},
	publisher = {Elsevier},
	doi = {10.1006/jcph.2001.6790},
}

@article{xu2010unified,
	title = {A unified gas-kinetic scheme for continuum and rarefied flows},
	author = {Xu, Kun and Huang, Juan-Chen},
	journal = {Journal of Computational Physics},
	volume = {229},
	number = {20},
	pages = {7747--7764},
	year = {2010},
	publisher = {Elsevier},
	doi = {10.1016/j.jcp.2010.06.032},
}

@book{xu2015,
	title = {Direct Modeling for Computational Fluid Dynamics},
	author = {Xu, Kun},
	series = {Advances in Computational Fluid Dynamics},
	year = {2015},
	publisher = {World Scientific},
	doi = {10.1142/9324},
}

@article{li2005kinetic,
	title = {Application of gas-kinetic scheme with kinetic boundary conditions in hypersonic flow},
	volume = {43},
	number = {10},
	journal = {AIAA Journal},
	author = {Li, Qibing and Fu, Song and Xu, Kun},
	year = {2005},
	pages = {2170--2176},
	doi = {10.2514/1.14130},
}

@article{chen2012moving,
	title = {A unified gas kinetic scheme with moving mesh and velocity space adaptation},
	author = {Chen, Songze and Xu, Kun and Lee, Cunbiao and Cai, Qingdong},
	journal = {Journal of Computational Physics},
	volume = {231},
	number = {20},
	pages = {6643--6664},
	year = {2012},
	doi = {10.1016/j.jcp.2012.05.019},
}

@article{liu2014diatomic,
	title = {Unified gas kinetic scheme for diatomic molecular simulations in all flow regimes},
	author = {Liu, Sha and Yu, Pengbo and Xu, Kun and Zhong, Zhaowei},
	journal = {Journal of Computational Physics},
	volume = {259},
	pages = {96--113},
	year = {2014},
	doi = {10.1016/j.jcp.2013.11.030},
}

@article{zhu2017multigrid,
	title = {Unified gas-kinetic scheme with multigrid convergence for rarefied flow study},
	author = {Zhu, Yajun and Zhong, Chengwen and Xu, Kun},
	journal = {Physics of Fluids},
	volume = {29},
	number = {9},
	pages = {096102},
	year = {2017},
	doi = {10.1063/1.4994020},
}

@article{xiao2020velocity,
	title = {A velocity-space adaptive unified gas kinetic scheme for continuum and rarefied flows},
	author = {Xiao, Tianbai and Liu, Chang and Xu, Kun and Cai, Qingdong},
	journal = {Journal of Computational Physics},
	volume = {415},
	pages = {109535},
	year = {2020},
	doi = {10.1016/j.jcp.2020.109535},
}

@article{liu2020unified,
	title = {Unified gas-kinetic wave-particle methods {I}: continuum and rarefied gas flow},
	author = {Liu, Chang and Zhu, Yajun and Xu, Kun},
	journal = {Journal of Computational Physics},
	volume = {401},
	pages = {108977},
	year = {2020},
	publisher = {Elsevier},
	doi = {10.1016/j.jcp.2019.108977},
}

@article{zhu2019unified,
	title = {Unified gas-kinetic wave-particle methods. {II}. multiscale simulation on unstructured mesh},
	author = {Zhu, Yajun and Liu, Chang and Zhong, Chengwen and Xu, Kun},
	journal = {Physics of Fluids},
	volume = {31},
	number = {6},
	pages = {067105},
	year = {2019},
	publisher = {AIP Publishing LLC},
	doi = {10.1063/1.5097645},
}

@article{long2024nonequilibrium,
	title = {Nonequilibrium flow simulations using unified gas-kinetic wave-particle method},
	author = {Long, Wenpei and Wei, Yufeng and Xu, Kun},
	journal = {AIAA Journal},
	volume = {62},
	number = {4},
	pages = {1411--1433},
	year = {2024},
	publisher = {American Institute of Aeronautics and Astronautics},
	doi = {10.2514/1.J063641},
}

@inproceedings{moss2006dsmc,
	title = {{DSMC} simulations of Apollo capsule aerodynamics for hypersonic rarefied conditions},
	author = {Moss, J. N. and Glass, C. E. and Greene, F. A.},
	booktitle = {Proceedings of the 9th AIAA/ASME Joint Thermophysics and Heat Transfer Conference},
	number = {AIAA Paper 2006-3577},
	year = {2006},
	doi = {10.2514/6.2006-3577},
}

@article{zhang2024conservative,
	title = {A conservative implicit scheme for three-dimensional steady flows of diatomic gases in all flow regimes using unstructured meshes in the physical and velocity spaces},
	author = {Zhang, Rui and Liu, Sha and Chen, Jianfeng and Zhuo, Congshan and Zhong, Chengwen},
	journal = {Physics of Fluids},
	volume = {36},
	number = {1},
	pages = {016114},
	year = {2024},
	doi = {10.1063/5.0186520},
}

@article{li2020unified,
	title = {Unified gas-kinetic wave-particle methods {III}: multiscale photon transport},
	author = {Li, Weiming and Liu, Chang and Zhu, Yajun and Zhang, Jiwei and Xu, Kun},
	journal = {Journal of Computational Physics},
	volume = {408},
	pages = {109280},
	year = {2020},
	publisher = {Elsevier},
	doi = {10.1016/j.jcp.2020.109280},
}

@article{liu2021unified,
	title = {Unified gas-kinetic wave-particle methods {IV}: multi-species gas mixture and plasma transport},
	author = {Liu, Chang and Xu, Kun},
	journal = {Advances in Aerodynamics},
	volume = {3},
	number = {1},
	pages = {9},
	year = {2021},
	publisher = {SpringerOpen},
	doi = {10.1186/s42774-021-00062-1},
}

@article{chen2020three,
	title = {A three-dimensional unified gas-kinetic wave-particle solver for flow computation in all regimes},
	author = {Chen, Yipei and Zhu, Yajun and Xu, Kun},
	journal = {Physics of Fluids},
	volume = {32},
	number = {9},
	pages = {096108},
	year = {2020},
	publisher = {AIP Publishing},
	doi = {10.1063/5.0021199},
}

@article{xu2021ugkwp_diatomic,
	title = {Unified gas-kinetic wave-particle methods {V}: diatomic molecular flow},
	author = {Xu, Xiaocong and Chen, Yipei and Liu, Chang and Li, Zhihui and Xu, Kun},
	journal = {Journal of Computational Physics},
	volume = {442},
	pages = {110496},
	year = {2021},
	doi = {10.1016/j.jcp.2021.110496},
}

@article{yang2022ugkwp_twophase,
	title = {Unified gas-kinetic wave-particle method for gas-particle two-phase flow from dilute to dense solid particle limit},
	author = {Yang, Xiaojian and Shyy, Wei and Xu, Kun},
	journal = {Physics of Fluids},
	volume = {34},
	number = {2},
	pages = {023312},
	year = {2022},
	doi = {10.1063/5.0081105},
}

@article{yang2022ugkwp_disperse,
	title = {Unified gas-kinetic wave-particle methods {VI}: disperse dilute gas-particle multiphase flow},
	author = {Yang, Xiaojian and Liu, Chang and Ji, Xing and Shyy, Wei and Xu, Kun},
	journal = {Communications in Computational Physics},
	volume = {31},
	number = {3},
	pages = {669--706},
	year = {2022},
	doi = {10.4208/cicp.OA-2021-0153},
}

@article{yang2023ugkwp_fluidized,
	title = {Unified gas-kinetic wave-particle method for three-dimensional simulation of gas-particle fluidized bed},
	author = {Yang, Xiaojian and Wei, Yufeng and Shyy, Wei and Xu, Kun},
	journal = {Chemical Engineering Journal},
	volume = {453},
	pages = {139541},
	year = {2023},
	doi = {10.1016/j.cej.2022.139541},
}

@article{liu2023ugkwp_radiative,
	title = {An implicit unified gas-kinetic wave-particle method for radiative transport process},
	author = {Liu, Chang and Li, Weiming and Wang, Yanli and Song, Peng and Xu, Kun},
	journal = {Physics of Fluids},
	volume = {35},
	number = {11},
	pages = {112013},
	year = {2023},
	doi = {10.1063/5.0174774},
}

@article{wei2023adaptive,
	title = {Adaptive wave-particle decomposition in {UGKWP} method for high-speed flow simulations},
	author = {Wei, Yufeng and Cao, Junzhe and Ji, Xing and Xu, Kun},
	journal = {Advances in Aerodynamics},
	volume = {5},
	pages = {25},
	year = {2023},
	doi = {10.1186/s42774-023-00156-y},
}

@article{allegre1997forces,
	author = {All{\`e}gre, J. and Bisch, D. and Lengrand, J.-C.},
	title = {Experimental rarefied aerodynamic forces at hypersonic conditions over 70-degree blunted cone},
	journal = {Journal of Spacecraft and Rockets},
	volume = {34},
	number = {6},
	pages = {719--723},
	year = {1997},
	doi = {10.2514/2.3301},
}

@article{long2024implicit,
	title = {An implicit adaptive unified gas-kinetic scheme for steady-state solutions of nonequilibrium flows},
	author = {Long, Wenpei and Wei, Yufeng and Xu, Kun},
	journal = {Physics of Fluids},
	volume = {36},
	pages = {106114},
	year = {2024},
	doi = {10.1063/5.0232275},
}

@article{wei2024adaptiveugks,
	title = {Adaptive unified gas-kinetic scheme for diatomic gases with rotational and vibrational nonequilibrium},
	author = {Wei, Yufeng and Long, Wenpei and Xu, Kun},
	journal = {Computer Physics Communications},
	volume = {305},
	pages = {109324},
	year = {2024},
	doi = {10.1016/j.cpc.2024.109324},
}

@article{zhang2025memory,
	title = {An efficiency and memory-saving programming paradigm for the unified gas-kinetic scheme},
	author = {Zhang, Yue and Wei, Yufeng and Long, Wenpei and Xu, Kun},
	journal = {Computer Physics Communications},
	volume = {314},
	pages = {109684},
	year = {2025},
	doi = {10.1016/j.cpc.2025.109684},
}

@article{wei2026reactive,
	title = {Unified gas-kinetic scheme for reactive flow with multi-scale transport and chemical non-equilibrium},
	author = {Wei, Yufeng and Cao, Junzhe and Xu, Kun},
	journal = {Journal of Computational Physics},
	volume = {546},
	pages = {114514},
	year = {2026},
	doi = {10.1016/j.jcp.2025.114514},
}

@article{wei2024diatomic,
	title = {Unified gas-kinetic wave-particle methods {VII}: diatomic gas with rotational and vibrational nonequilibrium},
	author = {Wei, Yufeng and Zhu, Yajun and Xu, Kun},
	journal = {Journal of Computational Physics},
	volume = {497},
	pages = {112610},
	year = {2024},
	doi = {10.1016/j.jcp.2023.112610},
}

@article{yang2024polydisperse,
	title = {Unified gas-kinetic wave-particle method for polydisperse gas-solid particle multiphase flow},
	author = {Yang, Xiaojian and Shyy, Wei and Xu, Kun},
	journal = {Journal of Fluid Mechanics},
	volume = {983},
	pages = {A37},
	year = {2024},
	doi = {10.1017/jfm.2024.80},
}

@article{pu2025plasma,
	title = {Unified gas-kinetic wave-particle method for multiscale flow simulation of partially ionized plasma},
	author = {Pu, Zhigang and Xu, Kun},
	journal = {Journal of Computational Physics},
	volume = {530},
	pages = {113918},
	year = {2025},
	doi = {10.1016/j.jcp.2025.113918},
}

@misc{yang2025waveparticle,
	title = {Wave-particle based multiscale modeling and simulation of non-equilibrium turbulent flows},
	author = {Yang, Xiaojian and Xu, Kun},
	year = {2025},
	eprint = {2503.07207},
	archivePrefix = {arXiv},
	primaryClass = {physics.comp-ph},
	url = {https://arxiv.org/abs/2503.07207},
	doi = {10.48550/arXiv.2503.07207},
}

@article{yang2025ugkwp_radiation,
	title = {Unified gas-kinetic wave-particle method for frequency-dependent radiation transport equation},
	author = {Yang, Xiaojian and Zhu, Yajun and Liu, Chang and Xu, Kun},
	journal = {Journal of Computational Physics},
	volume = {522},
	pages = {113587},
	year = {2025},
	doi = {10.1016/j.jcp.2024.113587},
}

@article{liu2025ugkwp_phonon,
	title = {Unified gas-kinetic wave-particle method for multiscale phonon transport},
	author = {Liu, Hongyu and Yang, Xiaojian and Zhang, Chuang and Ji, Xing and Xu, Kun},
	journal = {Physical Review E},
	volume = {112},
	pages = {065304},
	year = {2025},
	doi = {10.1103/hz9s-5qbm},
}

@article{cao2026adaptive,
	title = {Adaptive criterion and modification of wave-particle decomposition in {UGKWP} method for high-speed flow simulation},
	author = {Cao, Junzhe and Wei, Yufeng and Long, Wenpei and Zhong, Chengwen and Xu, Kun},
	journal = {Computers \& Fluids},
	volume = {305},
	pages = {106896},
	year = {2026},
	doi = {10.1016/j.compfluid.2025.106896},
}

@article{pu2026electromagnetic,
	title = {Electromagnetic flow control in hypersonic rarefied environment},
	author = {Pu, Zhigang and Xu, Kun},
	journal = {Journal of Fluid Mechanics},
	volume = {1033},
	pages = {A37},
	year = {2026},
	doi = {10.1017/jfm.2026.11462},
}

@article{ji2021cgks3d,
	title = {A gradient compression-based compact high-order gas-kinetic scheme on {3D} hybrid unstructured meshes},
	author = {Ji, Xing and Shyy, Wei and Xu, Kun},
	journal = {International Journal of Computational Fluid Dynamics},
	volume = {35},
	number = {7},
	pages = {485--509},
	year = {2021},
	doi = {10.1080/10618562.2021.1991329},
}

@article{zhang2023cgkssliding,
	title = {A high-order compact gas-kinetic scheme in a rotating coordinate frame and on sliding mesh},
	author = {Zhang, Yue and Ji, Xing and Xu, Kun},
	journal = {International Journal of Computational Fluid Dynamics},
	volume = {37},
	number = {2},
	pages = {105--128},
	year = {2023},
	doi = {10.1080/10618562.2023.2178647},
}

@article{palharini2015benchmark,
	title = {Benchmark numerical simulations of rarefied non-reacting gas flows using an open-source {DSMC} code},
	author = {Palharini, R. C. and White, C. and Scanlon, T. J. and Brown, R. E. and Borg, M. K. and Reese, J. M.},
	journal = {Computers \& Fluids},
	volume = {120},
	pages = {140--157},
	year = {2015},
	doi = {10.1016/j.compfluid.2015.07.021},
}

@inproceedings{moss1995dsmc,
	author = {Moss, James N. and Dogra, Virendra K. and Price, Joseph M. and Hash, David B.},
	title = {Comparison of {DSMC} and experimental results for hypersonic external flows},
	booktitle = {30th Thermophysics Conference},
	year = {1995},
	note = {{AIAA} Paper 95-2028},
	doi = {10.2514/6.1995-2028},
}

@article{cao2024nccr,
	title = {Multiple solutions of nonlinear coupled constitutive relation model and its rectification in non-equilibrium flow computation},
	author = {Cao, Junzhe and Liu, Sha and Zhong, Chengwen and Zhuo, Congshan and Xu, Kun},
	journal = {Computers \& Mathematics with Applications},
	volume = {174},
	pages = {1--17},
	year = {2024},
	doi = {10.1016/j.camwa.2024.08.017},
}

@article{schouler2020survey,
	title = {Survey of flight and numerical data of hypersonic rarefied flows encountered in {Earth} orbit and atmospheric reentry},
	author = {Schouler, Marc and Pr\'{e}vereaud, Ysolde and Mieussens, Luc},
	journal = {Progress in Aerospace Sciences},
	volume = {118},
	pages = {100638},
	year = {2020},
	doi = {10.1016/j.paerosci.2020.100638},
}

@article{boylan1967aerodynamics,
	title = {Aerodynamics of typical lifting bodies under conditions simulating very high altitudes},
	author = {Boylan, D. E. and Potter, J. L.},
	journal = {AIAA Journal},
	volume = {5},
	number = {2},
	pages = {226--232},
	year = {1967},
	doi = {10.2514/3.3946},
}

@inproceedings{palacios2014su2,
	title = {Stanford University Unstructured ({SU2}): Analysis and design technology for turbulent flows},
	author = {Palacios, Francisco and Economon, Thomas D. and Aranake, Aniket C. and Copeland, Sean R. and Lonkar, Amrita K. and Lukaczyk, Trent W. and Manosalvas, David E. and Naik, Kedar R. and Padr{\'o}n, Santiago A. and Tracey, Brendan D. and Variyar, Anil and Alonso, Juan J.},
	booktitle = {52nd AIAA Aerospace Sciences Meeting},
	number = {AIAA 2014-0243},
	year = {2014},
	doi = {10.2514/6.2014-0243},
}

@inproceedings{padilla2009assessment,
	title = {Assessment of Rarefied Hypersonic Aerodynamics Modeling and Windtunnel Data},
	author = {Padilla, Jose F. and Boyd, Iain D.},
	booktitle = {9th AIAA/ASME Joint Thermophysics and Heat Transfer Conference},
	number = {AIAA 2006-3390},
	year = {2006},
	doi = {10.2514/6.2006-3390},
}

@article{jin2026computational,
	title = {Computational and Experimental Study of Rarefied Aerodynamic Forces of Blunted Cone},
	author = {Jin, Xuhong and Liu, Chunfeng and Wang, Xuefeng and Cheng, Xiaoli and Wang, Qiang and Wang, Bing},
	journal = {AIAA Journal},
	volume = {64},
	number = {1},
	pages = {468--477},
	year = {2026},
	doi = {10.2514/1.J065362},
}

@article{greenshields2012rarefied,
	title = {Rarefied hypersonic flow simulations using the {Navier}--{Stokes} equations with non-equilibrium boundary conditions},
	author = {Greenshields, Christopher J. and Reese, Jason M.},
	journal = {Progress in Aerospace Sciences},
	volume = {52},
	pages = {80--87},
	year = {2012},
	doi = {10.1016/j.paerosci.2011.08.001},
}

\end{document}